\documentclass{aa}  
\usepackage{graphicx}
\usepackage{txfonts}
\usepackage{hyperref}
\usepackage{xcolor}

\defcitealias{tramonte23}{T23}

\begin{document} 

\title{A comparative test of different pressure profile models in clusters of galaxies using recent ACT data}

\author{Denis Tramonte\inst{1}}

\institute{\centering Department of Physics, Xi'an Jiaotong--Liverpool University, 111 Ren’ai Road, \\Suzhou Dushu Lake Science and Education Innovation District, Suzhou Industrial Park, Suzhou 215123, People's Republic of China \\\email{Denis.Tramonte@xjtlu.edu.cn}            }


 
  \abstract
   {The electron pressure profile is a convenient tool to characterize the thermodynamical state of a galaxy cluster, with several studies adopting a ``universal'' functional form.}
   {This study aims at using Sunyaev-Zel'dovich (SZ) data to test four different functional forms for the cluster pressure profile: generalized Navarro-Frenk-White (gNFW), $\beta$-model, polytropic, and exponential. The goal is to assess to what level they are universal over a population-level cluster sample.  }
   {A set of 3496 ACT--DR4 galaxy clusters, spanning the mass range 
   $[10^{14},10^{15.1}]\,\text{M}_{\odot}$ and the redshift range $[0,2]$, 
is stacked on the ACT--DR6 Compton parameter $y$ map over $\sim13,000\,\text{deg}^2$. 
An angular Compton profile is then extracted and modeled using the theoretical pressure recipes, whose free parameters are constrained against the measurement via a multi-stage MCMC approach. The analysis is repeated over cluster subsamples spanning smaller mass and redshift ranges.} 
   {All functional forms are effective in reproducing the measured $y$ profiles within their error bars, without a clearly favored model. While best-fit estimates are in broad agreement with previous findings, hints of residual subsample dependency are detected favoring higher amplitudes and steeper profiles in high-mass, low-redshift clusters.}
   {Population-level cluster studies based on SZ data alone are likely unable to accurately constrain different pressure profile models. Residual trends at population level and scatter at individual cluster level undermine the universal pressure model assumption whenever high precision is required.
   Finally, functional forms different from the gNFW prove equally effective while being more physically motivated.}


   \maketitle
%

\begin{figure*}
\includegraphics[trim= 0mm 0mm 0mm 0mm, scale=0.45]{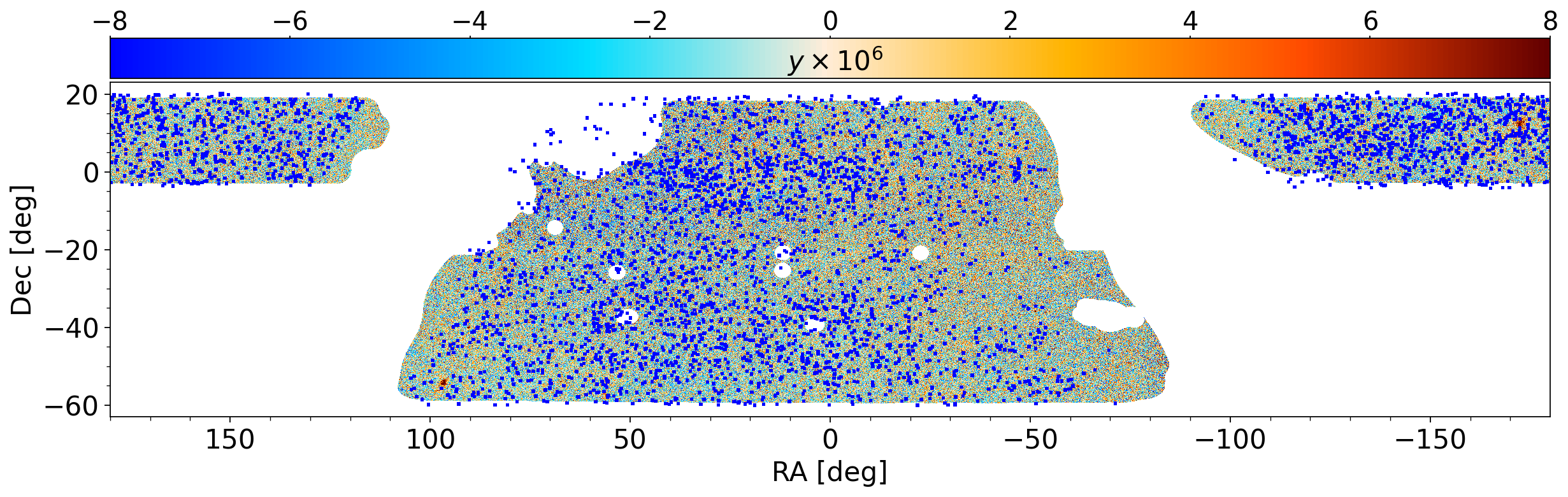}
\caption{Visual representation of the data set employed in this study. The ACT DR6 Compton 
	parameter map is plotted on an equirectangular projection, the white region 
	highlighting the mask footprint. Blue spots show the positions of the 4195 clusters 
	from the ACT DR5 catalog.}
\label{fig:data}
\end{figure*}
 
\section{Introduction}
\label{sec:introduction}

As the largest virialized structures in our Universe, clusters of galaxies encode a wealth 
of information about cosmology and astrophysics alike~\citep{voit05, allen11}. The task of 
linking cluster observational properties with theoretical models generally requires a proper 
characterization of the intracluster medium (ICM), which contains the majority of a cluster 
baryonic mass ($\lesssim90\%$) in the form of a hot ($10^7$--$10^8\,\text{K}$) and diffuse 
($10^{-4}$--$10^{-2}\,\text{cm}^{-3}$) plasma composed primarily of ionized hydrogen and 
helium~\citep{markevitch07}. 

The high temperature of the ICM has traditionally been exploited by observations in X-ray 
wavelengths~\citep{sarazin88, voges99, hicks08, mehrtens12, bulbul24}; in more recent years, 
cluster surveys in radio wavelengths have also gained relevance~\citep{birkinshaw99, carlstrom02}, 
by targeting the CMB spectral distortion signature of the Sunyaev-Zel'dovich (SZ) 
effect~\citep{sunyaev72}. The latter is proportional to the Compton parameter $y$, which 
quantifies the line-of-sight (LoS) integrated electron pressure. SZ observations have been used 
not only to compile cluster catalogs of increasing size~\citep{planck_15_xxvii, hilton21, kornoelje25}, 
but also to construct maps of the Compton parameter over extended regions of the 
sky~\citep{planck_15_xxii, coulton24}.

From a theoretical point of view, it is customary to model the ICM as a spherically symmetric 
gas distribution, whose properties depend on the nominal radial separation from the cluster 
center. By combining the radial dependence of the ICM temperature and density, and considering 
that the ICM is ionized, the electron pressure profile is the most suitable physical quantity 
to characterize a cluster thermodynamical state. While real systems are expected to show 
deviations from this spherical symmetry, due for example to asphericities or local 
turbulence~\citep{rainer17, adam25}, radial profiles can still provide a mean baseline effective 
in reproducing integrated quantities such as the total cluster mass or the total X-ray/SZ fluxes, 
and as such are fundamental ingredients in scaling relations between cluster masses and their 
observational proxies~\citep{kravtsov06, vikhlinin09, planck_ir_v, gallo24}. 
More crucially, if gravity dominates the cluster formation process, 
theory predicts the ICM to be approximately self-similar, implying that clusters of different masses and redshifts can 
be mapped onto each other through simple scaling relations for density, temperature, and pressure~\citep{kaiser86, arnaud10}.
It is therefore interesting to test this self-similarity scenario and 
assess whether the same functional form for the radial pressure profile can be adopted to model 
the ICM in clusters with different masses and at different redshifts. This would imply the existence 
of a ``universal'' recipe, in which the electron pressure depends exclusively on a scaled separation 
from the cluster center, and in which the dependence on the cluster mass and redshift 
can be conveniently factorized out as a global scaling function. 

Several studies in the literature have focused on a specific version of such a model, generally 
referred to as the ``universal pressure profile''~\citep[UPP,][]{nagai07, arnaud10} and obtained 
as a generalization of the Navarro-Frenk-White (NFW) parametrization used to model dark matter 
profiles from simulations~\citep{navarro97}; the aforementioned studies tested the UPP against
cluster observations to constrain the parameters entering its functional form. 
Works targeting reduced samples of high mass 
clusters~\citep{nagai07, arnaud10, planck_ir_v, pointecouteau21, he21} benefit from high quality 
data of well resolved systems; their results, however, are somewhat limited in their range of 
applicability and most likely affected by sample biases. More worringly, the existing best-fit 
estimates for the UPP parameters show a large scatter across different studies, hinting at the 
existence of strong degeneracies which hamper a solid physical interpretation of individual 
parameter values. On the other end of the scale, large population 
studies~\citep{gong19, tramonte23} can provide results with a broader range of applicability, 
but are more affected by possible systematics in the data set, and by a loss of precision 
when the average profiles are applied to individual objects. 

The study in~\citet[][hereafter T23]{tramonte23} was the first to test the universal pressure 
profile model on a complete and representative sample of galaxy clusters. The sample merged the 
contributions of different existing cluster catalogs derived from optical observations, for a total 
of 23,820 objects spanning the mass range $[10^{14},10^{15.1}]\,\text{M}_{\odot}$ and the redshift 
range $[0.02,0.98]$. The study measured the averaged cluster Compton parameter profile by stacking 
these clusters on the Compton parameter maps based on \textit{Planck}~\citep{planck_15_xxii} and 
Atacama Cosmology Telescope~\citep[ACT,][]{madhavacheril20} data. The UPP was then tested by 
comparing its predicted $y$ profiles with the measurements; the study also considered cluster 
subsamples to explore possible evolutions of the model with different cluster mass and redshift 
regimes. Overall, the fitted profile model was effective in reproducing the SZ measurements in 
all cases, the best-fit parameter values being broadly consistent with previous 
results in the literature. Although individual parameter values were found to 
depend on the chosen cluster subsample, the study did not provide any compelling evidence for 
a residual dependence on the cluster mass and redshift.

The present study is a new attempt at testing theoretical models for the ICM on Compton parameter 
profiles extracted from cluster stacks on $y$-maps, considering a population-level sample 
($\sim 10^3$ clusters) and recent ACT data. While the methodology is similar to the 
one followed in~\citetalias{tramonte23}, there are some important differences. First of all, the ACT 
$y$-map adopted in this study is considerably larger than the one used in~\citetalias{tramonte23}, 
by a factor $\sim6$ in the footprint area. Secondly, this study considers a homogeneous cluster 
catalog which was constructed via a blind search on the same SZ data employed to construct the 
$y$ map; the cluster sample used in~\citetalias{tramonte23}, on the contrary, was a heterogeneous 
combination of different catalogs built from optical data, and as such suffered from possible 
systematic effects in the mass definition\footnote{Although a substantial part of the analysis 
in~\citetalias{tramonte23} was dedicated to correct possible biases deriving from different mass 
definitions and to show that their potential effects are largely below the final error bars, the 
work still acknoweldged this as one potential source of systematics.}. Finally, the scope of this 
work is broader, as it considers not only the UPP parametrization, but also different recipes for 
modeling the electron pressure distribution in the ICM. Once more, the goal is first of all to test 
whether these models can be effective in reproducing observed features in the cluster SZ emission, 
and subsequently to assess possible residual dependencies of the model parameters on the cluster 
mass and redshift (as deviations from universality).

This paper is organized as follows. Sec.~\ref{sec:data} describes the data set used in this 
study. The $y$-profile measurement procedure and results are presented in Sec.~\ref{sec:measurement}, 
while the theoretical modeling is detailed in Sec.~\ref{sec:modeling}. Sec.~\ref{sec:inference} 
explains the chosen approach for parameter estimation with its results, whereas 
Sec.~\ref{sec:discussion} discusses the findings. Finally, conclusions are drawn in 
Sec.~\ref{sec:conclusions}. The present study adopts a spatially flat $\Lambda$CDM cosmological 
model with fiducial parameter values $h=0.674$, $\Omega_{\rm m}=0.315$, $\Omega_{\rm b}=0.0493$, 
$\sigma_8=0.811$ and $n_{\rm s}=0.965$~\citep{planck_18_vi}.

\begin{figure*}
\includegraphics[trim= 0mm 0mm 0mm 0mm, scale=0.36]{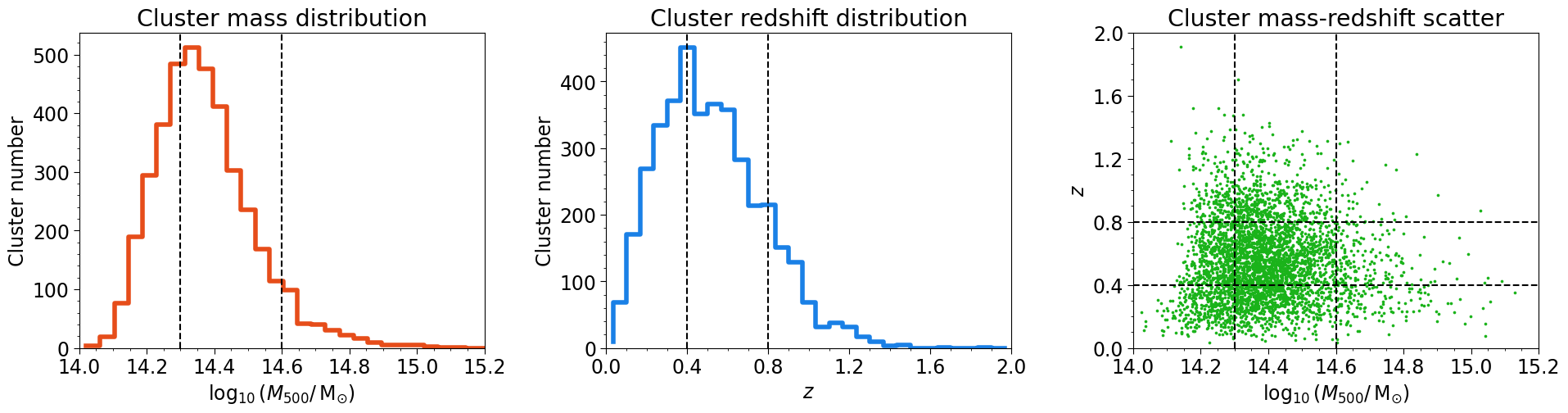}
\caption{Left and middle panel show, respectively, the distribution in mass and 
	redshift of the full cluster sample adopted in this study. The right panel 
	shows instead the catalog scatter across these two quantities. In all 
	panels, dashed lines mark the edges of the chosen bins in the corresponding quantity.}
\label{fig:distributions}
\end{figure*}


\section{Data}
\label{sec:data}
The Compton parameter map and cluster catalog considered in the present analysis 
are described in the following; Fig.~\ref{fig:data} provides a visual representation 
of the combined data set.

\subsection{Compton parameter map}
\label{ssec:ymap}
This study employs the Compton parameter map described in~\citet{coulton24}, which is 
publicly available at the NASA Legacy Archive for Microwave Background Data (LAMBDA) page for AdvACT 
products\footnote{\url{https://lambda.gsfc.nasa.gov/product/act/actadv_dr6_compton_maps_info.html}.}.
The map was generated via a blind component separation pipeline, an adapted version of 
the internal linear combination (ILC) method in wavelet frame, applied to individual 
frequency maps from \textit{Planck} NPIPE and from ACT DR4 and DR6, the two experiments 
being complementary in terms of spatial resolution. The final footprint is constrained by 
ACT observing area to approximately one third of the sky, or $\sim13,000\,\text{deg}^2$, and the 
final map resolution is $1.6\,\text{arcmin}$; the map is delivered in a flat \textit{plate carrée} 
(equirectangular) projection, on a 2D frame with axes parallel to the right ascension (RA) and 
declination (Dec) directions, and pixel uniform size of 0.5' in both RA and Dec. The present 
study also makes use of the associated sky mask, also provided as part of the data release, 
which combines the \textit{Planck} Galactic plane mask covering 30\% of the sky, an ACT 
footprint mask, and a mask of the bright extended sources in the individual frequency maps. 
 
As discussed in~\citet{coulton24}, cosmic infrared background (CIB) residuals in the Compton 
map can be a potential source of contamination. Such residuals can be modeled as 
modified blackbodies in their frequency spectral energy distribution (SED), and subtracted. 
The result is a set of CIB-deprojected Compton parameter maps, also available as part of the 
data release, which differ on the adopted values of the parameters entering the CIB SED functional 
form. The reference also warns, however, that existing uncertainty on the actual CIB SED can 
result in incomplete removal of this contaminant from the final map; besides, further 
deprojections come at the price of possibly removing part of the target tSZ signal. For these 
reasons, the standard $y$-map (without any CIB deprojection) is adopted as the fiducial map in 
the present study. The impact of possible CIB residuals on the results of this study is discussed 
in Appendix~\ref{sec:cib_effect}, where it is shown that, at the reconstructed profile level, 
using deprojected versions of the map yields differences within the overall uncertainties.

\subsection{Cluster catalog}
\label{ssec:catalog}
This study employs the galaxy cluster catalog described in~\citet{hilton21}, which is publicly 
available at the NASA LAMBDA page for ACTPol 
products\footnote{\url{https://lambda.gsfc.nasa.gov/product/act/actpol_dr5_szcluster_catalog_info.html}.}.
Cluster candidates were identified via a blind search on ACT DR5 coadded 98 and 150 GHz maps 
using a multifrequency matched filter. Masking of bright point sources and of regions with 
high dust emission (galactic latitudes $|b|<20^{\circ}$ plus others identified on \textit{Planck} 
353 GHz map), high-flux on the ACT 150 GHz maps, or deemed unfit after visual inspection, 
resulted in an effective search area of $13,211\,\text{deg}^2$. The search yielded 8878 cluster 
candidates detected with signal-to-noise ratio (S/N)>4; of these, 4195 objects were confirmed 
as clusters after cross-matching with optical/IR surveys, which also provided redshift estimates 
for the detections. Spectroscopic redshift estimates are available for 1648 ($\sim$39.3\%) 
objects, the remaining $\sim$60.7\% only quoting photometric redshift estimates. 

Cluster masses were estimated using the scaling relation from~\citet{arnaud10} to convert the 
SZ signal to mass. Estimates are quoted as spherical overdensity masses $M_{500}$, i.e. 
enclosing a region whose mean density is equal to 500 times the critical density of the Universe 
at that redshift. As the aforementioned scaling assumes hydrostatic equilibrium in the ICM, the 
resulting masses are known to underestimate the true cluster 
masses~\citep[see, e.g.][and reference therein]{miyatake19}. 
The catalog also provides bias-corrected mass estimates obtained using a common richness-based 
weak-lensing (WL) mass calibration factor, which shows that SZ-estimated masses underestimate 
WL-estimated ones by $\sim$30\%. The cluster 
mass values $M_{500}$ adopted in this study, however, are the ones quoted in the catalog as obtained 
directly from the SZ measurements, 
without correcting for the bias deriving from the assumptions of hydrostatic equilibrium.

As the optical/IR confirmation results in very high purity (the probability for a detection 
to be an actual cluster), the selection function of this catalog is determined by its 
completeness (the probability of a real cluster to be detected). The latter was evaluated to 
be larger than 90\% for masses $M_{500}>3.8\times10^{14}\text{M}_{\odot}$ at redshift $z=0.5$ 
(the median redshift of the sample). While this average value was found to have a very mild 
redshift dependence, the completeness evaluated in different regions of the cluster search 
areas showed high levels of inhomogeneity, which is a result of ACT observational scanning strategy. 
In the present study, the selection function is naturally accounted for in the theoretical modeling, 
as explained in Sec.~\ref{ssec:ysample}.

Before concluding this section, it is relevant to acknowledge the impact of Eddington bias on the 
catalog employed in this study, and as such on the cluster signals measured in Sec.~\ref{sec:measurement}. 
As the catalog was generated via a blind search on the same data set employed to generate the $y$-map, noise 
fluctuations in the map can potentially result in an overestimation of the measured cluster $y$ signal. The effect 
is especially relevant for low-mass systems, which if located on top of a positive noise fluctuation can 
still enter the catalog even though their true SZ signal is slightly below the detection threshold.  
A complete selection function to correct for this 
effect would need to factor in the effects of map noise, filtering, detection pipeline, intrinsic scatter in the SZ flux to 
mass scaling relation, detection thresholds, and survey inhomogeneity. Such a selection is not available, 
particularly not expressed explicitly as a function of the cluster mass and redshift. 
While the uncorrected Eddington bias does overestimate the amplitude of the reconstructed profiles, it also biases high 
the catalogued cluster mass values adopted to model the profiles theoretically in Sec.~\ref{ssec:ysample}; 
as already stated, these masses are indeed the ones obtained directly from the SZ measurement, uncorrected for any bias 
or selection effect. In other words, the amplitude overestimation 
affects equally the measured and the modeled $y$ profiles. The resulting 
best-fit parameter values are then reasonably robust despite the bias. Furthermore, the conclusions from this study 
are mostly focused on the comparison and viability of different theoretical models for the cluster pressure profile, 
rather than on the search for an ultimate universal set of parameter values.


\section{Measurements}
\label{sec:measurement}

As anticipated in Sec.~\ref{sec:introduction}, cluster Compton parameter profiles are 
the primary observable considered in this study for testing ICM theoretical models. This 
section describes the methodology adopted to reconstruct such profiles, and assesses the results.

\begin{figure*}
\includegraphics[trim= 0mm 0mm 0mm 0mm, scale=0.4]{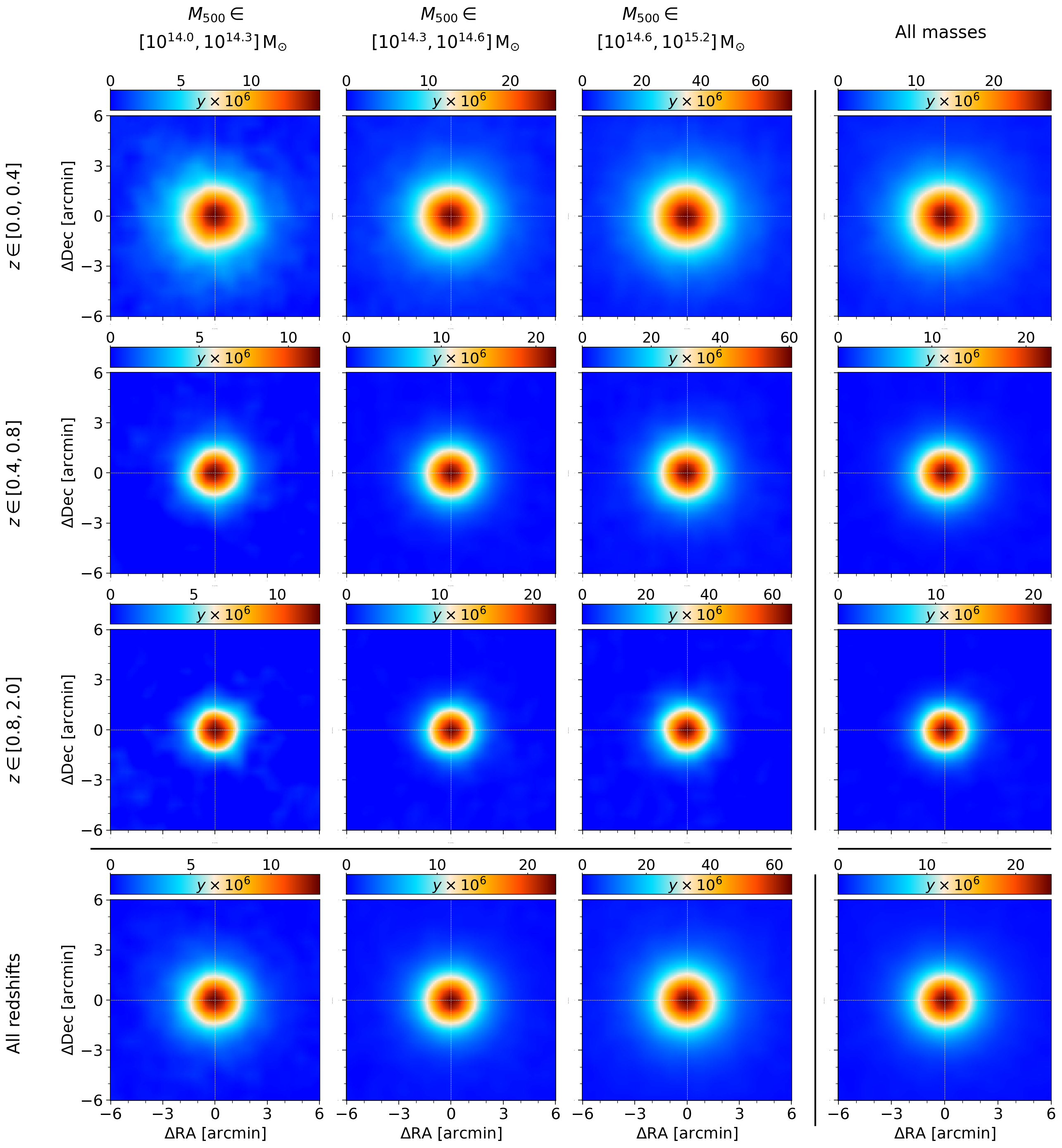}
	\caption{Summary plot showing the stacks on ACT $y$-map of all the 16 cluster 
	samples described in Sec.~\ref{ssec:samples}, with labels clarifying the 
	associated mass and redshift intervals. Although the final stack maps are 
	$30'\times30'$ in size, this figure zooms in the central $12'\times12'$ square to 
	better show details in the cluster signal.}
\label{fig:stacks}
\end{figure*}
\begin{figure}
\includegraphics[trim= 0mm 0mm 0mm 0mm, scale=0.18]{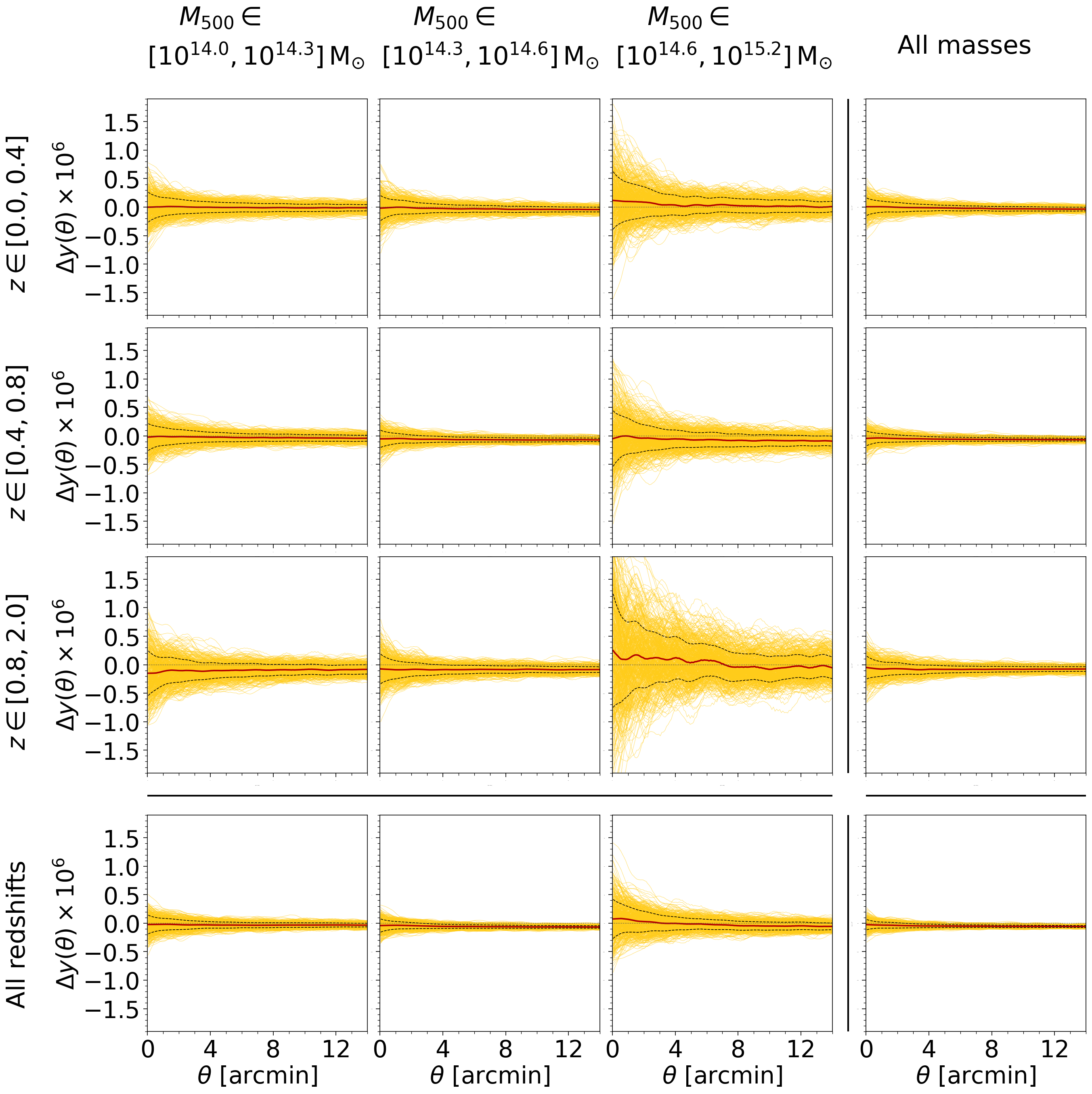}
	\caption{Estimation of the mean background level in the stacks. For each mass and 
	redshift bin, the colored bundle is the set of all 500 profiles extracted from the 
	shuffled stacks; the solid darker line shows instead their average, taken 
	as the zero-level profile, and the black dashed lines the 68\% confidence interval around 
	the mean.}
\label{fig:zerolevel}
\end{figure}

\subsection{Cluster samples}
\label{ssec:samples}
As it is visible from Fig.~\ref{fig:data}, some clusters from the ACT DR5 catalog lie outside 
of the map footprint. The catalog is then queried to select only the clusters that do not 
overlap with the mask described in Sec.~\ref{ssec:ymap}, resulting in a total of 3946 objects. 
This catalog is split into smaller samples spanning different mass and redshift ranges. 
This analysis considers three bins in mass, bounded by the edge points $[14.0,14.3,14.6,15.1]$ 
(in units $\log_{10}(M_{500}/\text{M}_{\odot})$, with $M_{500}$ the bias-uncorrected catalog masses), 
and three bins in redshift, bounded by the 
edge points $[0.0,0.4,0.8,2.0]$. The catalog is then binned in both mass and redshift over this 
grid, resulting in nine non-overlapping samples (one for each combination of a mass bin and a 
redshift bin), six marginalized cases (the full mass span of the catalog for each redshift bin, 
and the full redshift span of the catalog for each mass bin) and the complete unmasked catalog, 
for a total of 16 different samples. 

Fig.~\ref{fig:distributions} shows the catalog distributions in mass and redshift, highlighting 
the edges of the chosen bins. Due to the highly non-uniform trends of these distributions, each 
$(M,z)$ bin has a substantially different number of clusters. The right panel in 
Fig.~\ref{fig:distributions} also shows some degree of correlation 
between the two quantities. The most massive clusters are generally found at the lowest 
redshifts, as expected in a bottom-up scenario for structure formation. It is also possible to 
notice a slight rightward shift on the point distribution for higher redshifts, which is likely
an evidence of Malmquist bias affecting the 
cluster catalog. Again, these features are properly accounted for as selection effects in the 
theoretical modeling (Sec.~\ref{ssec:ysample}).

\subsection{Stacks}
\label{ssec:stacks}
Each of the 16 samples is stacked on the ACT map to obtain a measurement of the integrated 
cluster emission. The stacking procedure is ultimately a weighted average which employs the 
Compton parameter map, considered as the ``signal'' map, and the mask, considered as the ``weight'' 
map whose pixels have values either 0 or 1\footnote{Although the map is initially apodized,
every mask pixel with value lower than 1 is set equal to 0.}. For the generic $i$-th cluster, a 
square region centered on its nominal position is extracted from both the $y$ map and the weight 
map, resulting in two submaps $Y_i$ and $W_i$, respectively. The final, stacked map is obtained as:
\begin{equation}
	\label{eq:stacking}
	\bar{Y}=\frac{1}{W}\sum_{i=1}^{N} W_i\,Y_i, \qquad W=\sum_{i=1}^{N} W_i,
\end{equation}
where $N$ is the number of clusters in the considered sample and $W$ is the total weight of the stack.

\begin{figure}
\includegraphics[trim= 0mm 0mm 0mm 0mm, scale=0.19]{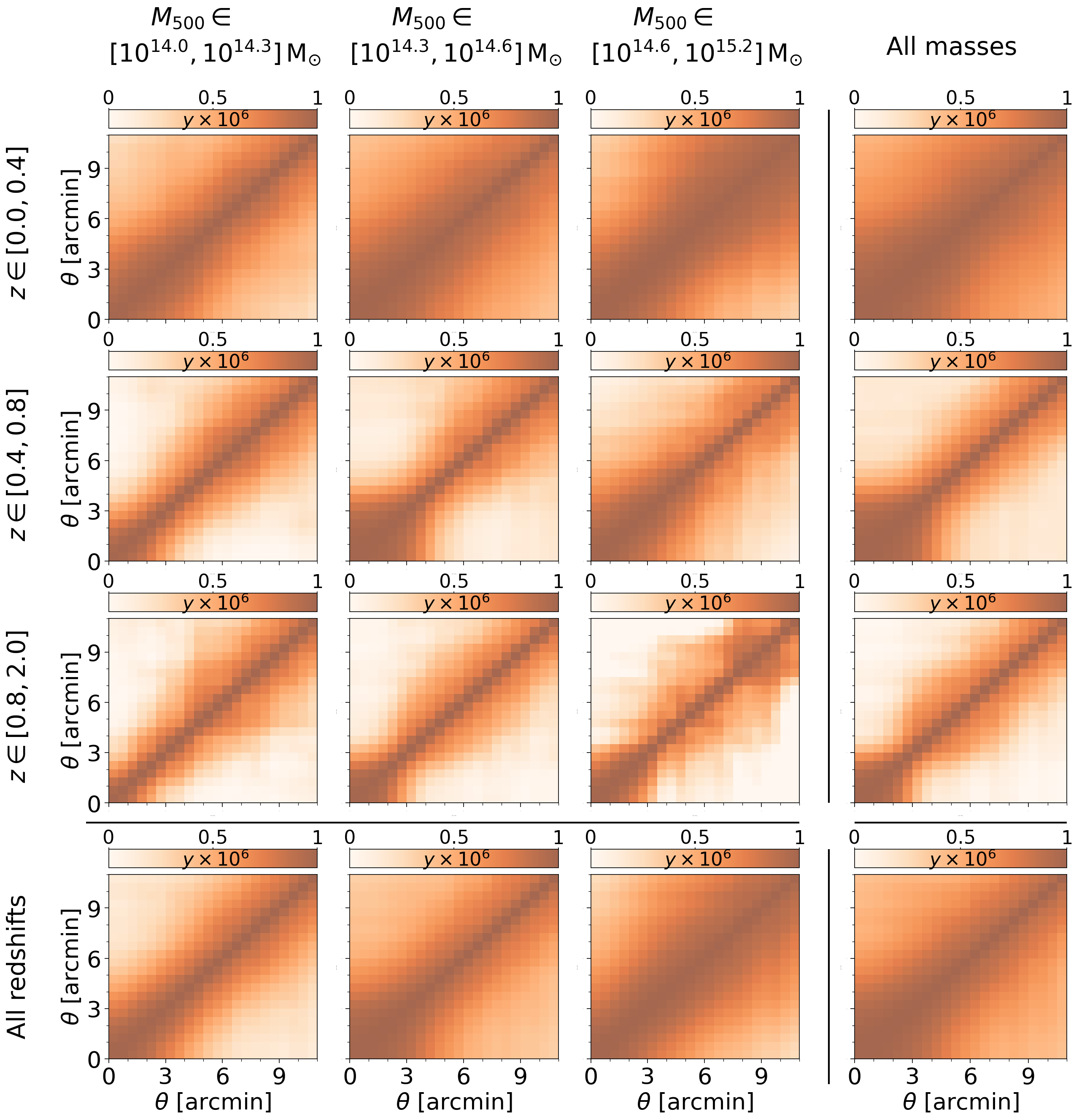}
	\caption{Correlation matrices associated to the reconstructed Compton parameter 
	profiles over the relevant angular span, for all mass and redshift bins.}
\label{fig:correlations}
\end{figure}

Because the ACT Compton parameter map is delivered in an equirectangular projection with a 
uniform pixelization in both coordinate axes, pixels located farther from the celestial equator 
subtend a smaller solid angle on the sky. More precisely, a pixel with generic declination $\delta$ 
spans an effective RA range shorter by a factor $\cos{\delta}$ compared to a pixel on the equator 
 ($\delta=0$). As the $y$ map extends down to latitudes of $\sim-60^{\circ}$, this implies that 
 pixels at the lower edge of the map would actually subtend a solid angle which is about half of 
 their nominal size. If not corrected for, this effect would produce spurious elongations along 
 the RA direction of any features within the final stacks; such ellipticities would in turn 
result in an artificial stretching of the associated angular profiles. 

This effect is corrected for at the level of the individual cluster submaps $Y_i$ and $W_i$, prior 
to the stack sums in eq.~\ref{eq:stacking}. In those submaps, the RA separation of each pixel 
from the cluster is multiplied by the cosine of the cluster declination; the submaps are then 
projected onto a common frame consisting of a square map centred on the cluster nominal coordinates, 
with size $30'\times30'$ and 320 pixels in each axis. The extension of this frame was chosen to 
ensure the final stacks are able to capture all features of the cluster emission, down to its 
outskirts; the resolution was chosen to be high enough not to lose information from the resolution 
of the parent ACT Compton map. Clearly, the shape of the region initially trimmed out of the ACT 
map was not a square, but a rectangle whose RA side was $30'/\cos{\delta_{\rm c}}$, where 
$\delta_c$ is the declination of the considered cluster. This ensured that, after 
the declination correction, the rectangle would fit exactly into the reference stack frame.

This approach is an important upgrade compared to the stacking procedure adopted 
in~\citetalias{tramonte23}, where all pixels were assumed to have equal area and individual 
cluster regions were simply trimmed out of the map and stacked; that study, however, employed the 
ACT-DR4 $y$ maps, which did not extend more than $\sim 19^{\circ}$ away from the equator; this 
results in the declination correction factor being always below $\sim 5\%$, which justifies the 
approximation adopted in that work. 

A summary table of the stacks obtained for all the 16 cluster samples is shown in 
Fig.~\ref{fig:stacks}. In order to better show the features in the averaged cluster signal, 
each color scale is saturated to the extreme values in the corresponding map, and the 
plot area focuses on a smaller $12'\times12'$ around the center. In all cases, the cluster 
emission is clearly detected with a high contrast against the background; we can notice that the 
contrast is the highest for the top mass bin, which yields the highest peak 
amplitude. We can also notice that higher redshift stacks appear smaller in angular size, an 
effect due to the increase of the angular diameter distance within the probed redshift range. 
Altough some irregularities affect the signal contours, the declination correction 
performed during the stack results in overall round profiles with no clear evidence of 
residual ellipticities. 

\begin{figure} 
\includegraphics[trim= 0mm 0mm 0mm 0mm, scale=0.34]{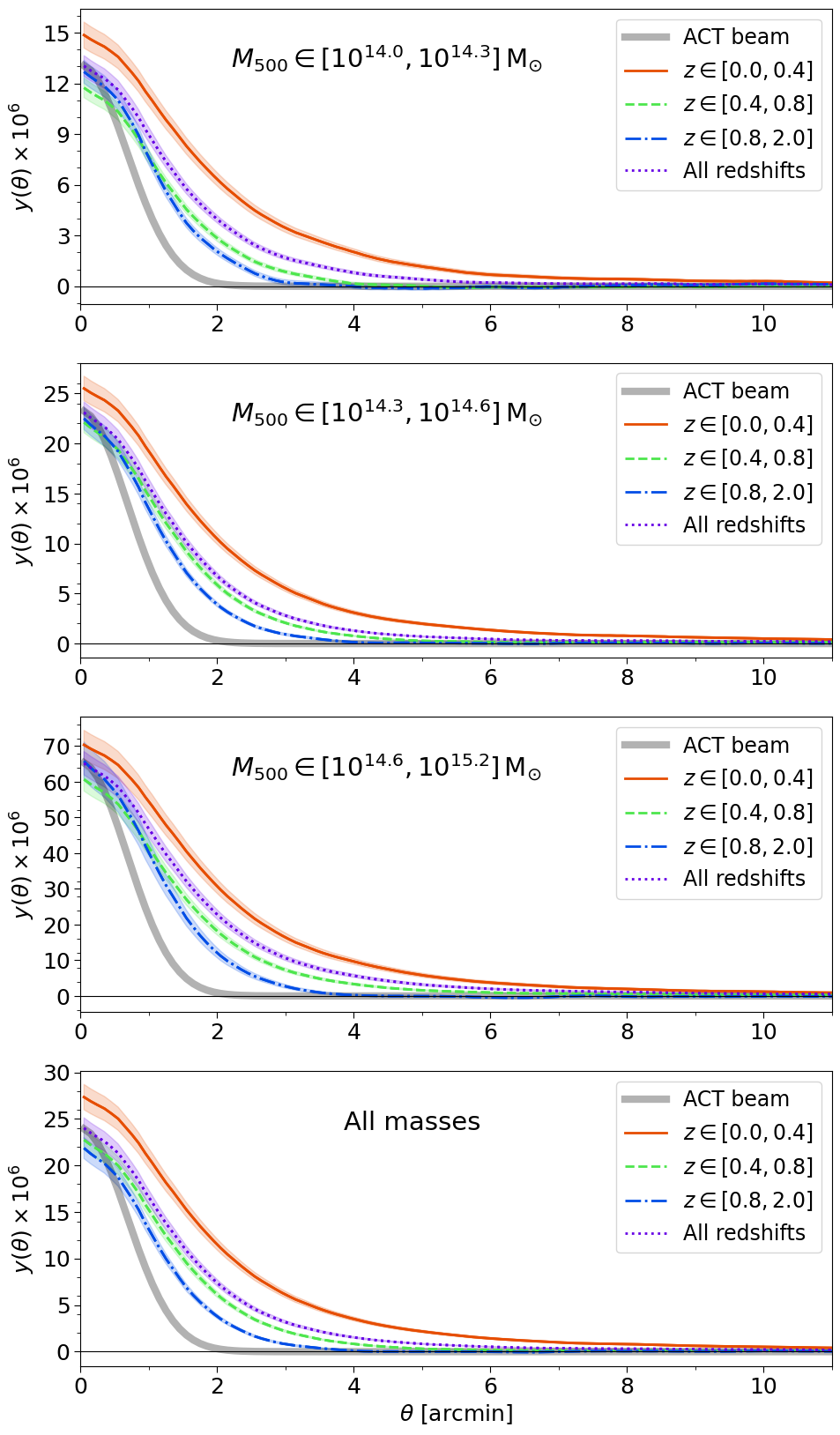}
	\caption{Circularly symmetrized Compton parameter angular profiles extracted from 
	the 16 stacks shown in Fig.~\ref{fig:stacks}, already background subtracted and 
	showing the 1-sigma uncertainty per bin as a shaded region. Profiles are grouped 
	according to their mass bin, showing different redshift bins in each panel; the 
	ACT beam profile is also included for comparison.}
\label{fig:profiles}
\end{figure}

\subsection{Profiles}
\label{ssec:profiles}

A circularly symmetrized angular Compton parameter profile $y(\theta)$ is extracted from 
each of the 16 stacks. Pixels in each stack map are assigned to bins in the radial separation 
from the center, and the average of all the pixels within the $i$-th ring is taken as the 
value of the Compton profile at the corresponding mean angular separation, $y(\theta_i)$. 
The size of each angular separation bin is set to 0.5', the same as the resolution of the 
original ACT Compton parameter map. 

As we are interested in the $y$ amplitude with respect to the mean map background level, 
from each of the 16 profiles a zero-level profile is subtracted, determined as follows. 
For each of the 16 samples, the associated stack is repeated 500 times (following the same 
procedure described in Sec.~\ref{ssec:stacks}), but this time shuffling the coordinates 
of the clusters within the allowed Compton map footprint; this results in 500 
randomized-amplitude profiles, whose average is taken as the zero-level profile to be 
subtracted from the real cluster profiles. Fig.~\ref{fig:zerolevel}
shows the result of this approach: for each sample, individual lines are the profiles extracted 
from each randomized stack, the thicker central line shows their average, and the dotted lines 
the 1-sigma confidence interval. The figure shows that all average profiles are well compatible 
with zero; as expected, we observe a larger scatter in the samples containing a lower number of 
clusters, an effect particularly evident in the largest mass bin. Although the average profiles 
are not exactly zero, and as such they are to be subtracted from the real cluster profiles, 
their mean value is negligible compared to the peak amplitude of the profiles 
extracted from the stacks. In summary, this test shows that there is no relevant zero-level 
systematics affecting our stacks. 

Uncertainties on the measured profiles, and correlation values between different angular bins, 
are evaluated via covariance matrices. For each cluster sample, another set of 500 randomized 
stacks is produced; this time each new cluster sample is generated via a bootstrap approach by 
randomly selecting, from the corresponding original $(M,z)$ sample, the same number of clusters 
with possibility of repetition. The covariance matrix $\rm{Cov}(\theta_i,\theta_j)$ between the 
two angular bins $\theta_i$ and $\theta_i$ is then obtained as:
\begin{equation}
	C(\theta_i,\theta_j) = \dfrac{1}{N_{\rm rand}} \sum_{k=1}^{N_{\rm rand}} [y_k(\theta_i)-\bar{y}(\theta_i)] [y_k(\theta_j)-\bar{y}(\theta_j)],
\end{equation}
where $N_{\rm rand}=500$ is the number of random realizations, the index $k$ identifies the individual 
random stack, and $\bar{y}(\theta_i)$ is the average value of the profile at the angular bin 
$\theta_i$, across all random realizations:
\begin{equation}
\bar{y}(\theta_i) = \dfrac{1}{N_{\rm rand}} \sum_{k=1}^{N_{\rm rand}} y_k(\theta_i).
\end{equation}
The correlation matrix $\text{Corr}(\theta_i,\theta_j)$ can be computed normalizing the covariance 
matrix as in:
\begin{equation}
	\text{Corr}(\theta_i,\theta_j) = \frac{\text{Cov}(\theta_i,\theta_j)}{\text{Cov}(\theta_i,\theta_i)\,\text{Cov}(\theta_j,\theta_j)}.
\end{equation}
The correlation matrices for all 16 stacks are shown in Fig.~\ref{fig:correlations}, and are a 
measurement of the level of statistical independence between bin pairs. The figure shows that residual 
correlations are still present even for largely separated bins, the effect being more important for 
the lowest redshift samples which yield the broader profiles. 

\begin{table*}
\centering
\caption{Summary table reporting, for each combination of a mass and a redshift bin, the number of clusters $N_{\rm cl}$ in the bin and the detection significance per bin $\chi^2_{\rm b}$ of the corresponding Compton parameter profile.}
\label{tab:stats}
\setlength{\tabcolsep}{1.3em}
\renewcommand{\arraystretch}{1.5}
\begin{tabular}{c|cccc}
\hline
 & $M_{\rm 500c} \in $ & $M_{\rm 500c} \in$ & $M_{\rm 500c} \in $ & All  \\ 
 & $[10^{14.0},10^{14.3}]\,\text{M}_{\odot}$ & $[10^{14.3},10^{14.6}]\,\text{M}_{\odot}$ & $[10^{14.6},10^{15.1}]\,\text{M}_{\odot}$ &  masses \\ 
\cline{2-5}
 & $N_{\rm cl},\quad \chi^2_{\rm b}$  &  $N_{\rm cl},\quad \chi^2_{\rm b}$  &  $N_{\rm cl},\quad \chi^2_{\rm b}$  &  $N_{\rm cl},\quad \chi^2_{\rm b}$  \\
\hline
$z \in [0.0,0.4]$ & $382,\quad 18.7$ & $ 701,\quad 19.7$ & $ 141,\quad 14.4$ & $1224,\quad 17.8$ \\ 
$z \in [0.4,0.8]$ & $ 477,\quad 19.7$ & $1381,\quad 20.8$ & $ 162,\quad 16.0$ & $2020,\quad 19.6$ \\ 
$z \in [0.8,2.0]$ & $ 193,\quad 17.7$ & $ 475,\quad 19.0$ & $  34,\quad 11.0$ & $ 702,\quad 17.1$ \\ 
All redshifts & $1052,\quad 20.7$ & $2557,\quad 21.2$ & $ 337,\quad 17.8$ & $3946,\quad 20.6$ \\ 
\hline
\end{tabular}
\end{table*}

The profiles for all 16 stacks are shown in Fig.~\ref{fig:profiles}, grouping different redshift 
bins for each individual mass bin; these profiles are already zero-level subtracted, and show the 
1-sigma uncertainty as a shaded region. The latter is estimated, for each bin $\theta_i$, as the 
square root of the corresponding diagonal entry in the covariance matrix, 
$\sigma(\theta_i)=\sqrt{\text{Cov}(\theta_i,\theta_i)}$. Exclusively for visualization purposes, 
these profiles (and their uncertainties) are evaluated with a finer angular resolution, choosing a 
bin size of $0.1'$ (the same consideration holds for the background estimation profiles in 
Fig.~\ref{fig:zerolevel}). Notice that, despite the stack maps extend out to angular separations 
of $\sim15'$ from the center, these plots show the profiles only up to 11', as in all cases this 
separation is enough to make the Compton parameter amplitude compatible with the zero level; this 
upper limit of 11' is maintained also in the modeling part of the present study. All panels in 
Fig.~\ref{fig:profiles} also include the ACT beam, whose angular profile is computed assuming a 
perfectly Gaussian shape with $\text{FWHM}=1.6'$, showing that all Compton profiles are resolved with 
the resolution of the ACT $y$-map.

To conclude this section, following the standard approach for correlated data, the significance of the measurement
of each profile can be quantified via a chi-square as in:
\begin{equation}
	\label{eq:significance}
	\chi^2 = \sum_{i=1}^{N_{\rm b}}\sum_{j=1}^{N_{\rm b}} y(\theta_i)\,I(\theta_i,\theta_j)\,y(\theta_j),
\end{equation}
where $N_{\rm b}=22$ is the total number of bins considered over the angular range from 
0 to 11', and $I$ is the inverse covariance matrix corrected by the Hartlap factor~\citep{hartlap07}:
\begin{equation}
	I(\theta_i,\theta_j) = \dfrac{N_{\rm rand}-N_{\rm b}-2}{N_{\rm rand}-1}\,C^{-1}(\theta_i,\theta_j).
\end{equation}
($N_{\rm rand}=500$ is the number of random realizations used to estimate the covariance matrices). 
In this expression, $y(\theta_i)$ is to be regarded as the difference between the measurement and the null hypothesis 
(i.e., no signal, $y=0$); higher $\chi^2$ values are therefore obtained for profiles that deviate more from zero, and 
are thus detected with higher significance (the same approach was adopted in \citetalias{tramonte23}).
Because the value of the $\chi^2$ from Eq.~\eqref{eq:significance} scales with $N_{\rm b}$, 
an effective significance per angular bin is calculated as $\chi^2_{\rm b}=\chi^2/N_{\rm b}$. 
Values of $\chi^2_{\rm b} > 1$ indicate that, on average, the measured profile deviates from zero 
(the null hypothesis) more than expected from the noise.
The per-bin significances are quoted in Table~\ref{tab:stats} for all 16 samples; the table shows that the profiles are reconstructed 
with a very high significance of 10--20 in all cases.

\section{Model}
\label{sec:modeling}

The approach adopted in this study to model the joint contribution of different clusters 
to the measured $y$-profiles is presented in Sec.~\ref{ssec:ysample}, while the specific 
theoretical recipes for the ICM pressure profiles are detailed in Sec.~\ref{ssec:theoprof}.

\subsection{The Compton parameter profile for a cluster sample}
\label{ssec:ysample}
Each Compton profile $y(\theta)$ reconstructed in Sec.~\ref{ssec:profiles} merges the 
contributions of clusters of different masses and redshifts.
In principle, if the Compton parameter profile $y_i(\theta)$ generated by the $i$-th cluster 
is known, the mean profile obtained with the stack of a given cluster sample can be predicted as:
\begin{equation}
	\label{eq:mean_y}
	y(\theta) = \frac{1}{N_{\rm cl}}\sum_{i=1}^{N_{\rm cl}}y_i(\theta),
\end{equation}
where $N_{\rm cl}$ is the number of clusters in the sample. In order to be comparable to 
the measured profiles, the smoothing effect introduced by ACT beam should be accounted for. 
In real space, this is implemented mathematically via a convolution between the emission 
profile $y(\theta)$ and the beam profile $B(\theta)$:
\begin{equation}
	\label{eq:convolve}
	y^{\rm (theo)}(\theta) = \int d\theta' \,y(\theta')\,B(\theta-\theta');
\end{equation}
we can reasonably assume the beam profile to be a perfect Gaussian (normalized to unit 
integral) as in:
\begin{equation}
	B(\theta) = \frac{1}{\sigma\sqrt{2\pi}}\exp{\left[-\frac{\theta^2}{2\sigma^2}\right]},
\end{equation}
where $\sigma=\text{FWHM}/\sqrt{8\ln{2}}$, and for ACT, FWHM=1.6'. As the stack weight maps $W$ 
are practically uniform in all stack cases (with amplitude differences $\lesssim 1\%$ in the 
central stack area) Eqs.~\eqref{eq:mean_y} and~\eqref{eq:convolve} are the mathematical equivalent 
of the stacking process, but for the theoretical profiles (strictly speaking, the stack is the 
average of the individual cluster profiles already convolved with the beam, but 
since both the stack and the beam smoothing are linear operations, the convolution of the average 
profile is the same as the average of the convolved profiles).

Despite its apparent simplicity, this approach for predicting the measured profiles is 
computationally demanding and not suitable for the statistical inference study presented in 
Sec.~\ref{sec:inference}. Indeed, for each cluster sample, this would require the computation 
of $\sim10^2$--$10^3$ $y$-profiles (depending on the chosen mass-redshift bin) for each 
step of the parameter estimation code, which requires about $\sim10^5$ iterations to converge 
(Sec.~\ref{sec:inference}). 

A more general and efficient approach to perform the average is to consider the generic 
profile $y(\theta;M_{500},z)$ of a cluster of mass $M_{500}$ at redshift $z$, and integrate it 
over mass and redshift as in:
\begin{align}
	\label{eq:yfull}
	y(\theta) &= \frac{1}{n_{\rm cl}} \int_{z_{\rm low}}^{z_{\rm top}}dz\, \frac{d^2V}{dzd\Omega}(z)\,\, \times \nonumber \\ 
	&\int_{M_{\rm low}}^{M_{\rm top}}dM_{500} \,\frac{dn}{dM_{500}}(M_{500},z)\,S(M_{500},z)\, y(\theta;M_{500},z),
\end{align}
where $d^2V/dzd\Omega(z)$ is the comoving volume element, $dn/dM_{500}(M_{500},z)$ is the halo 
mass function, and $S(M_{500},z)$ is a selection function quantifying any deviation of the 
sample cluster abundance from the theoretical mass function. The normalization factor 
$n_{\rm cl}$ is the total number density (per unit solid angle) of clusters within the sample 
mass span $[M_{\rm low},M_{\rm top}]$ and redshift span $[z_{\rm low},z_{\rm top}]$:
\begin{align}
	\label{eq:nfull}
	n_{\rm cl} = \int_{z_{\rm low}}^{z_{\rm top}}dz \,&\frac{d^2V}{dzd\Omega}(z) \,\,\times \nonumber \\
	&\int_{M_{\rm low}}^{M_{\rm top}}dM_{500} \,\frac{dn}{dM_{500}}(M_{500},z)\,S(M_{500},z).
\end{align}
In other words, the mean Compton profile of a cluster sample is the weighted average of the quantity 
$y(\theta;M_{500},z)$ over the sample mass and redshift spans, the weighting function being the product 
of the comoving volume element, the mass function and the selection function.
While the comoving 
volume element is uniquely defined for a given cosmology, and widely used parametrizations 
for the halo mass function exist in the literature, the selection function is a quite challenging 
component to quantify. An effective selection modeling should include a number of effects such as instrumental sensitivity 
and observational strategy, performance of the cluster finding algorithm, and subsequent catalog cuts; 
furthermore, the result should be expressed in terms of a suitable function of mass and 
redshift. This task is impractical, especially considering the importance 
of a proper characterization of the function $S(M,z)$ in order to not bias the model testing. 

A way to circumvent this issue, introduced in~\citet{gong19} and also adopted 
in~\citetalias{tramonte23}, is to split both the mass and redshift spans into a suitable set 
of smaller intervals, and approximating the Compton parameter profiles of all clusters within 
a given interval with the one of a hypothetical cluster whose mass and redshift are equal to the
interval midpoints. This way, 
Eqs.~\eqref{eq:yfull} and~\eqref{eq:nfull} become:
\begin{equation}
	\label{eq:yquick}
	y(\theta) = \frac{1}{N_{\rm cl}} \sum_{i=1}^{N_M} \sum_{j=1}^{N_z} N_{ij}\,y(\theta;\bar{M}_i,\bar{z}_j), \quad N_{\rm cl}=\sum_{i=1}^{N_M} \sum_{j=1}^{N_z} N_{ij},
\end{equation}
where $N_{M}$ and $N_{z}$ are respectively the number of mass and redshift intervals, 
$\bar{M}_i$ and $\bar{z}_j$ are the midpoints of the $i$-th mass and $j$-th redshift intervals, 
and $N_{ij}$ is the number of clusters in the corresponding cross mass-redshift interval
(the quantity $N_{\rm cl}$ is again the number of clusters in each of the stacked samples). 
Thanks to the weighting $N_{ij}$ factors, the average estimate of the Compton parameter profile 
via Eq.~\eqref{eq:yquick} naturally encodes the selection associated to each sample. 

Clearly, Eq.~\ref{eq:yquick} introduces an approximation in the theoretical prediction, 
which improves with larger values for $N_M$ and $N_z$; a compromise has to be made with 
computational efficiency, which favors instead lower values for $N_M$ and $N_z$. Given, 
for instance, a cluster sample with a fixed number of mass and redshift intervals, the accuracy 
in the estimation of the average profile depends on how many clusters are there in the 
sample and how their $M_{500}$ and $z$ values differ from the selected interval midpoints. 
It is then reasonable to expect that the interval numbers $N_M$ and $N_z$ depend on the 
considered cluster sample. 
This study adopts $N_M=4$, $N_M=3$, and $N_M=7$ for the lowest, intermediate, and highest 
mass bin, respectively, and $N_z=6$, $N_z=3$, and $N_z=4$ for the lowest, intermediate, and highest 
redshift bin, respectively. The marginalized bins use the union of all the intervals from 
individual bins, so $N_M=14$ and $N_z=13$ (now intervals have different sizes). These values 
were chosen to yield averaged profiles whose difference from the full computation in 
Eq.~\eqref{eq:mean_y} is well below the uncertainties in the measured $y$-profiles, over 
the full range of $\theta$.

This approach simplifies the theoretical prediction for a chosen 
cluster sample into the task of computing $N_M\times N_z$ Compton parameter profiles, i.e. 
$\sim10-100$ (depending on the chosen cluster sample) in order of magnitude, which is 
considerably faster compared to the 
$\sim10^3$ profiles in the most populated bins. This allows to incorporate the prediction into 
a suitable parameter estimation code. Besides, because Eq.~\ref{eq:yquick} does not 
use individual cluster masses, 
but rather the number of clusters within each interval, this strategy is also somewhat 
less sensitive to possible uncertainties affecting the mass estimates (as long as those 
uncertainties are smaller than the chosen interval size).

\begin{table*}
\centering
\caption{Best-fit estimates for the parameters entering the universal pressure profile (UPP) model, obtained with a multi-stage fit. Uncertainties are quoted at the 68\% C.L.; for $c_{500}$, the error bars have been rescaled using a Fisher-matrix correction as discussed in Sec.~\ref{ssec:mcmc}.}
\label{tab:upp_bestfits}
\renewcommand{\arraystretch}{1.4}
\begin{tabular}{cc|cccc|c}
\hline
\multicolumn{7}{c}{\textbf{UPP Parameter Estimates}} \\
\hline
\multicolumn{2}{c}{Priors:} &\quad $[1.0, 2.5]$ \quad &\quad $[0.3, 20.0]$ \quad &\quad $[0.8, 2.0]$ \quad &\quad $[3.0, 6.3]$ \quad \\
\hline
 Redshift & Mass  & $c_{500}$ & $P_0$ & $\alpha$ & $\beta$ & $\chi^2_{\rm r}$ \\
\hline
$z \in [0.0,0.4]$ & $M_{\rm 500c} \in [10^{14.0},10^{14.3}]\,\text{M}_{\odot}$ & $2.11^{+0.04}_{-0.04}$ & $6.4^{+3.6}_{-1.5}$ & $1.5^{+0.4}_{-0.4}$ & $3.30^{+0.08}_{-0.10}$ & 0.19 \\
$z \in [0.0,0.4]$ & $M_{\rm 500c} \in [10^{14.3},10^{14.6}]\,\text{M}_{\odot}$ & $1.75^{+0.02}_{-0.02}$ & $9.9^{+5.1}_{-3.2}$ & $1.1^{+0.2}_{-0.2}$ & $3.96^{+0.07}_{-0.06}$ & 0.64 \\
$z \in [0.0,0.4]$ & $M_{\rm 500c} \in [10^{14.6},10^{15.1}]\,\text{M}_{\odot}$ & $2.16^{+0.03}_{-0.02}$ & $9.1^{+4.8}_{-2.4}$ & $1.4^{+0.3}_{-0.3}$ & $4.04^{+0.09}_{-0.10}$ & 0.30 \\
$z \in [0.0,0.4]$ & All masses & $2.13^{+0.02}_{-0.02}$ & $9.2^{+5.0}_{-2.9}$ & $1.2^{+0.3}_{-0.2}$ & $3.66^{+0.08}_{-0.08}$ & 0.60 \\
\hline
$z \in [0.4,0.8]$ & $M_{\rm 500c} \in [10^{14.0},10^{14.3}]\,\text{M}_{\odot}$ & $2.33^{+0.04}_{-0.04}$ & $7.5^{+5.6}_{-2.5}$ & $1.3^{+0.4}_{-0.3}$ & $3.2^{+0.2}_{-0.1}$ & 0.46 \\
$z \in [0.4,0.8]$ & $M_{\rm 500c} \in [10^{14.3},10^{14.6}]\,\text{M}_{\odot}$ & $2.15^{+0.02}_{-0.02}$ & $6.1^{+4.2}_{-1.6}$ & $1.5^{+0.4}_{-0.4}$ & $3.5^{+0.1}_{-0.1}$ & 2.34 \\
$z \in [0.4,0.8]$ & $M_{\rm 500c} \in [10^{14.6},10^{15.1}]\,\text{M}_{\odot}$ & $2.15^{+0.03}_{-0.03}$ & $9.8^{+5.6}_{-3.6}$ & $1.2^{+0.4}_{-0.2}$ & $3.7^{+0.1}_{-0.1}$ & 1.92 \\
$z \in [0.4,0.8]$ & All masses & $2.18^{+0.02}_{-0.02}$ & $6.3^{+4.0}_{-1.6}$ & $1.4^{+0.3}_{-0.3}$ & $3.5^{+0.2}_{-0.1}$ & 2.15 \\
\hline
$z \in [0.8,2.0]$ & $M_{\rm 500c} \in [10^{14.0},10^{14.3}]\,\text{M}_{\odot}$ & $2.29^{+0.04}_{-0.04}$ & $8.3^{+5.6}_{-3.3}$ & $1.1^{+0.5}_{-0.2}$ & $3.2^{+0.2}_{-0.2}$ & 0.48 \\
$z \in [0.8,2.0]$ & $M_{\rm 500c} \in [10^{14.3},10^{14.6}]\,\text{M}_{\odot}$ & $2.25^{+0.02}_{-0.02}$ & $7.9^{+6.5}_{-3.4}$ & $1.2^{+0.4}_{-0.3}$ & $3.4^{+0.1}_{-0.2}$ & 0.58 \\
$z \in [0.8,2.0]$ & $M_{\rm 500c} \in [10^{14.6},10^{15.1}]\,\text{M}_{\odot}$ & $2.30^{+0.04}_{-0.03}$ & $8.9^{+5.4}_{-2.5}$ & $1.5^{+0.3}_{-0.4}$ & $4.0^{+0.2}_{-0.2}$ & 1.29 \\
$z \in [0.8,2.0]$ & All masses & $2.39^{+0.02}_{-0.02}$ & $7.9^{+3.7}_{-3.1}$ & $1.2^{+0.4}_{-0.2}$ & $3.3^{+0.1}_{-0.1}$ & 0.39 \\
\hline
All redshifts & $M_{\rm 500c} \in [10^{14.0},10^{14.3}]\,\text{M}_{\odot}$ & $2.13^{+0.03}_{-0.03}$ & $5.7^{+3.6}_{-1.4}$ & $1.4^{+0.4}_{-0.3}$ & $3.2^{+0.1}_{-0.1}$ & 1.05 \\
All redshifts & $M_{\rm 500c} \in [10^{14.3},10^{14.6}]\,\text{M}_{\odot}$ & $2.16^{+0.02}_{-0.02}$ & $9.8^{+6.0}_{-3.6}$ & $1.1^{+0.3}_{-0.2}$ & $3.46^{+0.10}_{-0.10}$ & 1.25 \\
All redshifts & $M_{\rm 500c} \in [10^{14.6},10^{15.1}]\,\text{M}_{\odot}$ & $2.10^{+0.02}_{-0.02}$ & $7.8^{+5.0}_{-1.9}$ & $1.4^{+0.3}_{-0.3}$ & $3.98^{+0.08}_{-0.09}$ & 0.29 \\
All redshifts & All masses & $2.22^{+0.01}_{-0.01}$ & $7.5^{+2.8}_{-2.5}$ & $1.3^{+0.4}_{-0.2}$ & $3.45^{+0.08}_{-0.13}$ & 0.49 \\
\hline
\end{tabular}
\end{table*}
 
\begin{table*}
\centering
\caption{Best-fit estimates for the parameters entering the beta-model profile (BMP) model, obtained with a multi-stage fit. Uncertainties are quoted at the 68\% C.L.; for $c_{500}$, the error bars have been rescaled using a Fisher-matrix correction as discussed in Sec.~\ref{ssec:mcmc}.}
\label{tab:bmp_bestfits}
\renewcommand{\arraystretch}{1.4}
\begin{tabular}{cc|ccc|c}
\hline
\multicolumn{6}{c}{\textbf{BMP Parameter Estimates}} \\
\hline
\multicolumn{2}{c}{Priors:} &\quad $[1.0, 3.0]$ \quad &\quad $[0.1, 20.0]$ \quad &\quad $[0.1, 2.0]$ \quad \\
\hline
 Redshift & Mass  & $c_{500}$ & $P_0$ & $\beta$ & $\chi^2_{\rm r}$ \\
\hline
$z \in [0.0,0.4]$ & $M_{\rm 500c} \in [10^{14.0},10^{14.3}]\,\text{M}_{\odot}$ & $2.28^{+0.05}_{-0.04}$ & $5.8^{+0.4}_{-0.3}$ & $1.07^{+0.03}_{-0.03}$ & 0.07 \\
$z \in [0.0,0.4]$ & $M_{\rm 500c} \in [10^{14.3},10^{14.6}]\,\text{M}_{\odot}$ & $2.42^{+0.03}_{-0.03}$ & $6.0^{+0.3}_{-0.3}$ & $1.08^{+0.02}_{-0.02}$ & 1.02 \\
$z \in [0.0,0.4]$ & $M_{\rm 500c} \in [10^{14.6},10^{15.1}]\,\text{M}_{\odot}$ & $2.49^{+0.03}_{-0.03}$ & $8.1^{+0.5}_{-0.5}$ & $1.29^{+0.03}_{-0.03}$ & 0.11 \\
$z \in [0.0,0.4]$ & All masses & $2.52^{+0.02}_{-0.02}$ & $6.7^{+0.3}_{-0.3}$ & $1.10^{+0.02}_{-0.02}$ & 0.45 \\
\hline
$z \in [0.4,0.8]$ & $M_{\rm 500c} \in [10^{14.0},10^{14.3}]\,\text{M}_{\odot}$ & $2.49^{+0.04}_{-0.04}$ & $5.0^{+0.4}_{-0.4}$ & $0.99^{+0.03}_{-0.03}$ & 0.38 \\
$z \in [0.4,0.8]$ & $M_{\rm 500c} \in [10^{14.3},10^{14.6}]\,\text{M}_{\odot}$ & $2.08^{+0.02}_{-0.02}$ & $4.6^{+0.2}_{-0.2}$ & $1.18^{+0.02}_{-0.02}$ & 1.73 \\
$z \in [0.4,0.8]$ & $M_{\rm 500c} \in [10^{14.6},10^{15.1}]\,\text{M}_{\odot}$ & $2.18^{+0.03}_{-0.03}$ & $5.4^{+0.3}_{-0.3}$ & $1.19^{+0.02}_{-0.02}$ & 2.29 \\
$z \in [0.4,0.8]$ & All masses & $2.11^{+0.02}_{-0.02}$ & $4.9^{+0.2}_{-0.2}$ & $1.18^{+0.02}_{-0.02}$ & 1.76 \\
\hline
$z \in [0.8,2.0]$ & $M_{\rm 500c} \in [10^{14.0},10^{14.3}]\,\text{M}_{\odot}$ & $2.50^{+0.04}_{-0.04}$ & $3.8^{+0.7}_{-0.6}$ & $0.90^{+0.06}_{-0.06}$ & 0.41 \\
$z \in [0.8,2.0]$ & $M_{\rm 500c} \in [10^{14.3},10^{14.6}]\,\text{M}_{\odot}$ & $2.28^{+0.03}_{-0.03}$ & $4.2^{+0.3}_{-0.3}$ & $1.08^{+0.03}_{-0.03}$ & 0.57 \\
$z \in [0.8,2.0]$ & $M_{\rm 500c} \in [10^{14.6},10^{15.1}]\,\text{M}_{\odot}$ & $2.64^{+0.04}_{-0.04}$ & $7.6^{+1.0}_{-0.9}$ & $1.18^{+0.05}_{-0.05}$ & 0.92 \\
$z \in [0.8,2.0]$ & All masses & $2.50^{+0.02}_{-0.02}$ & $4.5^{+0.4}_{-0.4}$ & $1.00^{+0.03}_{-0.03}$ & 0.65 \\
\hline
All redshifts & $M_{\rm 500c} \in [10^{14.0},10^{14.3}]\,\text{M}_{\odot}$ & $2.36^{+0.03}_{-0.03}$ & $4.8^{+0.4}_{-0.4}$ & $1.00^{+0.03}_{-0.03}$ & 0.16 \\
All redshifts & $M_{\rm 500c} \in [10^{14.3},10^{14.6}]\,\text{M}_{\odot}$ & $2.01^{+0.01}_{-0.01}$ & $4.5^{+0.2}_{-0.2}$ & $1.19^{+0.02}_{-0.02}$ & 1.01 \\
All redshifts & $M_{\rm 500c} \in [10^{14.6},10^{15.1}]\,\text{M}_{\odot}$ & $2.37^{+0.02}_{-0.02}$ & $7.1^{+0.4}_{-0.4}$ & $1.26^{+0.03}_{-0.03}$ & 0.07 \\
All redshifts & All masses & $2.16^{+0.01}_{-0.01}$ & $5.2^{+0.3}_{-0.3}$ & $1.17^{+0.02}_{-0.02}$ & 0.27 \\
\hline
\end{tabular}
\end{table*}
 
\begin{table*}
\centering
\caption{Best-fit estimates for the parameters entering the polytropic profile (PTP) model, obtained with a multi-stage fit. Uncertainties are quoted at the 68\% C.L.; for $c_{500}$, the error bars have been rescaled using a Fisher-matrix correction as discussed in Sec.~\ref{ssec:mcmc}.}
\label{tab:ptp_bestfits}
\renewcommand{\arraystretch}{1.4}
\begin{tabular}{cc|ccc|c}
\hline
\multicolumn{6}{c}{\textbf{PTP Parameter Estimates}} \\
\hline
\multicolumn{2}{c}{Priors:} &\quad $[1.0, 3.0]$ \quad &\quad $[0.1, 40.0]$ \quad &\quad $[1.0, 9.0]$ \quad \\
\hline
 Redshift & Mass  & $c_{500}$ & $P_0$ & $n$ & $\chi^2_{\rm r}$ \\
\hline
$z \in [0.0,0.4]$ & $M_{\rm 500c} \in [10^{14.0},10^{14.3}]\,\text{M}_{\odot}$ & $2.30^{+0.04}_{-0.04}$ & $15.6^{+1.2}_{-1.2}$ & $4.9^{+0.1}_{-0.1}$ & 0.08 \\
$z \in [0.0,0.4]$ & $M_{\rm 500c} \in [10^{14.3},10^{14.6}]\,\text{M}_{\odot}$ & $2.27^{+0.03}_{-0.03}$ & $16.5^{+1.1}_{-1.0}$ & $5.4^{+0.1}_{-0.1}$ & 0.53 \\
$z \in [0.0,0.4]$ & $M_{\rm 500c} \in [10^{14.6},10^{15.1}]\,\text{M}_{\odot}$ & $2.00^{+0.02}_{-0.02}$ & $22.5^{+1.6}_{-1.5}$ & $7.0^{+0.2}_{-0.2}$ & 0.19 \\
$z \in [0.0,0.4]$ & All masses & $2.07^{+0.02}_{-0.02}$ & $17.9^{+1.0}_{-1.0}$ & $6.1^{+0.1}_{-0.1}$ & 0.32 \\
\hline
$z \in [0.4,0.8]$ & $M_{\rm 500c} \in [10^{14.0},10^{14.3}]\,\text{M}_{\odot}$ & $2.62^{+0.05}_{-0.05}$ & $13.1^{+1.6}_{-1.5}$ & $4.5^{+0.2}_{-0.2}$ & 0.44 \\
$z \in [0.4,0.8]$ & $M_{\rm 500c} \in [10^{14.3},10^{14.6}]\,\text{M}_{\odot}$ & $2.34^{+0.02}_{-0.02}$ & $14.0^{+1.0}_{-0.9}$ & $5.1^{+0.1}_{-0.1}$ & 3.40 \\
$z \in [0.4,0.8]$ & $M_{\rm 500c} \in [10^{14.6},10^{15.1}]\,\text{M}_{\odot}$ & $2.24^{+0.03}_{-0.03}$ & $17.3^{+1.2}_{-1.2}$ & $5.7^{+0.1}_{-0.1}$ & 2.54 \\
$z \in [0.4,0.8]$ & All masses & $2.39^{+0.02}_{-0.02}$ & $14.4^{+1.0}_{-1.0}$ & $5.0^{+0.1}_{-0.1}$ & 3.54 \\
\hline
$z \in [0.8,2.0]$ & $M_{\rm 500c} \in [10^{14.0},10^{14.3}]\,\text{M}_{\odot}$ & $2.51^{+0.04}_{-0.04}$ & $9.7^{+2.4}_{-1.9}$ & $4.2^{+0.4}_{-0.4}$ & 0.42 \\
$z \in [0.8,2.0]$ & $M_{\rm 500c} \in [10^{14.3},10^{14.6}]\,\text{M}_{\odot}$ & $2.39^{+0.03}_{-0.02}$ & $12.1^{+1.4}_{-1.3}$ & $4.9^{+0.2}_{-0.2}$ & 0.80 \\
$z \in [0.8,2.0]$ & $M_{\rm 500c} \in [10^{14.6},10^{15.1}]\,\text{M}_{\odot}$ & $2.16^{+0.03}_{-0.03}$ & $26.1^{+3.9}_{-3.5}$ & $6.9^{+0.3}_{-0.3}$ & 1.29 \\
$z \in [0.8,2.0]$ & All masses & $2.58^{+0.02}_{-0.02}$ & $11.9^{+1.6}_{-1.4}$ & $4.6^{+0.2}_{-0.2}$ & 0.72 \\
\hline
All redshifts & $M_{\rm 500c} \in [10^{14.0},10^{14.3}]\,\text{M}_{\odot}$ & $2.40^{+0.03}_{-0.03}$ & $12.5^{+1.3}_{-1.2}$ & $4.6^{+0.2}_{-0.2}$ & 0.17 \\
All redshifts & $M_{\rm 500c} \in [10^{14.3},10^{14.6}]\,\text{M}_{\odot}$ & $2.37^{+0.02}_{-0.02}$ & $13.1^{+1.0}_{-0.9}$ & $4.9^{+0.1}_{-0.1}$ & 1.68 \\
All redshifts & $M_{\rm 500c} \in [10^{14.6},10^{15.1}]\,\text{M}_{\odot}$ & $1.98^{+0.02}_{-0.02}$ & $21.1^{+1.4}_{-1.4}$ & $6.8^{+0.1}_{-0.1}$ & 0.26 \\
All redshifts & All masses & $2.16^{+0.01}_{-0.01}$ & $15.7^{+1.1}_{-1.0}$ & $5.6^{+0.1}_{-0.1}$ & 0.35 \\
\hline
\end{tabular}
\end{table*}
 
\begin{table*}
\centering
\caption{Best-fit estimates for the parameters entering the exponential universal profile (EUP) model, obtained with a multi-stage fit. Uncertainties are quoted at the 68\% C.L.; for $c_{500}$, the error bars have been rescaled using a Fisher-matrix correction as discussed in Sec.~\ref{ssec:mcmc}.}
\label{tab:eup_bestfits}
\renewcommand{\arraystretch}{1.4}
\begin{tabular}{cc|cccc|c}
\hline
\multicolumn{7}{c}{\textbf{EUP Parameter Estimates}} \\
\hline
\multicolumn{2}{c}{Priors:} &\quad $[0.0, 3.0]$ \quad &\quad $[0.1, 40.0]$ \quad &\quad $[0.0, 3.0]$ \quad &\quad $[0.1, 6.0]$ \quad \\
\hline
 Redshift & Mass  & $c_{500}$ & $P_0$ & $\gamma$ & $\zeta$ & $\chi^2_{\rm r}$ \\
\hline
$z \in [0.0,0.4]$ & $M_{\rm 500c} \in [10^{14.0},10^{14.3}]\,\text{M}_{\odot}$ & $2.21^{+0.04}_{-0.04}$ & $15.0^{+2.7}_{-3.2}$ & $0.3^{+0.3}_{-0.2}$ & $2.8^{+0.5}_{-0.7}$ & 0.08 \\
$z \in [0.0,0.4]$ & $M_{\rm 500c} \in [10^{14.3},10^{14.6}]\,\text{M}_{\odot}$ & $2.16^{+0.03}_{-0.02}$ & $18.4^{+1.5}_{-1.9}$ & $0.1^{+0.1}_{-0.0}$ & $3.6^{+0.2}_{-0.3}$ & 0.42 \\
$z \in [0.0,0.4]$ & $M_{\rm 500c} \in [10^{14.6},10^{15.1}]\,\text{M}_{\odot}$ & $2.02^{+0.02}_{-0.02}$ & $23.1^{+3.0}_{-3.7}$ & $0.3^{+0.3}_{-0.2}$ & $3.8^{+0.5}_{-0.8}$ & 0.17 \\
$z \in [0.0,0.4]$ & All masses & $2.23^{+0.02}_{-0.02}$ & $20.4^{+2.0}_{-2.2}$ & $0.1^{+0.1}_{-0.1}$ & $3.6^{+0.2}_{-0.4}$ & 0.31 \\
\hline
$z \in [0.4,0.8]$ & $M_{\rm 500c} \in [10^{14.0},10^{14.3}]\,\text{M}_{\odot}$ & $2.37^{+0.04}_{-0.04}$ & $10.3^{+4.6}_{-3.6}$ & $0.5^{+0.3}_{-0.3}$ & $2.0^{+1.0}_{-1.1}$ & 0.43 \\
$z \in [0.4,0.8]$ & $M_{\rm 500c} \in [10^{14.3},10^{14.6}]\,\text{M}_{\odot}$ & $1.98^{+0.02}_{-0.02}$ & $15.0^{+2.8}_{-2.3}$ & $0.3^{+0.2}_{-0.2}$ & $3.3^{+0.5}_{-0.5}$ & 1.77 \\
$z \in [0.4,0.8]$ & $M_{\rm 500c} \in [10^{14.6},10^{15.1}]\,\text{M}_{\odot}$ & $2.18^{+0.03}_{-0.03}$ & $20.6^{+1.6}_{-1.9}$ & $0.08^{+0.08}_{-0.04}$ & $3.8^{+0.1}_{-0.2}$ & 2.49 \\
$z \in [0.4,0.8]$ & All masses & $2.36^{+0.02}_{-0.01}$ & $17.2^{+1.9}_{-1.7}$ & $0.07^{+0.06}_{-0.04}$ & $3.4^{+0.1}_{-0.2}$ & 4.16 \\
\hline
$z \in [0.8,2.0]$ & $M_{\rm 500c} \in [10^{14.0},10^{14.3}]\,\text{M}_{\odot}$ & $2.39^{+0.04}_{-0.04}$ & $8.3^{+3.6}_{-3.0}$ & $0.4^{+0.4}_{-0.2}$ & $2.1^{+0.8}_{-1.1}$ & 0.47 \\
$z \in [0.8,2.0]$ & $M_{\rm 500c} \in [10^{14.3},10^{14.6}]\,\text{M}_{\odot}$ & $2.10^{+0.02}_{-0.02}$ & $10.5^{+4.7}_{-3.9}$ & $0.5^{+0.4}_{-0.3}$ & $2.5^{+0.9}_{-1.2}$ & 0.62 \\
$z \in [0.8,2.0]$ & $M_{\rm 500c} \in [10^{14.6},10^{15.1}]\,\text{M}_{\odot}$ & $2.38^{+0.04}_{-0.04}$ & $12.3^{+4.4}_{-3.6}$ & $1.0^{+0.3}_{-0.3}$ & $1.5^{+0.9}_{-0.9}$ & 0.97 \\
$z \in [0.8,2.0]$ & All masses & $2.18^{+0.02}_{-0.02}$ & $9.6^{+5.2}_{-3.4}$ & $0.6^{+0.4}_{-0.3}$ & $2.1^{+1.1}_{-1.1}$ & 0.75 \\
\hline
All redshifts & $M_{\rm 500c} \in [10^{14.0},10^{14.3}]\,\text{M}_{\odot}$ & $2.34^{+0.03}_{-0.03}$ & $10.8^{+3.4}_{-3.9}$ & $0.4^{+0.4}_{-0.2}$ & $2.2^{+0.7}_{-1.2}$ & 0.15 \\
All redshifts & $M_{\rm 500c} \in [10^{14.3},10^{14.6}]\,\text{M}_{\odot}$ & $2.00^{+0.01}_{-0.01}$ & $15.5^{+1.8}_{-2.3}$ & $0.1^{+0.2}_{-0.1}$ & $3.6^{+0.3}_{-0.5}$ & 1.18 \\
All redshifts & $M_{\rm 500c} \in [10^{14.6},10^{15.1}]\,\text{M}_{\odot}$ & $2.33^{+0.02}_{-0.02}$ & $21.4^{+3.8}_{-4.0}$ & $0.3^{+0.3}_{-0.2}$ & $3.3^{+0.5}_{-0.7}$ & 0.25 \\
All redshifts & All masses & $2.34^{+0.01}_{-0.01}$ & $16.1^{+2.3}_{-2.8}$ & $0.2^{+0.2}_{-0.1}$ & $3.1^{+0.4}_{-0.6}$ & 0.81 \\
\hline
\end{tabular}
\end{table*}

\subsection{The Compton parameter profile for individual clusters}
\label{ssec:theoprof}

We are now left with the problem of computing the Compton parameter profile $y(\theta;M_{500},z)$ 
for chosen values of the cluster mass $M_{500}$ and redshift $z$. As anticipated in 
Sec.~\ref{sec:introduction}, the Compton parameter can be obtained via the LoS integral of the 
electron pressure $P_{\rm e}$; more precisely, for a cluster of mass $M_{500}$ at redshift $z$, 
the Compton parameter value at a separation $\theta$ from the center can be computed as:
\begin{equation}
	\label{eq:ymz}
	y(\theta;M,z) = \frac{2\sigma_{\rm T}}{m_{\rm e}c^2}\int_0^{r_{\rm out}}d\ell\, P_{\rm e}\left(\sqrt{D^2_{\rm A}(z)\theta^2+\ell^2};M_{500},z\right). 
\end{equation}
In the equation above, $\ell$ is the LoS distance variable, $D_{\rm A}(z)$ is the angular diameter 
distance to the redshift $z$ of the cluster, and the LoS integration extends out to:
\begin{equation}
	r_{\rm out} = \sqrt{R^2_{\rm cl}-D_{\rm A}^2(z)\theta^2}, 
\end{equation}
where $R_{\rm cl}$ is the radius of the cluster. It is customary to set 
$R_{\rm cl}=5\,R_{500}$, where the overdensity radius can be computed from the cataloged 
masses\footnote{Notice that, for consistency in the usage of the mass definition in this study, 
the mass employed in Eq.~\eqref{eq:r500} is the one derived assuming hydrostatic equilibrium, 
thus bias-uncorrected.} $M_{500}$ as:
\begin{equation}
	\label{eq:r500}
	R_{500} = \left[ \frac{3M_{500}}{4\pi\cdot 500\,\rho_{\rm c}(z)} \right]^{1/3},
\end{equation}
with $\rho_{\rm c}(z)$ the critical density of the Universe at redshift $z$.  
It is possible to verify that extending the range of integration in~\ref{eq:ymz} using values 
of $R_{\rm cl}$ larger than $5\,R_{500}$ adds only a negligible contribution to the integral.

The problem now has narrowed to the choice of a suitable functional form for the electron 
pressure profile $P_{\rm e}(r;M_{500},z)$, with $r$ the radial separation from the cluster 
center. If we assume that different clusters share the same functional form for the 
profile shape, it should be possible to factorize out the dependence on $r$ as in:
\begin{equation}
	\label{eq:pe}
	P_{\rm e}(r;M_{500},z)= \xi(M_{500},z)\,\mathbb{P}(r/r_{\rm s}).
\end{equation}
The function $\mathbb{P}(x)$ is supposed to reproduce the profile shape for clusters of any mass 
and redshift; clearly, in order to account for the different ICM spatial extension in clusters of 
different mass, $\mathbb{P}$ should be a function of a scaled radial separation $x=r/r_{\rm s}$, with 
$r_{\rm s}$ a suitable scale radius. The latter can be defined as:
\begin{equation}
	\label{eq:c500}
	r_{\rm s}(M_{500},z) = \frac{R_{500}(M_{500},z)}{c_{500}},
\end{equation}
where $c_{500}$ is a universal concentration parameter quantifying the deviation of the scale 
radius $r_{\rm s}$ from the overdensity radius $R_{500}$, with smaller values of $c_{500}$ resulting 
in broader profiles. As it is clear from Eq.~\eqref{eq:c500}, $c_{500}$ is initially taken 
to be independent of mass and redshift; this assumption is put to a test in Sec.~\ref{sec:inference}.

The first factor in Eq.~\eqref{eq:pe} is an overall amplitude independent from the 
separation from the cluster center, which depends on the cluster mass and redshift. 
A convenient choice for this factor is a generalization of the characteristic cluster 
pressure $P_{500}$ in the self-similar model:
\begin{equation}
	\label{eq:xi}
	\xi(M_{500},z) = \xi_0\, h^2 \,E^{8/3}(z)\left[\frac{M_{500}}{2.1h\times10^{14}\,\text{M}_{\odot}}\right]^{2/3+\alpha}, 
\end{equation}
where $\xi_0=3.367\times10^{-3}\,\text{keV cm}^{-3}$, $E(z)=H(z)/H_0$ encodes the redshift 
evolution of the Hubble parameter, and $h=H_0/100\text{km s}^{-1}\text{Mpc}^{-1}$ is the 
dimensionless Hubble constant. For the case $\alpha=0$, the expression above is the 
characteristic pressure $P_{500}$, which quantifies the pressure at radius $R_{500}$ required 
to prevent the gravitational collapse of a cluster of mass $M_{500}$ at redshift $z$. The study 
conducted in~\citet{arnaud10} showed that, in order to remove residual mass dependencies in the 
universal pressure model, an extra contribution $\alpha=0.12$ should be added to the 
mass exponent of $M_{500}$. This study shall include the same additional scaling, as in principle 
the function $\xi(M_{500},z)$ should include all the mass and redshift dependence; once more, this 
hypothesis is tested in Sec.~\ref{sec:inference}. Notice that, as the scaled profile 
$\mathbb{P}(x)$ will also contain an adjustable amplitude factor, the actual absolute 
value of the function $\xi(M_{500},z)$ is not crucial; what is important, however, is the functional 
dependence on mass and redshift shown in Eq.~\eqref{eq:xi}.

The core of the present study is actually to test different parametrizations for the scaled 
profile $\mathbb{P}(x)$, by comparing the predicted mean $y$-profile against the ones extracted 
from ACT data. Based on its definition, it is already clear that $\mathbb{P}(x)$ depends 
at least on two 
parameters: an overall amplitude and the concentration $c_{500}$ used to define the scale radius. 
Any additional parameters entering the computation of $\mathbb{P}(x)$ depend on its functional 
form, for which the present analysis shall consider the following four cases.

\subsubsection{The universal pressure profile (UPP)}
The universal pressure profile\footnote{Although all the functional forms for $\mathbb{P}$ 
explored in this study are universal, in the sense that they lose the dependence on the cluster 
mass and redshift, the expression ``universal profile'' is used only for this particular 
case in order to be consisted with the terminology used in the literature.} (hereafter UPP) reads: 
\begin{equation}
	\label{eq:upp}
	\mathbb{P}(x)=\frac{P_0}{x^{\gamma}\left[1+x^{\alpha}\right]^{(\beta-\gamma)/\alpha}},
\end{equation}
where apart from the overall normalization $P_0$, the parameters $\gamma$, $\alpha$ and $\beta$ 
quantify the profile slopes at small ($x \ll 1$), intermediate ($x\simeq 1$) and large ($x \gg 1$) 
separations from the cluster center. 

This generalized Navarro-Frenk-White (gNFW) functional form was first proposed in~\citet{nagai07} 
to fit cluster pressure profiles reconstructed merging simulation results with Chandra-based 
X-ray data. The UPP was later the subject of a number of different studies aimed at testing its 
validity range or provinding novel estimates on its 
parameters~\citep[see, e.g.][and references therein]{arnaud10, planck_ir_v, sayers16, gong19, ma21, he21, pointecouteau21, gallo24}. 
The UPP was also the base model adopted in~\citetalias{tramonte23}, where its parameter values 
showed no detectable residual dependencies on $M_{500}$ or $z$ within the error bars. 

Despite proving effective in reproducing observed X-ray and SZ cluster emissions, the UPP 
functional form presents a few drawbacks, such as its unphysical divergence at $x=0$, or the 
strong degeneracies existing between the different slope parameters. As a result of this last 
feature, the best-fit parameter values quoted in the aforementioned studies exhibit 
a strong scatter (see for instance Table 1 in~\citetalias{tramonte23}); in other words, while 
the combined effect of all parameter values in Eq.~\eqref{eq:upp} provides a consistent prediction, 
the value assigned to each parameter individually alone cannot be trusted, thus preventing (to 
some extent) a proper physical characterization of the ICM pressure distribution. 

In the present study, the free parameters considered for this functional form are $P_0$, 
$\alpha$ and $\beta$; as customary in other works testing the UPP against SZ observables, 
the parameter $\gamma$ is fixed to a nominal value of 0.31.

\subsubsection{The $\beta$-model profile (BMP)}
Under the assumption of hydrostatic equilibrium and isothermality in the ICM, the electron 
number density profile can be conveniently fitted using the functional form:
\begin{equation}
	\label{eq:iso}
	n_{\rm e} = n_{\rm e0}\left[1+\left(\frac{r}{r_{\rm c}}\right)^2 \right]^{-\frac{3}{2}\beta},
\end{equation}
which is known as the isothermal $\beta$-model~\citep{cavaliere76,cavaliere78}; $r_{\rm c}$ is 
a suitable core radius, $n_{\rm e0}$ the central density, and the slope parameter $\beta$ has typical 
values around $\sim0.7$. This is a well established model for cluster density profiles, 
and has long been used to fit X-ray surface brightness in clusters~\citep{birkinshaw99, sarazin88}.
The theoretical motivations for this model, and its evident simplicity, come at the price of some 
inaccuracies when fitting extended data sets~\citep{arnaud09}. The recent study 
in~\citet{braspenning25} proposes a more effective, modified version of this profile,
which however comes at the price of a substantial increase in complexity and a set of 
seven free parameters.

This work shall adopt a simpler approach based on the original profile in Eq.~\eqref{eq:iso}. 
Under the assumption of isothermality, the electron pressure profile follows the same functional 
form as the density profile; it is then reasonable to propose the following 
functional form for the scaled cluster pressure profile:
\begin{equation}
	\label{eq:bmp}
	\mathbb{P}(x) = P_0 \left[1+x^2 \right]^{-\frac{3}{2}\beta},
\end{equation}
where the normalization $P_0$ and the slope $\beta$ are free parameters. To some extent, 
this profile choice relaxes the assumption of isothermality, as the free slope $\beta$ can 
accommodate possible variations in the temperature; nonetheless, temperature is still 
assumed to share the same scale radius $r_{\rm s}$ as the density and to follow a similar 
decreasing trend for increasing separations from the center. Hereafter this profile shall be referred to 
as the $\beta$-model profile (BMP).

\subsubsection{The polytropic profile (PTP)}

The polytropic assumption implies a simple power-law relation between a gas pressure $P$ and 
density $\rho$, in the form $P\propto \rho^{\Gamma}$, where $\Gamma$ is known as the 
polytropic index. Similarly to the case of the $\beta$-model approach, the polytropic one 
also assumes hydrostatic 
equilibrium, but it is less restrictive as it does not require isothermality. 
This assumption was explored in the work by~\citet{bulbul10} in combination with a NFW profile 
for the density distribution generalized by a parameter $\beta$, 
which quantifies the steepness of the gravitational potential. 
This results in a $\beta$-dependent functional form for the pressure profile, 
which in the limit $\beta\rightarrow 2$ (corresponding to a standard NFW density profile) reads:
\begin{equation}
	\label{eq:ptp}
	\mathbb{P}(x)=P_0\left[\frac{1}{x}\ln(1+x) \right]^{n+1},
\end{equation}
where the exponent term $n$ is related to the polytropic index via $n=1/(\Gamma-1)$ 
(notice that in the equation above the singularity at $x=0$ is only apparent because 
for $x\rightarrow0$ the function has limit $P_0$). Hereafter the 
functional form in~\eqref{eq:ptp} shall be referred to as the polytropic profile (PTP). 
Despite fixing $\beta$ 
may result in a partial loss of applicability (particularly near cluster cores) the present study 
shall use Eq.~\ref{eq:ptp} as an additional model for the cluster pressure profile, 
again with the goal of using less parameters to mitigate their degeneracy; the free parameters 
considered for this functional form are $P_0$ and $n$.

\subsubsection{The exponential universal profile (EUP)}

This study also considers a new ansatz to model the scaled profile, in the form:
\begin{equation}
	\label{eq:eup}
	\mathbb{P}(x)={P_0}\,e^{-\gamma x} \left(1+x\right)^{-\zeta},
\end{equation}
which for clear reasons shall be labeled the ``exponential universal profile"" (EUP). 
While not being justified by any simulation-based or analytical result, this profile is 
intended to be a simplified extension of the UPP. In this framework, 
the exponential factor $e^{-\gamma x}$ 
replaces the UPP power factor $x^{-\gamma}$, thus solving the central divergence, 
and the function $(1+x)^{-\zeta}$ mimics the UPP bracket factor, but with a 
simplified form which 
can help remove the tight degeneracy between the UPP exponents $\alpha$ and $\beta$. 
In the present study, the free parameters considered for this functional form are $P_0$, 
$\gamma$ and $\zeta$.

\begin{figure*} 
\includegraphics[trim= 0mm 0mm 50mm 0mm, scale=0.26]{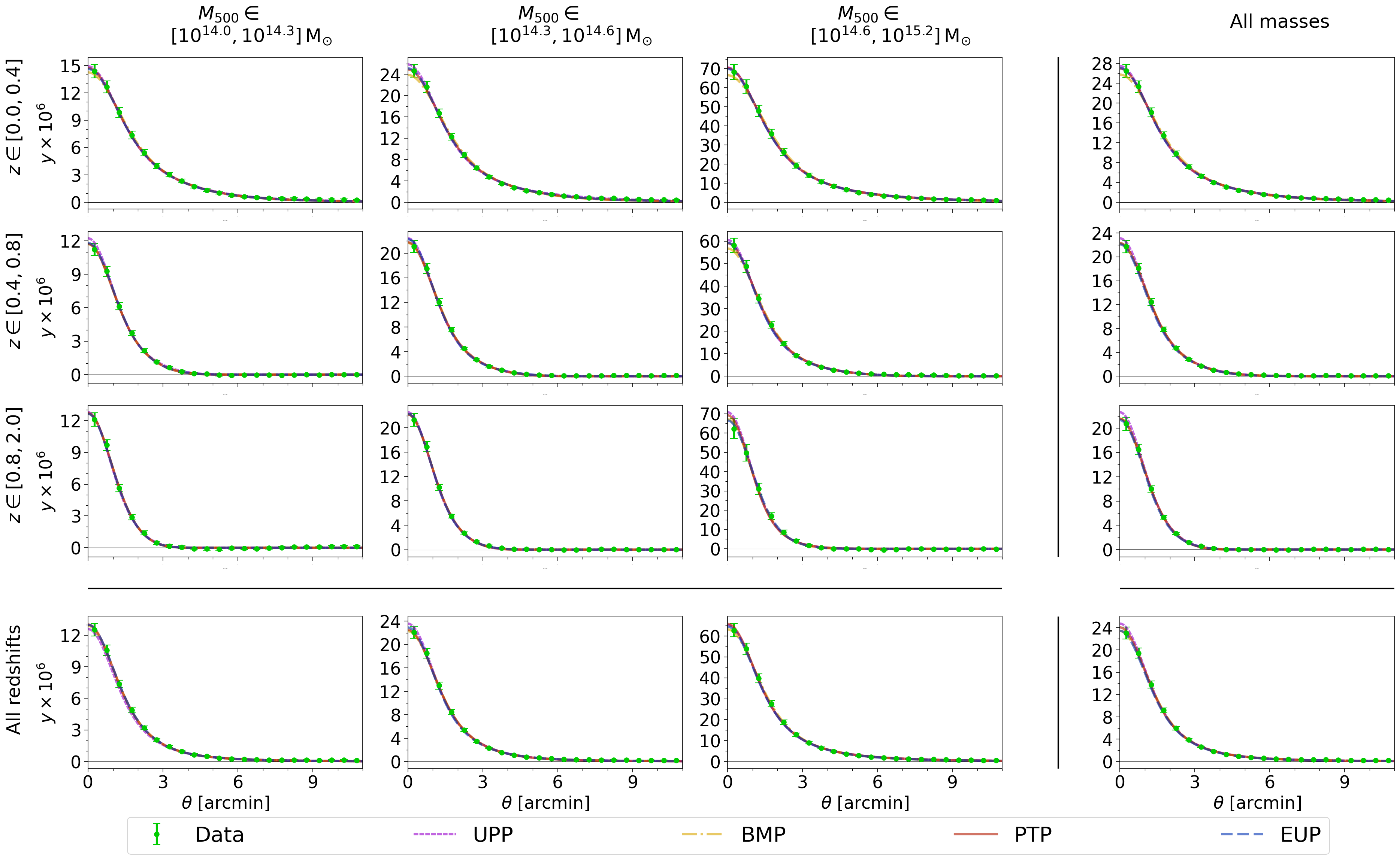}
	\caption{Comparison between the best-fit theoretical predictions of the Compton 
	parameter profiles and the measured data values, for all cluster samples considered 
	in this study; the legend at the bottom specifies the meaning of each plot component.}
\label{fig:fitted_profiles}
\end{figure*}   
\begin{figure*} 
\includegraphics[trim= 0mm 0mm 0mm 0mm, scale=0.26]{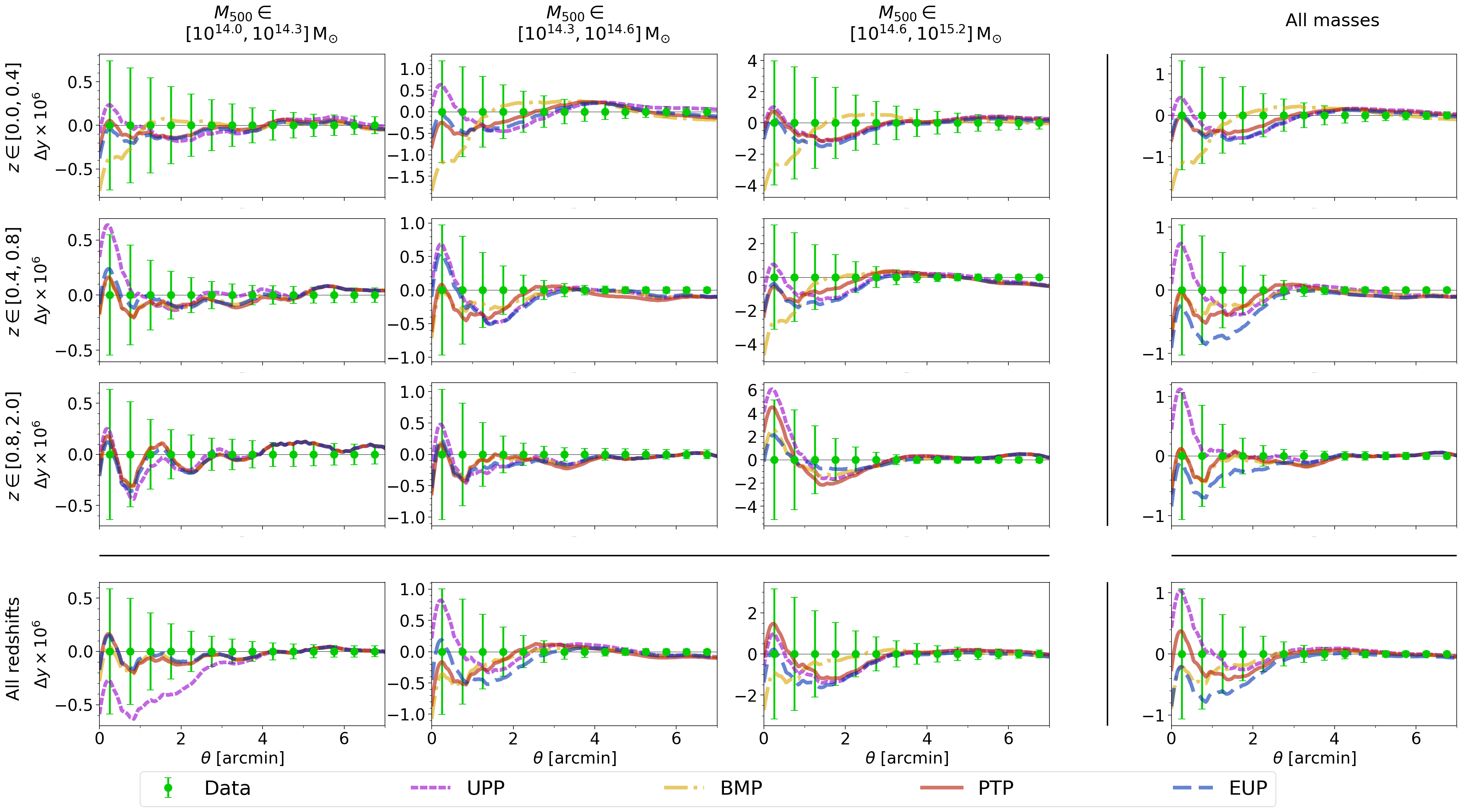}
	\caption{Same as in Fig.~\ref{fig:fitted_profiles} but showing this time the 
	residuals $\Delta y$ between the theoretical predictions and the data values.}
\label{fig:residuals}
\end{figure*}

\section{Model testing}
\label{sec:inference}

This section is dedicated to a direct comparison between the angular Compton parameter 
profiles measured from the ACT $y$-map in Sec.~\ref{sec:measurement}, and those
predicted using the theoretical recipe described in Sec.~\ref{sec:modeling}, using 
each of the universal models listed in Sec.~\ref{ssec:theoprof}. The goal is first of all 
to infer best-fit estimates on the model parameters to match the measurements, 
and secondly to assess which of the models can best reproduce the profile features 
reconstructed from observations. The results can also reveal drifts (if any) of the parameter 
best-fit values with mass and redshift, as this comparison is conducted for all 
of the 16 cluster samples considered in this study.

\subsection{Parameter estimation method}

For each profile model, its free parameters plus the concentration $c_{500}$ are estimated 
with a Monte Carlo Markov Chain (MCMC) approach implemented via the Python package 
\texttt{emcee}\footnote{\url{https://emcee.readthedocs.io/en/stable/}.}, based on the affine 
invariant sampler from~\citet{goodman10}.
This method enables to extensively sample the parameter space and reconstruct the posterior 
distributions on individual and pairs of parameters, thus allowing to infer their 
best-fit values. 

The MCMC runs are based on the following chi-squared definition:
\begin{equation}
	\label{eq:chi2}
	\chi^2(\Theta) = \sum_{i=1}^{N_{\rm b}}\sum_{j=1}^{N_{\rm b}} \left[y_i(\Theta)-y_i^{\rm (m)}\right]\,C^{-1}_{ij}\,\left[y_j(\Theta)-y_j^{\rm (m)}\right],
\end{equation}
where $N_{\rm b}$ is the number of angular bins, the indices $i$ and $j$ select a specific 
bin, $y(\Theta)$ is the Compton profile predicted using the parameter set 
$\Theta$ (and the corresponding theoretical model), $y^{\rm (m)}$ is the measured Compton 
profile for the chosen cluster sample, and $C^{-1}$ is the inverse of the covariance matrix 
for the same cluster sample (Sec.~\ref{ssec:profiles}).

For each parameter, a flat prior probability distribution is adopted, within reasonable ranges 
justified by previous results in the literature. The procedure is repeated for each profile model 
and each cluster sample, for a resulting total of 64 MCMC runs consisting of 12 walkers and 500,000 
steps, enough to ensure convergence. Chains are then trimmed to remove the first 10\% (assumed to 
be a standard burn-in) and thinned according to the autocorrelation length estimated by the code 
(typically in the range of a few hundreds).

\subsection{Practical implementation}
\label{ssec:practical_implementation}

While the measured angular profiles plotted in Fig.~\ref{fig:fitted_profiles} extend out to 11 arcmin, 
interesting profile features only happen at quite smaller angular scales. 
An initial MCMC run employing the full 11 arcmin range performed quite poorly 
in reproducing the profile shape near the origin, as error bars are considerably smaller at 
large angular separations. As a result, large $\theta$ points have a high statistical weight and 
tend to anchor the profile at its tail, while producing non-optimal agreement in the center. For 
this reason, the data profiles employed in the MCMC runs were trimmed to a maximum 
angular separation of 7 arcmin. 

A first fit run conducted by letting all parameters free to move yielded best-fit estimates capable
of reproducing the measured profiles within their error bars. However, due to the strong degeneracies 
existing between model parameters, the posteriors probability distributions appeared broad and poorly 
constrained, strongly 
hinting at the possibility of the results being prior driven; this was especially true for 
the parameter $c_{500}$, which appeared with error bars in the range 0.2--0.5 over a prior range 
of 2 to 3 (depending on the considered model). 

In order to mitigate the effect of parameter degeneracy, a second round of fits was conducted. 
Initially, $c_{500}$ was fitted alone while fixing all other model parameters to their
best-fit values obtained in the first fit round. Subsequently, these new estimates of $c_{500}$ 
were kept fixed while fitting all remaining parameters. This approach noticeably tightened the 
final constraints and improved the quality of the posterior distributions, as 
discussed in Sec.~\ref{ssec:mcmc}.

Another strong limiting factor to a robust MCMC convergence is the substantial correlation
existing between different profile bins, as clearly noticeable in the 
measured covariance matrices (Fig.~\ref{fig:correlations}). These off-diagonal contributions 
arise from the averaging procedure, smoothing of the maps, and instrumental effects; when combined 
with the strong parameter degeneracy in a MCMC run, they lead to highly multimodal and unstable 
posteriors, particularly for the outer radial bins where nominal errors are very small. 
In the second round of fits, a diagonal-only covariance was then adopted for stability, 
allowing for robust estimation of the main profile parameters while limiting sensitivity to 
potentially spurious correlations in the outer bins.

\subsection{Results}

The resulting parameter best-fit values, for all the combinations 
of model profiles and cluster samples, are listed in Tables~\ref{tab:upp_bestfits} 
to~\ref{tab:eup_bestfits}. In these tables, the values of $c_{500}$ are the ones obtained from the 
fit of the concentration alone, while the values of the remaining parameter are the ones 
obtained while keeping $c_{500}$ fixed. The tables also quote the prior range adopted for each 
parameter, and the final reduced chi-squared 
values used to assess the goodness of fit. The reduced chi-squared is defined as 
$\chi^2_{\rm r}=\chi^2(\Theta_{\rm bf})/N_{\rm dof}$, where $\chi^2(\Theta_{\rm bf})$ is the 
chi-squared computed using Eq.~\ref{eq:chi2} inputting the best-fit parameter set 
$\Theta_{\rm bf}$, and the number of degrees of freedom $N_{\rm dof}$ is computed as the number 
of fitted data points minus the number of free parameters. These chi-squared values refer to 
the final fit in which $c_{500}$ was kept fixed to the tabulated values, and the other parameters
were allowed to move. 

In order not to clutter the main text, all plots for the posterior distributions on the 
model parameters are shown separately in Appendix~\ref{sec:contours}.
A direct comparison between the data and 
all the best-fit predictions is instead shown in Fig.~\ref{fig:fitted_profiles}, where for each 
cluster sample the measured profile points are plotted together with the theoretical profiles. 
In order to better show the level of agreement between data and theory, 
Fig.~\ref{fig:residuals} shows instead the 
residuals computed as differences between each predicted profile and the measured profile. 

\begin{figure*}
\includegraphics[trim= 40mm 30mm 0mm 0mm, scale=0.3]{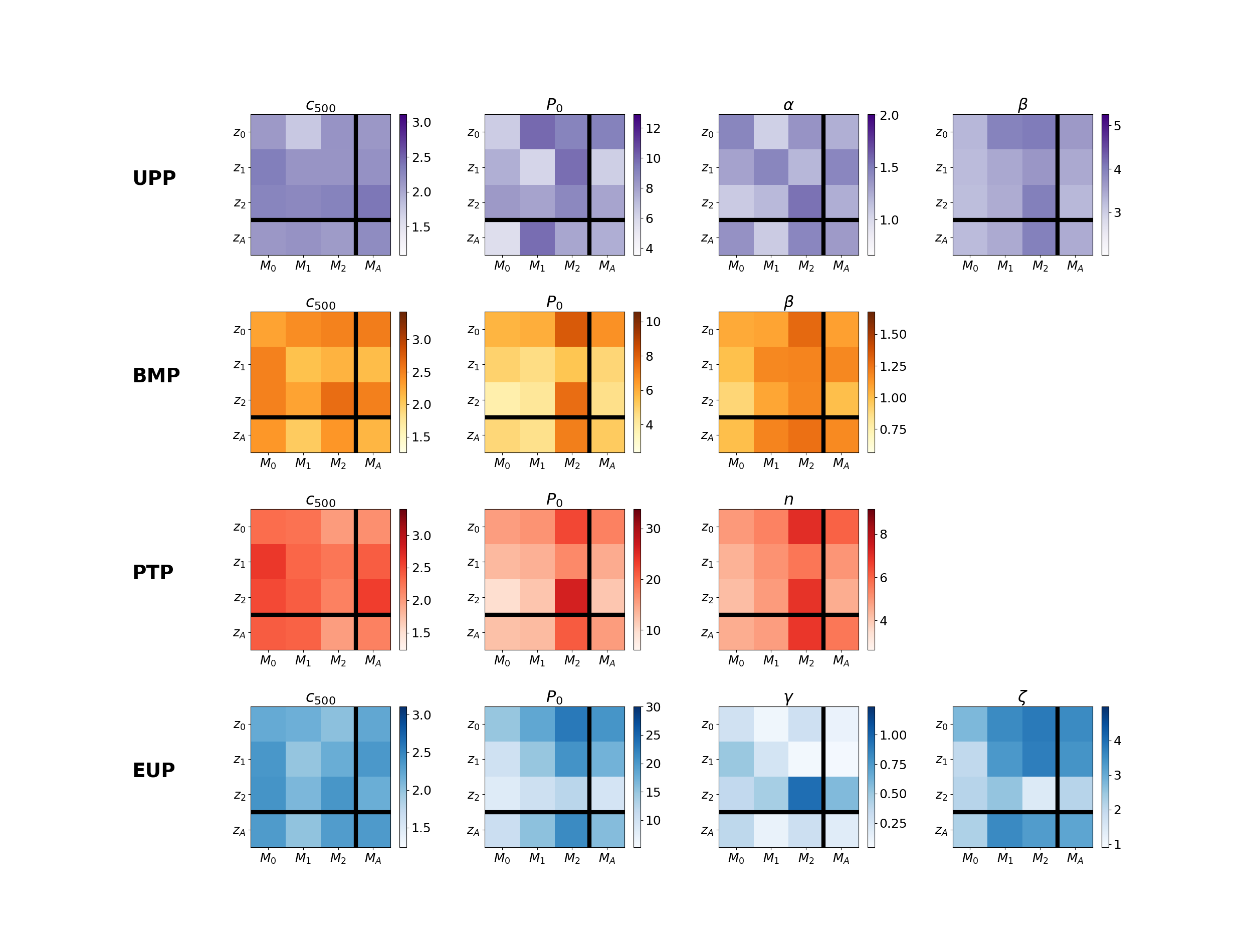}
	\caption{Visual representation of the constrained parameters dependence on the 
	chosen mass and redshift bin. Each row refers to a different theoretical model, and 
	each plot to a specific parameter; mass and redshift bins are 
	labeled with a number subscript or with ``A'' for ``All''.}
\label{fig:ptrend}
\end{figure*}

A final goal of this study is to assess 
possible marginal dependencies of the parameters on the 
chosen mass and redshift bin; to this aim, the best-fit values of the same parameter for different 
cluster samples are visually compared in Fig.~\ref{fig:ptrend}. The figure summarizes the constrained
magnitudes of all parameters for all theoretical models, and is intended to easily reveal 
the existence of possible trends.

\section{Discussion}
\label{sec:discussion}

The following is dedicated to the assessment and interpretation of the results presented in 
Sec.~\ref{sec:inference}.

\subsection{About the output of the MCMC runs}
\label{ssec:mcmc}

An evaluation of the results from the model testing can begin with an overview of the 
posterior contours shown in Appendix~\ref{sec:contours}. The final best-fit values for 
the concentration $c_{500}$ are very similar 
to those obtained in the first fitting round, which is expected since the other
parameters were kept fixed. The associated uncertainties, however, became 
considerably smaller. The final posterior distributions on $c_{500}$ are shown in 
Fig.~\ref{fig:c500_posteriors}, where all four pressure profile models are overplotted in each 
data-sample panel. The figure shows these posteriors are narrow and yield final error bars 
of order 0.01--0.02; such uncertainties are in fact underestimated, as they arise from the artificially 
broken degeneracy with the remaining parameters and from the use of diagonalized covariances in the 
MCMC runs. To provide a more realistic uncertainty estimate, Fisher-matrix errors for $c_{500}$ are 
computed using both the full and the diagonal covariance matrices evaluated at the best-fit model; 
the ratio of these errors defines an inflation factor that is applied to the MCMC-derived uncertainties. 
This procedure preserves the best-fit parameter values while restoring an uncertainty that reflects 
the effective number of independent modes in the full covariance. These corrected uncertainties are
the ones quoted for $c_{500}$ in Tables~\ref{tab:upp_bestfits} to~\ref{tab:eup_bestfits}. The 
calculated inflation factors were found in the range 1.1--2.3: although this confirms the presence of
non-negligible correlations, it does not drastically increase the size of the error bars. 
In summary, the uncertainties reported for $c_{500}$, while corrected to provide a more conservative 
estimate, may still underestimate the fully marginalized constraints that would result from sampling 
all parameters simultaneously with the full covariance.

The posteriors on the remaining parameters obtained from the last MCMC stage are
shown in Figs.~\ref{fig:upp_contours} to~\ref{fig:eup_contours}. For a given 
pressure-profile model, the contours derived from the different cluster samples exhibit broadly
consistent shapes; this is especially true for the BMP and PTP cases, for which only two parameters
are fitted. The UPP and EUP models, which involve three free parameters, display a larger
degree of variation across samples. This behavior reflects the residual parameter degeneracies 
that remain even in a reduced parameter setting, and is particularly evident for the UPP case.

In fact, similar features can be noted in the findings of previous works that fitted the UPP on SZ 
profiles, for instance in~\citet[][Figs. 5 and 7]{gong19}, in~\citet[][Fig. 5]{planck_ir_v}, and 
in~\citetalias[][Figs. 19 to 22]{tramonte23}, the latter providing an extensive 
discussion about this issue.
The present study offers an advantage with respect to the UPP-based analysis 
of~\citetalias{tramonte23} by relying on a smaller set of free parameters.
Cluster positions in~\citetalias{tramonte23} were based on optical data, 
and could then be offset from the peak of the SZ signal, thus requiring to also fit for two 
parameters quantifying the statistical properties of such miscentering. This analysis,
on the contrary, employs cluster positions determined via matched-filter search directly on 
ACT coadded maps, which are then expected to be representative of the SZ signal 
peaks. Furthermore, the theoretical modeling in~\citetalias{tramonte23} adopted cluster masses 
obtained via scaling relations calibrated on weak-lensing data; this required the introduction 
of a bias factor for conversion into the~\citet{arnaud10} mass definition, which assumed hydrostatic 
equilibrium in the ICM, when testing the UPP profile. The present study 
employs already such ``uncorrected'' masses (as quoted in the ACT cluster catalog), so that the 
introduction of the hydrostatic bias is not necessary. 
The combination of these two reductions in model complexity, together with the effective 
breaking of the degeneracy with $c_{500}$, leads to more stable UPP fits in this work compared to 
the results of \citetalias{tramonte23}.
In summary, the modest variations observed among the contours in Appendix~\ref{sec:contours} 
are consistent with expectations given the different data samples and profile models.

The shapes of the posterior contours for the UPP model are similar to the ones shown 
in~\citet{planck_ir_v} and~\citet{gong19}, although in our case the removal of the degeneracy 
with $c_{500}$ shows overall narrower posteriors between the other pairs of parameters. 
Posteriors obtained for the BMP and PTP pressure profiles show a clear positive correlation 
between $P_0$ and $\beta$ in the former, and between $P_0$ and $n$ in the latter. This is 
expected, as the fitted observable is the LoS-integrated pressure: hence, steeper
profiles (larger $\beta$ or $n$) require higher peak amplitudes (larger $P_0$) to yield the 
same integral over the cluster span. In the case of the UPP or EUP, interpreting degeneracy directions
is complicated by the presence of multiple exponents in the functional form. 
Mild anticorrelation is observed between $\alpha$ and $\beta$, which is expected from the 
UPP functional form with $\gamma$ fixed; similarly, a strong anticorrelation is visible between 
$\gamma$ and $\zeta$ in the EUP, as they both contribute to steepen the profile. For both profile 
models, the amplitude
$P_0$ is generally found to be positively correlated with the ``stronger'' exponent ($\beta$ 
in the UPP, $\zeta$ in the EUP) while being anticorrelated with the other.

\subsection{About the agreement between theory and data}
Fig.~\ref{fig:fitted_profiles} shows a good agreement between the measured Compton  
profiles and those predicted theoretically using the best-fit parameters, for all different 
profile models. Fig.~\ref{fig:residuals} shows the residual differences between predictions and 
measurements, and enables a more detailed view of their level of agreement up to $7'$ from the 
cluster center, which is the effective range employed in the fits. 
The residual plots confirm a substantial agreement between data and 
models, with more visible fluctuations near the core where error bars are larger. 
Once again, such deviations are not unusual in this context; similar deviations 
have been found in previous fits of pressure profile models against cluster 
data as in~\citet{gong19} and~\citetalias{tramonte23}.

The goodness of fit can be assessed more quantitatively with the $\chi^2_{\rm r}$ values from 
Tables~\ref{tab:upp_bestfits} to~\ref{tab:eup_bestfits}; values depend on the chosen cluster 
sample, but are mostly below unity. In fact, some cases are substantially below unity, hinting 
at overfitting possibly due to the diagonalization of the covariance. Other cases, instead,
denote poorer levels of agreement with the data; this almost always happens 
for the intermediate redshift bin, which yields $\chi^2_{\rm r}$ values mostly above unity.
An inspection of Fig.~\ref{fig:residuals} reveals in fact a poorer agreement between fitted curves 
and data points at large radii (being clearly visible at $\theta\gtrsim 5'$), 
signaling the models are struggling to accommodate the Compton profile decrease in the cluster 
outskirts in this redshift range. Specifically, models in this region tend to underestimate the 
amplitude, potentially suggesting that even after subtracting the estimated zero-level in each 
stack, small residual large-scale signals in the maps could still contribute to a modest excess in the outskirts.
For all other cases, the $\chi^2_{\rm r}$ estimates overall confirm the 
initial assessment of all profile models being effective in reproducing the angular dependence 
of the cluster Compton signal. 

There is also no compelling indication of what model should be favored; 
$\chi^2_{\rm r}$ fluctuations across different cluster samples for the same model profile are 
larger than average $\chi^2_{\rm r}$ differences between different models. The goodness of fit, 
in each case, is mostly determined by the performance of the MCMC in retrieving a global minimum, not in 
different efficiency of the underlying theoretical model. The BMP and UPP model could perhaps be deemed 
the most and least favored ones, respectively, as $\chi^2_{\rm r}$ are overall smaller for the former 
and larger for the latter; this conclusion, however, only holds by a relatively small difference. 
It is also interesting to notice that the UPP and EUP profiles, 
despite having one additional free parameter, do not always yield lower $\chi^2_{\rm r}$ values 
compared to the BMP and PTP profiles. As already discussed, including more parameters in the 
profile fitting comes at the cost of increasing the model degeneracy, thus making a robust MCMC 
convergence more challenging. 

\subsection{About trends in the best-fit parameters}

The set of 16 different samples allows for an assessment of possible residual dependencies of 
the profile parameters on mass and/or redshift. The error bars 
quoted in Tables~\ref{tab:upp_bestfits} to~\ref{tab:eup_bestfits} indicate that,
for the same profile model, best-fit parameters obtained from different data samples can 
differ beyond the 68\% C.L. This tension is likely due to the error bars being overall underestimated, 
as a result of the measures implemented to achieve a robust MCMC fitting, as discussed
in Sec.~\ref{ssec:mcmc} and~\ref{ssec:practical_implementation}. 
The variations in best-fit parameter estimates across different mass and redshift bins are visually 
summarized in Fig.~\ref{fig:ptrend}, which can also be useful to recognize possible patterns. 

The normalization parameter $P_0$ hints at an increase for higher masses and lower redshifts; this is 
very clear for the EUP model and marginally visible for the UPP model. In the case of the BMP and PTP profiles, 
a noticeable exception is the highest-mass, highest-redshift bin, whose amplitude is considerably
larger than this trend would suggest. The $\chi^2_{\rm r}$ for this data sample, however, is relatively
on the high side compared to the same redshift bin values in Tables~\ref{tab:bmp_bestfits} and~\ref{tab:ptp_bestfits}, 
and the error bars quoted on $P_0$ are the highest. Hence, despite being a clear outlier from the 
proposed trend, this bin also provides looser constraints and lower agreement between theory and data. 
The work in~\citet{gong19} also detected a marginal redshift dependence 
in the profile amplitude, which the author corrected by adding an additional exponent $\eta<0$ to 
the $E(z)$ factor in Eq.~\eqref{eq:xi}, thus mitigating the profile amplitude increase with 
redshift. The finding in the present study goes in the same direction, favoring a higher 
normalization for lower redshifts. 

As for the other parameters, it is not possible 
to identify any clear trend for the concentration $c_{500}$ in any of the considered models; 
similar considerations apply to the UPP parameter $\alpha$ and the EUP parameter $\gamma$.
The slopes UPP $\beta$, BMP $\beta$, PTP $n$, and EUP $\zeta$, seem to follow a trend similar to 
the profile normalization (with a varying extent depending on the considered model), 
which can be expected from their positive correlation with $P_0$ 
as it is visible in the parameter posterior probability contours. 
This hints at a steeper profile decrease at higher masses and lower redshifts. 

In summary, the 
observed marginal trends 
in the amplitude and slope suggest that ICM pressure in the most massive and lowest-redshift 
clusters has higher peak values and steeper decreases at their outskirts. This finding may indeed 
have a physical origin, as this description fits the most evolved systems that have reached 
equilibrium (typically cool-core clusters). On the contrary, lower mass and 
higher redshift clusters tend to be more disturbed systems, with higher chances of hot gas 
being spread out as a result of mergers and turbulence~\citep{zhang11, lovisari20}.
In principle, this scenario should result in a similar trend for the concentration, with higher 
values of $c_{500}$ at high masses and low redshifts, but the plots in Fig.~\ref{fig:ptrend} 
show no compelling evidence for this. Once more, this could be a result of parameter degeneracy, 
as the concentration values are largely determined by the first MCMC round which fits for all 
parameter simultaneously; in this scenario, the normalization and slope parameters might be
``absorbing'' the changes in $c_{500}$, effectively erasing any quantifiable concentration trend
in the best-fit estimates.

\subsection{About the constrained parameter values}

To conclude this section, it is worth commenting on the values obtained for individual parameters. 
Starting from the concentration $c_{500}$, which is common to all profile models, its best-fit 
values are generally above 2. In fact, if the calculation of the overdensity radius in 
Eq.~\eqref{eq:r500} had used the bias-corrected masses from the ACT catalog (on average 30\% larger), 
then the resulting $R_{500}$ would have been $\sim9\%$ larger, and so would have the best-fit 
concentration parameters to yield the same fit (thus adding one or two decimal units to the 
retrieved values of $c_{500}$). Still, this is well within the range covered by estimates from 
the literature. While lower concentration values are favored by previous studies of the UPP 
profile~\citep[e.g.,][]{arnaud10, gong19}, other work adopting either UPP or polytropic functional forms
 quotes or hints at values of $c_{500}$ larger than 
2~\citep[e.g.,][]{bulbul10, mcdonald14, ma21, ghirardini19, ettori19, tramonte23}.

\begin{figure}
\includegraphics[trim= 0mm 0mm 0mm 0mm, scale=0.3]{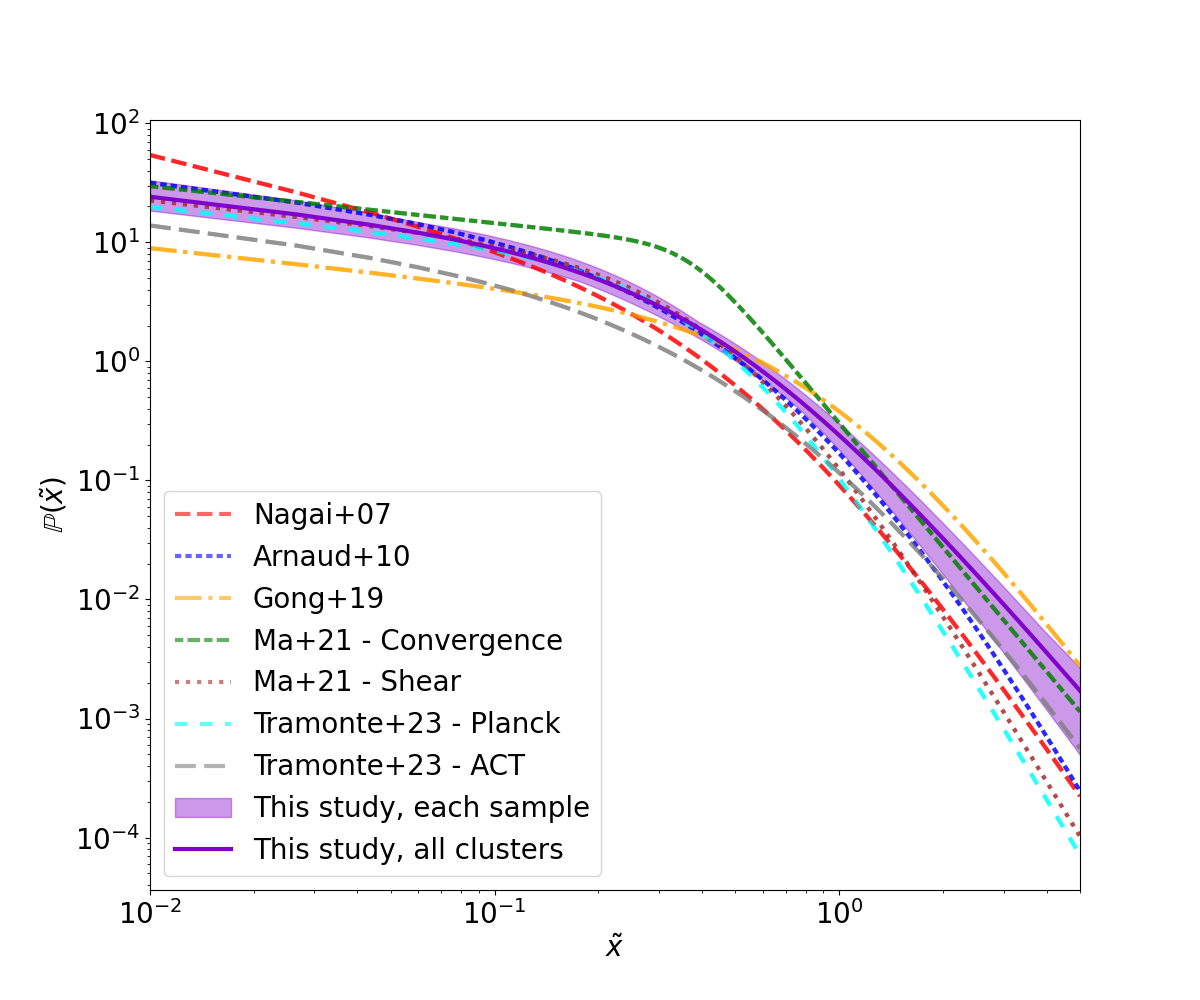}
	\caption{Comparison between the universal pressure profile $\mathbb{P}(x)$ 
	predictions calculated using the parameters quoted by different works. For 
	the present study, the solid line shows the prediction from the parameters 
	fitted on the full cluster sample, while the shaded region encompasses the 
	predictions from all different samples.}
\label{fig:upp_check}
\end{figure}

The remaining UPP parameter estimates are in broad agreement 
with literature findings. Figure~\ref{fig:upp_check} shows a direct comparison between the UPP 
profile $\mathbb{P}(x)$ computed using the best fit parameters from this study (for all 
cluster sample cases) and the same profile computed using the parameters quoted in previous 
works of different kind. These include the studies 
in~\citet{nagai07} and~\citet{arnaud10}, which fitted the profile on a reduced sample of high-significance 
clusters, the study in~\citet{ma21}, which fitted the profile on cross-correlations between SZ and 
lensing data (the latter including both shear and convergence), and the study in~\citetalias{tramonte23}, 
which fitted the profile on a population-level sample of galaxy clusters using both~\textit{Planck}
and ACT maps. The independent variable in the figure is defined as $\tilde{x}=x/c_{500}=r/{R_{500}}$, 
where $x$ is the independent variable used so far in this study, in order to include the effect of 
the concentration parameter from different references. The comparison shows that the predictions 
obtained with the current analysis are in substantial agreement with the ones from the literature; 
some marginal difference can be found in the outskirts, where the profile fitted in this study
tends to decrease more slowly. 

In the case of the BMP model, the best-fit values for the slope $\beta$ are in the range $[0.9,1.3]$; 
these are slightly larger than previous estimates based on studies of clusters with X-ray data, typically 
pointing towards $\beta$ values in the range $\sim0.5-0.8$~\citep{birkinshaw99, croston08}, 
although values $\beta>1$ have occasionally been reported\footnote{These reports, however, 
refer to extended versions of the $\beta$-model, which include additional parameters, so this 
comparison is only marginally relevant.}~\citep{ettori00, vikhlinin06}. 
The higher slopes obtained in this study might be due to the 
scaling of the radial separation by the radius $r_{\rm s}$ defined via the 
concentration parameter $c_{500}$; the referenced works, instead, all scaled the separation using the ``core'' radius 
of the cluster $r_{\rm c}$, defined as the very central region of the cluster 
where its density can be approximated as uniform. There is no fixed relation between  
$R_{500}$ and $r_{\rm c}$ in a cluster, with some cases having ratios of a few units and others 
larger than 10~\citep{vikhlinin06}, with higher values typically assigned to relaxed, cool-core 
systems. In general, it is safe to say that $r_{\rm c}<r_{\rm s}$ is always satisfied; as our BMP model 
is scaled by $r_{\rm s}$, and not by $r_{\rm c}$, it is therefore 
reasonable to expect higher values of $\beta$. 
In fact, the values quoted in Table~\ref{tab:bmp_bestfits} are even not as high as it could be 
expected from using $r_{\rm s}$ instead of $r_{\rm c}$. Since $\beta$ and $c_{500}$ are expected to be anticorrelated, 
as they change the profile amplitude in the same direction, it is possible that lower slope 
values are allowed by the slightly higher concentration estimates obtained for the BMP ($c_{500}\sim 2.0-2.6$).  

As for the PTP model, the best-fit estimates are also consistent with previous findings. 
The works in~\citet{eckert13} and~\citet{ghirardini19} point to polytropic index values 
$\Gamma\sim1.2$ (more broadly, in the range $\Gamma\sim1.1-1.3$); the latter translates into 
an exponent $n\sim5$, in agreement with the values listed in Table~\ref{tab:ptp_bestfits}.
The study in~\citet{ghirardini19} also provides constraints on the concentration at 
$c_{500}\sim2.6$, supporting the present findings of concentrations $c_{500}>2$. 

Finally, the constraints from the EUP model do not have any term of comparison in the literature. 
It is perhaps worth commenting that the parameter $\gamma$ is constrained to quite low 
values ($<1$) to mitigate the exponential cutoff of the profile amplitude at large radii. 
The estimates for the polynomial exponent $\zeta$ are instead in between the values constrained 
for the parameters $\alpha$ and $\beta$ for the UPP model, hinting at a similar power-law 
decrease in the intermediate-to-large radial range. 

\section{Conclusions}
\label{sec:conclusions}

This study measured the integrated Compton parameter profile of a population-level sample of 3496 galaxy 
clusters extracted from the ACT-DR4 catalog, by means of stacks on the ACT-DR6 $y$-map which covers 
a sky area of $\sim13,000\,\text{deg}^2$ with a resolution of $1.6'$. Clusters span a mass range 
$[10^{14},10^{15.1}]\,\text{M}_{\odot}$ and a redshift range $[0,2]$, each further split into 
three smaller bins to yield nine disjoint cluster sub-samples and six sub-samples obtained marginalizing 
over one variable at a time. For all samples, the averaged Compton parameter profiles were measured with 
high significance per bin in the range $\chi^2_{\rm b} \sim 10 - 20$. 

These profiles were modeled theoretically integrating the cluster pressure along the line of 
sight and performing the relevant sample averages. Pressure profiles were modeled in a ``universal'' 
approach, whereby the cluster mass and redshift dependence is factorized out as a common factor, and 
the pressure only depends on the separation from the cluster center scaled via a suitable concentration 
parameter $c_{500}$. As for the actual pressure profile form, four different models were considered: 
a gNFW model (UPP), a $\beta$-profile model (BMP), a polytropic model (PTP), and a new ansatz including 
a pressure exponential decay (EUP). Each model came with a different sets of free parameters to be 
constrained against observations. A set of multi-stage MCMC runs, designed to 
break the strong degeneracy of the model-specific parameters with $c_{500}$, was employed to compare 
the predicted and measured $y$ profiles and reconstruct the posterior probability distribution on 
such parameters.
Best-fit estimates were obtained and tabulated, and the associated predictions compared to the data. 

All models proved effective in reproducing the observed angular dependence of the averaged 
Compton profiles for all cluster samples, within the error bars. The constrained values of individual 
parameters were in broad agreement with literature findings; in particular, the predicted 
UPP pressure profile was found to be compatible with the predictions from previous studies. 
For a given pressure model, individual parameters were found to exhibit some 
tension across different cluster samples; error bars, however, are likely overall underestimated, 
as a result of the staged MCMC fitting and the removal of correlations in different angular 
bins implemented to ensure robust convergence. For some parameters these residual dependencies on 
mass and redshift hinted at an overall trend, suggesting high-mass, low-redshift clusters 
(typically the most dynamically relaxed systems) 
favor higher amplitudes and steeper profiles than predicted by the universal model. 

While similar to the analysis conducted in~\citepalias{tramonte23}, this study differs in the 
detailed implementation of the analysis, and in its overall goals. The cluster SZ stack in this study 
was conducted on a more extended $y$ map and employed a refined pipeline which corrects for the 
pixel area distortion. Furthermore, the cluster sample used in this study was homogeneous and 
extracted from ACT coadded maps, unlike the sample in~\citepalias{tramonte23} which was a 
heterogeneous merge of catalogs based on optical observations.
This mitigated the effect of residual biases in the mass definition and allowed to forego 
the usage of miscentering parameters in the cluster SZ modeling. Besides, this study did not 
employ the hydrostatic mass bias at any point of the analysis, always referring to the 
SZ-based mass estimates from the catalog. The result is a reduction in the number of free 
parameters, which contributed to mitigate degeneracy effects and
likely allowed to detect residual trends in some of their best-fit values. 
Finally, while~\citepalias{tramonte23} only focused on the UPP parametrization, this study 
aimed at comparing the efficiency of different pressure profile models. 

Compared to the study of individual and well-resolved clusters, stacks have the advantage of 
allowing for the inclusion of lower mass objects and to perform population-level studies, which 
may highlight possible drifts of the model parameters for different regimes, as indeed  
proven in the present work. At the same time, however, it has also shown how very different 
pressure profile parametrizations are almost equally effective in fitting the data; this may suggest 
that population-level studies of the cluster pressure profile, using SZ data alone, lack the 
required constraining power to discriminate among competing models. A clear drawback of using 
either SZ or X-ray cluster data is that the ``true'', underlying three-dimensional pressure profile 
is not directly observable. Measurement always contains LoS-integrated quantities, which 
allow for different functional forms, and different parameter combinations, 
to yield the same prediction. A joint study of both cluster SZ emission 
(proportional to the integrated density) and X-ray emission (proportional to the integrated square 
of the density), could help break the degeneracy, although requiring suitable 
ansaetze for the radial temperature dependence. This approach has already been 
explored~\citep{planck_ir_v, sayers16}, although it has not been applied to a population-level
sample.  

In this context, while the UPP gNFW parametrization has become a default choice in several astrophysical 
and cosmological studies involving the ICM profile, it seems there is no substantial reason to 
prefer it over other, more physically motivated, profiles such as the $\beta$-model or the polytropic model.
While a strongly steep profile in the center may be favored by numerical simulation of dark matter 
halos~\citet{navarro97}, this has not always been confirmed by observations~\citep{sand04, newman13, li23}, 
and even more so, could not be applicable to the baryonic component. This might be the reason why 
flatter-core profiles, such as the BMP or the EUP in the present study, are still effective in 
fitting the measured cluster SZ signal. 
The usage of the gNFW parametrization can still be justified as a matter of convenience and continuity 
with previous studies; however, given its \textit{ad-hoc} fitting-oriented nature, and 
the heavy degeneracies intrinsic to its functional form,
 any further attempt at constraining its parameter values would not provide any real insight 
into the physics of the ICM. 

On a different note, the fact that some parameters seem to be marginally dependent on the cluster mass 
and redshift may question to what extent the assumption of a universal model is applicable.  
Clusters of galaxies span up to 3 orders of magnitude in mass and are continuously evolving systems, 
particularly at relatively higher redshifts. While the idea of a common, underlying pressure profile 
form is theoretically 
appealing, no universal model would ever be able to exactly capture variations in individual systems or even 
in different mass and/or redshift regimes, at least not without the price of increasing the number 
of model parameters to the point their individual values become detached from any clear physical meaning. 
Even the previously cited works that adopt physically motivated 
profiles, such as the BMP or the PTP, typically use them in combination with others (e.g. gNFW forms) or 
in more elaborate parametrizations with an extended parameter set. This does not mean, however, that 
universal forms should be abandoned: they are still capable of capturing average cluster properties with a 
large range of applicability, 
and are especially useful in cosmology-oriented studies.

A final conclusion of this work is perhaps that, whenever an analytical description of the ICM
pressure is required, a compromise has to be reached between 
precision in the modeling and physical interpretability and universality of the model itself.
Future studies can further investigate this tension using increasingly extended data sets; for example, 
the ACT collaboration has recently announced its largest cluster sample to date~\citep{aguena25}, which 
is expected to be released publicly as part of Data Release 6, and which can be employed for a natural 
extension of the present analysis.


\section*{Acknowledgements}
The author acknowledges financial support from the XJTLU Research Development Fund (RDF) grant with number RDF-22-02-068.

\bibliographystyle{aa}
\bibliography{bibliography}

@ARTICLE{adam25,
       author = {{Adam}, R. and {Eynard-Machet}, T. and {Bartalucci}, I. and {Cherouvrier}, D. and {Clerc}, N. and {Di Mascolo}, L. and {Dupourqu{\'e}}, S. and {Ferrari}, C. and {Mac{\'\i}as-P{\'e}rez}, J. -F. and {Pointecouteau}, E. and {Pratt}, G.~W.},
        title = "{PITSZI: Probing intra-cluster medium turbulence with Sunyaev{\textendash}Zel'dovich imaging: Application to the triple merging cluster MACS J0717.5+3745}",
      journal = {\aap},
         year = 2025,
        month = feb,
       volume = {694},
          eid = {A182},
        pages = {A182},
          doi = {10.1051/0004-6361/202452342},
archivePrefix = {arXiv},
       eprint = {2409.14804},
 primaryClass = {astro-ph.CO},
       adsurl = {https://ui.adsabs.harvard.edu/abs/2025A&A...694A.182A}
}

@ARTICLE{aguena25,
       author = {{ACTDESHSC Collaboration} and {Aguena}, M. and {Aiola}, S. and {Allam}, S. and {Andrade-Oliveira}, F. and {Bacon}, D. and {Bahcall}, N. and {Battaglia}, N. and {Battistelli}, E.~S. and {Bocquet}, S. and {Bolliet}, B. and {Bond}, J.~R. and {Brooks}, et al.},
        title = "{The Atacama Cosmology Telescope: DR6 Sunyaev-Zel'dovich Selected Galaxy Clusters Catalog}",
      journal = {arXiv e-prints},
         year = 2025,
        month = jul,
          eid = {arXiv:2507.21459},
        pages = {arXiv:2507.21459},
          doi = {10.48550/arXiv.2507.21459},
archivePrefix = {arXiv},
       eprint = {2507.21459},
 primaryClass = {astro-ph.CO},
       adsurl = {https://ui.adsabs.harvard.edu/abs/2025arXiv250721459A}
}

@ARTICLE{allen11,
       author = {{Allen}, Steven W. and {Evrard}, August E. and {Mantz}, Adam B.},
        title = "{Cosmological Parameters from Observations of Galaxy Clusters}",
      journal = {\araa},
         year = 2011,
        month = sep,
       volume = {49},
       number = {1},
        pages = {409-470},
          doi = {10.1146/annurev-astro-081710-102514},
archivePrefix = {arXiv},
       eprint = {1103.4829},
 primaryClass = {astro-ph.CO},
       adsurl = {https://ui.adsabs.harvard.edu/abs/2011ARA&A..49..409A}
}

@ARTICLE{arnaud09,
       author = {{Arnaud}, M.},
        title = "{The {\ensuremath{\beta}}-model of the intracluster medium. Commentary on: Cavaliere A. and Fusco-Femiano R., 1976, A\&A, 49, 137}",
      journal = {\aap},
         year = 2009,
        month = jun,
       volume = {500},
       number = {1},
        pages = {103-104},
          doi = {10.1051/0004-6361/200912150},
       adsurl = {https://ui.adsabs.harvard.edu/abs/2009A&A...500..103A}
}

@ARTICLE{arnaud10,
       author = {{Arnaud}, M. and {Pratt}, G.~W. and {Piffaretti}, R. and
         {B{\"o}hringer}, H. and {Croston}, J.~H. and {Pointecouteau}, E.},
        title = "{The universal galaxy cluster pressure profile from a representative sample of nearby systems (REXCESS) and the Y$_{SZ}$ - M$_{500}$ relation}",
      journal = {\aap},
         year = 2010,
        month = jul,
       volume = {517},
          eid = {A92},
        pages = {A92},
          doi = {10.1051/0004-6361/200913416},
archivePrefix = {arXiv},
       eprint = {0910.1234},
 primaryClass = {astro-ph.CO},
       adsurl = {https://ui.adsabs.harvard.edu/abs/2010A&A...517A..92A}
}

@ARTICLE{birkinshaw99,
       author = {{Birkinshaw}, M.},
        title = "{The Sunyaev-Zel'dovich effect}",
      journal = {\physrep},
         year = 1999,
        month = mar,
       volume = {310},
       number = {2-3},
        pages = {97-195},
          doi = {10.1016/S0370-1573(98)00080-5},
archivePrefix = {arXiv},
       eprint = {astro-ph/9808050},
 primaryClass = {astro-ph},
       adsurl = {https://ui.adsabs.harvard.edu/abs/1999PhR...310...97B}
}

@ARTICLE{braspenning25,
       author = {{Braspenning}, Joey and {Schaye}, Joop and {Schaller}, Matthieu and {Kugel}, Roi and {Kay}, Scott T.},
        title = "{Hydrostatic mass bias for galaxy groups and clusters in the FLAMINGO simulations}",
      journal = {\mnras},
         year = 2025,
        month = feb,
       volume = {536},
       number = {4},
        pages = {3784-3802},
          doi = {10.1093/mnras/stae2798},
archivePrefix = {arXiv},
       eprint = {2409.07849},
 primaryClass = {astro-ph.CO},
       adsurl = {https://ui.adsabs.harvard.edu/abs/2025MNRAS.536.3784B}
}

@ARTICLE{bulbul10,
       author = {{Bulbul}, G. Esra and {Hasler}, Nicole and {Bonamente}, Massimiliano and {Joy}, Marshall},
        title = "{An Analytic Model of the Physical Properties of Galaxy Clusters}",
      journal = {\apj},
         year = 2010,
        month = sep,
       volume = {720},
       number = {2},
        pages = {1038-1044},
          doi = {10.1088/0004-637X/720/2/1038},
archivePrefix = {arXiv},
       eprint = {0911.2827},
 primaryClass = {astro-ph.CO},
       adsurl = {https://ui.adsabs.harvard.edu/abs/2010ApJ...720.1038B}
}

@ARTICLE{bulbul24,
       author = {{Bulbul}, E. and {Liu}, A. and {Kluge}, M. and {Zhang}, X. and {Sanders}, J.~S. and {Bahar}, Y.~E. and {Ghirardini}, V. and {Artis}, E. and {Seppi}, R. and {Garrel}, C. and {Ramos-Ceja}, M.~E. and {Comparat}, J. and {Balzer}, F. and {B{\"o}ckmann}, K. and {Br{\"u}ggen}, M. and {Clerc}, N. and {Dennerl}, K. and {Dolag}, K. and {Freyberg}, M. and {Grandis}, S. and {Gruen}, D. and {Kleinebreil}, F. and {Krippendorf}, S. and {Lamer}, G. and {Merloni}, A. and {Migkas}, K. and {Nandra}, K. and {Pacaud}, F. and {Predehl}, P. and {Reiprich}, T.~H. and {Schrabback}, T. and {Veronica}, A. and {Weller}, J. and {Zelmer}, S.},
        title = "{The SRG/eROSITA All-Sky Survey. The first catalog of galaxy clusters and groups in the Western Galactic Hemisphere}",
      journal = {\aap},
         year = 2024,
        month = may,
       volume = {685},
          eid = {A106},
        pages = {A106},
          doi = {10.1051/0004-6361/202348264},
archivePrefix = {arXiv},
       eprint = {2402.08452},
 primaryClass = {astro-ph.CO},
       adsurl = {https://ui.adsabs.harvard.edu/abs/2024A&A...685A.106B}
}

@ARTICLE{carlstrom02,
   author = {{Carlstrom}, J.~E. and {Holder}, G.~P. and {Reese}, E.~D.},
    title = "{Cosmology with the Sunyaev-Zel'dovich Effect}",
  journal = {\araa},
   eprint = {astro-ph/0208192},
     year = 2002,
   volume = 40,
    pages = {643-680},
      doi = {10.1146/annurev.astro.40.060401.093803},
   adsurl = {http://adsabs.harvard.edu/abs/2002ARA%26A..40..643C}
}

@ARTICLE{cavaliere76,
       author = {{Cavaliere}, A. and {Fusco-Femiano}, R.},
        title = "{X-rays from hot plasma in clusters of galaxies.}",
      journal = {\aap},
         year = 1976,
        month = may,
       volume = {49},
        pages = {137-144},
       adsurl = {https://ui.adsabs.harvard.edu/abs/1976A&A....49..137C}
}

@ARTICLE{cavaliere78,
       author = {{Cavaliere}, A. and {Fusco-Femiano}, R.},
        title = "{The Distribution of Hot Gas in Clusters of Galaxies}",
      journal = {\aap},
         year = 1978,
        month = nov,
       volume = {70},
        pages = {677},
       adsurl = {https://ui.adsabs.harvard.edu/abs/1978A&A....70..677C}
}

@ARTICLE{coulton24,
       author = {{Coulton}, William and {Madhavacheril}, Mathew S. and {Duivenvoorden}, Adriaan J. and {Hill}, J. Colin and {Abril-Cabezas}, Irene and {Ade}, Peter A.~R. and {Aiola}, Simone et al.},
        title = "{Atacama Cosmology Telescope: High-resolution component-separated maps across one third of the sky}",
      journal = {\prd},
         year = 2024,
        month = mar,
       volume = {109},
       number = {6},
          eid = {063530},
        pages = {063530},
          doi = {10.1103/PhysRevD.109.063530},
archivePrefix = {arXiv},
       eprint = {2307.01258},
 primaryClass = {astro-ph.CO},
       adsurl = {https://ui.adsabs.harvard.edu/abs/2024PhRvD.109f3530C}
}

@ARTICLE{eckert13,
       author = {{Eckert}, D. and {Molendi}, S. and {Vazza}, F. and {Ettori}, S. and {Paltani}, S.},
        title = "{The X-ray/SZ view of the virial region. I. Thermodynamic properties}",
      journal = {\aap},
         year = 2013,
        month = mar,
       volume = {551},
          eid = {A22},
        pages = {A22},
          doi = {10.1051/0004-6361/201220402},
archivePrefix = {arXiv},
       eprint = {1301.0617},
 primaryClass = {astro-ph.CO},
       adsurl = {https://ui.adsabs.harvard.edu/abs/2013A&A...551A..22E}
}

@ARTICLE{ettori00,
       author = {{Ettori}, Stefano},
        title = "{{\ensuremath{\beta}}-model and cooling flows in X-ray clusters of galaxies}",
      journal = {\mnras},
         year = 2000,
        month = nov,
       volume = {318},
       number = {4},
        pages = {1041-1046},
          doi = {10.1046/j.1365-8711.2000.03664.x},
archivePrefix = {arXiv},
       eprint = {astro-ph/0005224},
 primaryClass = {astro-ph},
       adsurl = {https://ui.adsabs.harvard.edu/abs/2000MNRAS.318.1041E}
}

@ARTICLE{ettori19,
       author = {{Ettori}, S. and {Ghirardini}, V. and {Eckert}, D. and {Pointecouteau}, E. and {Gastaldello}, F. and {Sereno}, M. and {Gaspari}, M. and {Ghizzardi}, S. and {Roncarelli}, M. and {Rossetti}, M.},
        title = "{Hydrostatic mass profiles in X-COP galaxy clusters}",
      journal = {\aap},
         year = 2019,
        month = jan,
       volume = {621},
          eid = {A39},
        pages = {A39},
          doi = {10.1051/0004-6361/201833323},
archivePrefix = {arXiv},
       eprint = {1805.00035},
 primaryClass = {astro-ph.CO},
       adsurl = {https://ui.adsabs.harvard.edu/abs/2019A&A...621A..39E}
}

@ARTICLE{croston08,
       author = {{Croston}, J.~H. and {Pratt}, G.~W. and {B{\"o}hringer}, H. and {Arnaud}, M. and {Pointecouteau}, E. and {Ponman}, T.~J. and {Sanderson}, A.~J.~R. and {Temple}, R.~F. and {Bower}, R.~G. and {Donahue}, M.},
        title = "{Galaxy-cluster gas-density distributions of the representative XMM-Newton cluster structure survey (REXCESS)}",
      journal = {\aap},
         year = 2008,
        month = aug,
       volume = {487},
       number = {2},
        pages = {431-443},
          doi = {10.1051/0004-6361:20079154},
archivePrefix = {arXiv},
       eprint = {0801.3430},
 primaryClass = {astro-ph},
       adsurl = {https://ui.adsabs.harvard.edu/abs/2008A&A...487..431C}
}

@ARTICLE{gallo24,
       author = {{Gallo}, S. and {Douspis}, M. and {Soubri{\'e}}, E. and {Salvati}, L.},
        title = "{Characterising galaxy clusters' completeness function in Planck with hydrodynamical simulations}",
      journal = {\aap},
         year = 2024,
        month = jun,
       volume = {686},
          eid = {A15},
        pages = {A15},
          doi = {10.1051/0004-6361/202347678},
       adsurl = {https://ui.adsabs.harvard.edu/abs/2024A&A...686A..15G}
}

@ARTICLE{ghirardini19,
       author = {{Ghirardini}, V. and {Ettori}, S. and {Eckert}, D. and {Molendi}, S.},
        title = "{Polytropic state of the intracluster medium in the X-COP cluster sample}",
      journal = {\aap},
         year = 2019,
        month = jul,
       volume = {627},
          eid = {A19},
        pages = {A19},
          doi = {10.1051/0004-6361/201834875},
archivePrefix = {arXiv},
       eprint = {1906.00977},
 primaryClass = {astro-ph.CO},
       adsurl = {https://ui.adsabs.harvard.edu/abs/2019A&A...627A..19G}
}

@ARTICLE{gong19,
       author = {{Gong}, Yan and {Ma}, Yin-Zhe and {Tanimura}, Hideki},
        title = "{Probing galaxy cluster and intra-cluster gas with luminous red galaxies}",
      journal = {\mnras},
         year = 2019,
        month = jul,
       volume = {486},
       number = {4},
        pages = {4904-4916},
          doi = {10.1093/mnras/stz1177},
archivePrefix = {arXiv},
       eprint = {1904.12089},
 primaryClass = {astro-ph.CO},
       adsurl = {https://ui.adsabs.harvard.edu/abs/2019MNRAS.486.4904G}
}

@ARTICLE{goodman10,
       author = {{Goodman}, Jonathan and {Weare}, Jonathan},
        title = "{Ensemble samplers with affine invariance}",
      journal = {Communications in Applied Mathematics and Computational Science},
         year = 2010,
        month = jan,
       volume = {5},
       number = {1},
        pages = {65-80},
          doi = {10.2140/camcos.2010.5.65},
       adsurl = {https://ui.adsabs.harvard.edu/abs/2010CAMCS...5...65G}
}

@ARTICLE{hartlap07,
       author = {{Hartlap}, J. and {Simon}, P. and {Schneider}, P.},
        title = "{Why your model parameter confidences might be too optimistic. Unbiased estimation of the inverse covariance matrix}",
      journal = {\aap},
         year = 2007,
        month = mar,
       volume = {464},
       number = {1},
        pages = {399-404},
          doi = {10.1051/0004-6361:20066170},
archivePrefix = {arXiv},
       eprint = {astro-ph/0608064},
 primaryClass = {astro-ph},
       adsurl = {https://ui.adsabs.harvard.edu/abs/2007A&A...464..399H}
}

@ARTICLE{he21,
       author = {{He}, Yizhou and {Mansfield}, Philip and {Rau}, Markus Michael and {Trac}, Hy and {Battaglia}, Nicholas},
        title = "{Debiased Galaxy Cluster Pressure Profiles from X-Ray Observations and Simulations}",
      journal = {\apj},
         year = 2021,
        month = feb,
       volume = {908},
       number = {1},
          eid = {91},
        pages = {91},
          doi = {10.3847/1538-4357/abd0ff},
archivePrefix = {arXiv},
       eprint = {2008.04334},
 primaryClass = {astro-ph.CO},
       adsurl = {https://ui.adsabs.harvard.edu/abs/2021ApJ...908...91H}
}

@ARTICLE{hicks08,
   author = {{Hicks}, A.~K. and {Ellingson}, E. and {Bautz}, M. and {Cain}, B. and 
	{Gilbank}, D.~G. and {Gladders}, M.~G. and {Hoekstra}, H. and 
	{Yee}, H.~K.~C. and {Garmire}, G.},
    title = "{Chandra X-Ray Observations of the 0.6 $\lt$ z $\lt$ 1.1 Red-Sequence Cluster Survey Sample}",
  journal = {\apj},
archivePrefix = "arXiv",
   eprint = {0710.5513},
     year = 2008,
    month = jun,
   volume = 680,
      eid = {1022-1041},
    pages = {1022-1041},
      doi = {10.1086/587682},
   adsurl = {http://adsabs.harvard.edu/abs/2008ApJ...680.1022H}
}

@ARTICLE{hilton21,
       author = {{Hilton}, M. and {Sif{\'o}n}, C. and {Naess}, S. and {Madhavacheril}, M. and {Oguri}, M. and {Rozo}, E. and {Rykoff}, E. and {Abbott}, T.~M.~C. and {Adhikari}, S. and {Aguena}, M. et al.},
        title = "{The Atacama Cosmology Telescope: A Catalog of >4000 Sunyaev-Zel{\textquoteright}dovich Galaxy Clusters}",
      journal = {\apjs},
         year = 2021,
        month = mar,
       volume = {253},
       number = {1},
          eid = {3},
        pages = {3},
          doi = {10.3847/1538-4365/abd023},
archivePrefix = {arXiv},
       eprint = {2009.11043},
 primaryClass = {astro-ph.CO},
       adsurl = {https://ui.adsabs.harvard.edu/abs/2021ApJS..253....3H}
}

@ARTICLE{kaiser86,
       author = {{Kaiser}, N.},
        title = "{Evolution and clustering of rich clusters.}",
      journal = {\mnras},
         year = 1986,
        month = sep,
       volume = {222},
        pages = {323-345},
          doi = {10.1093/mnras/222.2.323},
       adsurl = {https://ui.adsabs.harvard.edu/abs/1986MNRAS.222..323K}
}

@ARTICLE{kornoelje25,
       author = {{Kornoelje}, K. and {Bleem}, L.~E. and {Rykoff}, E.~S. and {Abbott}, T.~M.~C. and {Ade}, P.~A.~R. and {Aguena}, M. and {Alves}, O. and {Anderson}, A.~J. and {Andrade-Oliveira}, et al.},
        title = "{The SPT-Deep Cluster Catalog: Sunyaev-Zel'dovich Selected Clusters from Combined SPT-3G and SPTpol Measurements over 100 Square Degrees}",
      journal = {arXiv e-prints},
         year = 2025,
        month = mar,
          eid = {arXiv:2503.17271},
        pages = {arXiv:2503.17271},
          doi = {10.48550/arXiv.2503.17271},
archivePrefix = {arXiv},
       eprint = {2503.17271},
 primaryClass = {astro-ph.CO},
       adsurl = {https://ui.adsabs.harvard.edu/abs/2025arXiv250317271K}
}

@ARTICLE{kravtsov06,
       author = {{Kravtsov}, Andrey V. and {Vikhlinin}, Alexey and {Nagai}, Daisuke},
        title = "{A New Robust Low-Scatter X-Ray Mass Indicator for Clusters of Galaxies}",
      journal = {\apj},
         year = 2006,
        month = oct,
       volume = {650},
       number = {1},
        pages = {128-136},
          doi = {10.1086/506319},
archivePrefix = {arXiv},
       eprint = {astro-ph/0603205},
 primaryClass = {astro-ph},
       adsurl = {https://ui.adsabs.harvard.edu/abs/2006ApJ...650..128K}
}

@ARTICLE{li23,
       author = {{Li}, Pengfei and {Tian}, Yong and {J{\'u}lio}, Mariana P. and {Pawlowski}, Marcel S. and {Lelli}, Federico and {McGaugh}, Stacy S. and {Schombert}, James M. and {Read}, Justin I. and {Yu}, Po-Chieh and {Ko}, Chung-Ming},
        title = "{Measuring galaxy cluster mass profiles into the low-acceleration regime with galaxy kinematics}",
      journal = {\aap},
         year = 2023,
        month = sep,
       volume = {677},
          eid = {A24},
        pages = {A24},
          doi = {10.1051/0004-6361/202346431},
archivePrefix = {arXiv},
       eprint = {2303.10175},
 primaryClass = {astro-ph.CO},
       adsurl = {https://ui.adsabs.harvard.edu/abs/2023A&A...677A..24L}
}

@ARTICLE{lovisari20,
       author = {{Lovisari}, Lorenzo and {Schellenberger}, Gerrit and {Sereno}, Mauro and {Ettori}, Stefano and {Pratt}, Gabriel W. and {Forman}, William R. and {Jones}, Christine and {Andrade-Santos}, Felipe and {Randall}, Scott and {Kraft}, Ralph},
        title = "{X-Ray Scaling Relations for a Representative Sample of Planck-selected Clusters Observed with XMM-Newton}",
      journal = {\apj},
         year = 2020,
        month = apr,
       volume = {892},
       number = {2},
          eid = {102},
        pages = {102},
          doi = {10.3847/1538-4357/ab7997},
archivePrefix = {arXiv},
       eprint = {2002.11740},
 primaryClass = {astro-ph.CO},
       adsurl = {https://ui.adsabs.harvard.edu/abs/2020ApJ...892..102L}
}

@ARTICLE{ma21,
       author = {{Ma}, Yin-Zhe and {Gong}, Yan and {Tr{\"o}ster}, Tilman and {Van Waerbeke}, Ludovic},
        title = "{Probing the cluster pressure profile with thermal Sunyaev-Zeldovich effect and weak lensing cross-correlation}",
      journal = {\mnras},
         year = 2021,
        month = jan,
       volume = {500},
       number = {2},
        pages = {1806-1816},
          doi = {10.1093/mnras/staa3369},
archivePrefix = {arXiv},
       eprint = {2010.15064},
 primaryClass = {astro-ph.CO},
       adsurl = {https://ui.adsabs.harvard.edu/abs/2021MNRAS.500.1806M}
}

@ARTICLE{madhavacheril20,
       author = {{Madhavacheril}, Mathew S. and {Hill}, J. Colin and {N{\ae}ss}, Sigurd and {Addison}, Graeme E. and {Aiola}, Simone and {Baildon}, Taylor and {Battaglia}, Nicholas and {Bean}, Rachel and {Bond}, J. et al.},
        title = "{Atacama Cosmology Telescope: Component-separated maps of CMB temperature and the thermal Sunyaev-Zel'dovich effect}",
      journal = {\prd},
         year = 2020,
        month = jul,
       volume = {102},
       number = {2},
          eid = {023534},
        pages = {023534},
          doi = {10.1103/PhysRevD.102.023534},
archivePrefix = {arXiv},
       eprint = {1911.05717},
 primaryClass = {astro-ph.CO},
       adsurl = {https://ui.adsabs.harvard.edu/abs/2020PhRvD.102b3534M}
}

@ARTICLE{markevitch07,
       author = {{Markevitch}, Maxim and {Vikhlinin}, Alexey},
        title = "{Shocks and cold fronts in galaxy clusters}",
      journal = {\physrep},
         year = 2007,
        month = may,
       volume = {443},
       number = {1},
        pages = {1-53},
          doi = {10.1016/j.physrep.2007.01.001},
archivePrefix = {arXiv},
       eprint = {astro-ph/0701821},
 primaryClass = {astro-ph},
       adsurl = {https://ui.adsabs.harvard.edu/abs/2007PhR...443....1M}
}

@ARTICLE{mccarthy24,
       author = {{McCarthy}, Fiona and {Hill}, J. Colin},
        title = "{Component-separated, CIB-cleaned thermal Sunyaev-Zel'dovich maps from Planck PR4 data with a flexible public needlet ILC pipeline}",
      journal = {\prd},
         year = 2024,
        month = jan,
       volume = {109},
       number = {2},
          eid = {023528},
        pages = {023528},
          doi = {10.1103/PhysRevD.109.023528},
archivePrefix = {arXiv},
       eprint = {2307.01043},
 primaryClass = {astro-ph.CO},
       adsurl = {https://ui.adsabs.harvard.edu/abs/2024PhRvD.109b3528M}
}

@ARTICLE{mcdonald14,
       author = {{McDonald}, M. and {Benson}, B.~A. and {Vikhlinin}, A. and {Aird}, K.~A. and {Allen}, S.~W. and {Bautz}, M. and {Bayliss}, M. and {Bleem}, L.~E. et al.},
        title = "{The Redshift Evolution of the Mean Temperature, Pressure, and Entropy Profiles in 80 SPT-Selected Galaxy Clusters}",
      journal = {\apj},
         year = 2014,
        month = oct,
       volume = {794},
       number = {1},
          eid = {67},
        pages = {67},
          doi = {10.1088/0004-637X/794/1/67},
archivePrefix = {arXiv},
       eprint = {1404.6250},
 primaryClass = {astro-ph.HE},
       adsurl = {https://ui.adsabs.harvard.edu/abs/2014ApJ...794...67M}
}

@ARTICLE{mehrtens12,
   author = {{Mehrtens}, N. and {Romer}, A.~K. and {Hilton}, M. and {Lloyd-Davies}, E.~J. and 
	{Miller}, C.~J. and {Stanford}, S.~A. and {Hosmer}, M. and {Hoyle}, B. and 
	{Collins}, C.~A. and {Liddle}, A.~R. and {Viana}, P.~T.~P. and 
	{Nichol}, R.~C. and {Stott}, J.~P. and {Dubois}, E.~N. and {Kay}, S.~T. and 
	{Sahl{\'e}n}, M. and {Young}, O. and {Short}, C.~J. and {Christodoulou}, L. and 
	{Watson}, W.~A. and {Davidson}, M. and {Harrison}, C.~D. and 
	{Baruah}, L. and {Smith}, M. and {Burke}, C. and {Mayers}, J.~A. and 
	{Deadman}, P.-J. and {Rooney}, P.~J. and {Edmondson}, E.~M. and 
	{West}, M. and {Campbell}, H.~C. and {Edge}, A.~C. and {Mann}, R.~G. and 
	{Sabirli}, K. and {Wake}, D. and {Benoist}, C. and {da Costa}, L. and 
	{Maia}, M.~A.~G. and {Ogando}, R.},
    title = "{The XMM Cluster Survey: optical analysis methodology and the first data release}",
  journal = {\mnras},
archivePrefix = "arXiv",
   eprint = {1106.3056},
     year = 2012,
    month = jun,
   volume = 423,
    pages = {1024-1052},
      doi = {10.1111/j.1365-2966.2012.20931.x},
   adsurl = {http://adsabs.harvard.edu/abs/2012MNRAS.423.1024M}
}

@ARTICLE{miyatake19,
       author = {{Miyatake}, Hironao and {Battaglia}, Nicholas and {Hilton}, Matt and {Medezinski}, Elinor and {Nishizawa}, Atsushi J. and {More}, Surhud and {Aiola}, Simone and {Bahcall}, Neta and {Bond}, et al.},
        title = "{Weak-lensing Mass Calibration of ACTPol Sunyaev-Zel{\textquoteright}dovich Clusters with the Hyper Suprime-Cam Survey}",
      journal = {\apj},
         year = 2019,
        month = apr,
       volume = {875},
       number = {1},
          eid = {63},
        pages = {63},
          doi = {10.3847/1538-4357/ab0af0},
archivePrefix = {arXiv},
       eprint = {1804.05873},
 primaryClass = {astro-ph.CO},
       adsurl = {https://ui.adsabs.harvard.edu/abs/2019ApJ...875...63M}
}

@ARTICLE{nagai07,
   author = {{Nagai}, D. and {Kravtsov}, A.~V. and {Vikhlinin}, A.},
    title = "{Effects of Galaxy Formation on Thermodynamics of the Intracluster Medium}",
  journal = {\apj},
   eprint = {astro-ph/0703661},
     year = 2007,
    month = oct,
   volume = 668,
    pages = {1-14},
      doi = {10.1086/521328},
   adsurl = {http://adsabs.harvard.edu/abs/2007ApJ...668....1N}
}

@ARTICLE{navarro97,
   author = {{Navarro}, J.~F. and {Frenk}, C.~S. and {White}, S.~D.~M.},
    title = "{A Universal Density Profile from Hierarchical Clustering}",
  journal = {\apj},
   eprint = {astro-ph/9611107},
     year = 1997,
    month = dec,
   volume = 490,
    pages = {493-508},
      doi = {10.1086/304888},
   adsurl = {http://adsabs.harvard.edu/abs/1997ApJ...490..493N}
}

@ARTICLE{newman13,
       author = {{Newman}, Andrew B. and {Treu}, Tommaso and {Ellis}, Richard S. and {Sand}, David J. and {Nipoti}, Carlo and {Richard}, Johan and {Jullo}, Eric},
        title = "{The Density Profiles of Massive, Relaxed Galaxy Clusters. I. The Total Density Over Three Decades in Radius}",
      journal = {\apj},
         year = 2013,
        month = mar,
       volume = {765},
       number = {1},
          eid = {24},
        pages = {24},
          doi = {10.1088/0004-637X/765/1/24},
archivePrefix = {arXiv},
       eprint = {1209.1391},
 primaryClass = {astro-ph.CO},
       adsurl = {https://ui.adsabs.harvard.edu/abs/2013ApJ...765...24N}
}

@ARTICLE{planck_ir_v,
       author = {{Planck Collaboration} and {Ade}, P.~A.~R. and {Aghanim}, N. and {Arnaud}, M. and {Ashdown}, M. and {Atrio-Barandela}, F. and {Aumont}, J. and {Baccigalupi}, C. and {Balbi}, A. and {Banday}, A.~J. and {Barreiro}, R.~B. and {Bartlett}, J.~G. and {Battaner}, E. and {Benabed}, K. },
       title = "{Planck intermediate results. V. Pressure profiles of galaxy clusters from the Sunyaev-Zeldovich effect}",
      journal = {\aap},
         year = 2013,
        month = feb,
       volume = {550},
          eid = {A131},
        pages = {A131},
          doi = {10.1051/0004-6361/201220040},
archivePrefix = {arXiv},
       eprint = {1207.4061},
 primaryClass = {astro-ph.CO},
       adsurl = {https://ui.adsabs.harvard.edu/abs/2013A&A...550A.131P}
}

@ARTICLE{planck_15_xxvii,
       author = {{Planck Collaboration} and {Ade}, P.~A.~R. and {Aghanim}, N. and {Arnaud}, M. and {Ashdown}, M. and {Aumont}, J. and {Baccigalupi}, C. and {Banday}, A.~J. and {Barreiro}, R.~B. and {Barrena}, R. and {Bartlett}, J.~G. et al.},
        title = "{Planck 2015 results. XXVII. The second Planck catalogue of Sunyaev-Zeldovich sources}",
      journal = {\aap},
         year = 2016,
        month = sep,
       volume = {594},
          eid = {A27},
        pages = {A27},
          doi = {10.1051/0004-6361/201525823},
archivePrefix = {arXiv},
       eprint = {1502.01598},
 primaryClass = {astro-ph.CO},
       adsurl = {https://ui.adsabs.harvard.edu/abs/2016A&A...594A..27P}
}

@ARTICLE{planck_15_xxii,
       author = {{Planck Collaboration} and {Aghanim}, N. and {Arnaud}, M. and {Ashdown}, M. and {Aumont}, J. and {Baccigalupi}, C. and {Banday}, A.~J. and {Barreiro}, R.~B. and {Bartlett}, J.~G. and {Bartolo}, N. and {Battaner}, E. et al.},
        title = "{Planck 2015 results. XXII. A map of the thermal Sunyaev-Zeldovich effect}",
      journal = {\aap},
         year = 2016,
        month = sep,
       volume = {594},
          eid = {A22},
        pages = {A22},
          doi = {10.1051/0004-6361/201525826},
archivePrefix = {arXiv},
       eprint = {1502.01596},
 primaryClass = {astro-ph.CO},
       adsurl = {https://ui.adsabs.harvard.edu/abs/2016A&A...594A..22P}
}

@ARTICLE{planck_18_vi,
       author = {{Planck Collaboration} and {Aghanim}, N. and {Akrami}, Y. and {Ashdown}, M. and {Aumont}, J. and {Baccigalupi}, C. and {Ballardini}, M. and {Banday}, A.~J. and {Barreiro}, R.~B. and {Bartolo}, N. and {Basak}, S. et al.},
        title = "{Planck 2018 results. VI. Cosmological parameters}",
      journal = {\aap},
         year = 2020,
        month = sep,
       volume = {641},
          eid = {A6},
        pages = {A6},
          doi = {10.1051/0004-6361/201833910},
archivePrefix = {arXiv},
       eprint = {1807.06209},
 primaryClass = {astro-ph.CO},
       adsurl = {https://ui.adsabs.harvard.edu/abs/2020A&A...641A...6P}
}

@ARTICLE{pointecouteau21,
       author = {{Pointecouteau}, E. and {Santiago-Bautista}, I. and {Douspis}, M. and {Aghanim}, N. and {Crichton}, D. and {Diego}, J. -M. and {Hurier}, G. and {Macias-Perez}, J. and {Marriage}, T.~A. and {Remazeilles}, M. and {Caretta}, C.~A. and {Bravo-Alfaro}, H.},
        title = "{PACT. II. Pressure profiles of galaxy clusters using Planck and ACT}",
      journal = {\aap},
         year = 2021,
        month = jul,
       volume = {651},
          eid = {A73},
        pages = {A73},
          doi = {10.1051/0004-6361/202040213},
       adsurl = {https://ui.adsabs.harvard.edu/abs/2021A&A...651A..73P}
}

@ARTICLE{rainer17,
       author = {{Weinberger}, Rainer and {Ehlert}, Kristian and {Pfrommer}, Christoph and {Pakmor}, R{\"u}diger and {Springel}, Volker},
        title = "{Simulating the interaction of jets with the intracluster medium}",
      journal = {\mnras},
         year = 2017,
        month = oct,
       volume = {470},
       number = {4},
        pages = {4530-4546},
          doi = {10.1093/mnras/stx1409},
archivePrefix = {arXiv},
       eprint = {1703.09223},
 primaryClass = {astro-ph.GA},
       adsurl = {https://ui.adsabs.harvard.edu/abs/2017MNRAS.470.4530W}
}

@ARTICLE{sand04,
       author = {{Sand}, David J. and {Treu}, Tommaso and {Smith}, Graham P. and {Ellis}, Richard S.},
        title = "{The Dark Matter Distribution in the Central Regions of Galaxy Clusters: Implications for Cold Dark Matter}",
      journal = {\apj},
         year = 2004,
        month = mar,
       volume = {604},
       number = {1},
        pages = {88-107},
          doi = {10.1086/382146},
archivePrefix = {arXiv},
       eprint = {astro-ph/0309465},
 primaryClass = {astro-ph},
       adsurl = {https://ui.adsabs.harvard.edu/abs/2004ApJ...604...88S}
}

@BOOK{sarazin88,
       author = {{Sarazin}, Craig L.},
        title = "{X-ray emission from clusters of galaxies}",
         year = 1988,
       adsurl = {https://ui.adsabs.harvard.edu/abs/1988xrec.book.....S}
}

@ARTICLE{sayers16,
       author = {{Sayers}, Jack and {Golwala}, Sunil R. and {Mantz}, Adam B. and {Merten}, Julian and {Molnar}, Sandor M. and {Naka}, Michael and {Pailet}, Gregory and {Pierpaoli}, Elena and {Siegel}, Seth R. and {Wolman}, Ben},
        title = "{A Comparison and Joint Analysis of Sunyaev-Zel{\textquoteright}dovich Effect Measurements from Planck and Bolocam for a Set of 47 Massive Galaxy Clusters}",
      journal = {\apj},
         year = 2016,
        month = nov,
       volume = {832},
       number = {1},
          eid = {26},
        pages = {26},
          doi = {10.3847/0004-637X/832/1/26},
archivePrefix = {arXiv},
       eprint = {1605.03541},
 primaryClass = {astro-ph.CO},
       adsurl = {https://ui.adsabs.harvard.edu/abs/2016ApJ...832...26S}
}

@ARTICLE{sunyaev72,
   author = {{Sunyaev}, R.~A. and {Zeldovich}, Y.~B.},
    title = "{The Observations of Relic Radiation as a Test of the Nature of X-Ray Radiation from the Clusters of Galaxies}",
  journal = {Comments on Astrophysics and Space Physics},
     year = 1972,
    month = nov,
   volume = 4,
    pages = {173},
   adsurl = {http://adsabs.harvard.edu/abs/1972CoASP...4..173S}
}

@ARTICLE{tramonte23,
       author = {{Tramonte}, Denis and {Ma}, Yin-Zhe and {Yan}, Ziang and {Maturi}, Matteo and {Castignani}, Gianluca and {Sereno}, Mauro and {Bardelli}, Sandro and {Giocoli}, Carlo and {Marulli}, Federico and {Moscardini}, Lauro and {Puddu}, Emanuella and {Radovich}, Mario and {Van Waerbeke}, Ludovic and {Wright}, Angus H.},
        title = "{Exploring the Mass and Redshift Dependencies of the Cluster Pressure Profile with Stacks on Thermal Sunyaev-Zel'dovich Maps}",
      journal = {\apjs},
         year = 2023,
        month = apr,
       volume = {265},
       number = {2},
          eid = {55},
        pages = {55},
          doi = {10.3847/1538-4365/acbcca},
archivePrefix = {arXiv},
       eprint = {2302.06266},
 primaryClass = {astro-ph.CO},
       adsurl = {https://ui.adsabs.harvard.edu/abs/2023ApJS..265...55T}
}

@ARTICLE{vikhlinin06,
       author = {{Vikhlinin}, A. and {Kravtsov}, A. and {Forman}, W. and {Jones}, C. and {Markevitch}, M. and {Murray}, S.~S. and {Van Speybroeck}, L.},
        title = "{Chandra Sample of Nearby Relaxed Galaxy Clusters: Mass, Gas Fraction, and Mass-Temperature Relation}",
      journal = {\apj},
         year = 2006,
        month = apr,
       volume = {640},
       number = {2},
        pages = {691-709},
          doi = {10.1086/500288},
archivePrefix = {arXiv},
       eprint = {astro-ph/0507092},
 primaryClass = {astro-ph},
       adsurl = {https://ui.adsabs.harvard.edu/abs/2006ApJ...640..691V}
}

@ARTICLE{vikhlinin09,
       author = {{Vikhlinin}, A. and {Burenin}, R.~A. and {Ebeling}, H. and {Forman}, W.~R. and {Hornstrup}, A. and {Jones}, C. and {Kravtsov}, A.~V. and {Murray}, S.~S. and {Nagai}, D. and {Quintana}, H. and {Voevodkin}, A.},
        title = "{Chandra Cluster Cosmology Project. II. Samples and X-Ray Data Reduction}",
      journal = {\apj},
         year = 2009,
        month = feb,
       volume = {692},
       number = {2},
        pages = {1033-1059},
          doi = {10.1088/0004-637X/692/2/1033},
archivePrefix = {arXiv},
       eprint = {0805.2207},
 primaryClass = {astro-ph},
       adsurl = {https://ui.adsabs.harvard.edu/abs/2009ApJ...692.1033V}
}

@ARTICLE{voges99,
   author = {{Voges}, W. and {Aschenbach}, B. and {Boller}, T. and {Br{\"a}uninger}, H. and 
	{Briel}, U. and {Burkert}, W. and {Dennerl}, K. and {Englhauser}, J. and 
	{Gruber}, R. and {Haberl}, F. and {Hartner}, G. and {Hasinger}, G. and 
	{K{\"u}rster}, M. and {Pfeffermann}, E. and {Pietsch}, W. and 
	{Predehl}, P. and {Rosso}, C. and {Schmitt}, J.~H.~M.~M. and 
	{Tr{\"u}mper}, J. and {Zimmermann}, H.~U.},
    title = "{The ROSAT all-sky survey bright source catalogue}",
  journal = {\aap},
   eprint = {astro-ph/9909315},
     year = 1999,
    month = sep,
   volume = 349,
    pages = {389-405},
   adsurl = {http://adsabs.harvard.edu/abs/1999A%26A...349..389V}
}

@ARTICLE{voit05,
       author = {{Voit}, G. Mark},
        title = "{Tracing cosmic evolution with clusters of galaxies}",
      journal = {Reviews of Modern Physics},
         year = 2005,
        month = apr,
       volume = {77},
       number = {1},
        pages = {207-258},
          doi = {10.1103/RevModPhys.77.207},
archivePrefix = {arXiv},
       eprint = {astro-ph/0410173},
 primaryClass = {astro-ph},
       adsurl = {https://ui.adsabs.harvard.edu/abs/2005RvMP...77..207V}
}

@ARTICLE{zhang11,
       author = {{Zhang}, Y. -Y. and {Andernach}, H. and {Caretta}, C.~A. and {Reiprich}, T.~H. and {B{\"o}hringer}, H. and {Puchwein}, E. and {Sijacki}, D. and {Girardi}, M.},
        title = "{HIFLUGCS: Galaxy cluster scaling relations between X-ray luminosity, gas mass, cluster radius, and velocity dispersion}",
      journal = {\aap},
         year = 2011,
        month = feb,
       volume = {526},
          eid = {A105},
        pages = {A105},
          doi = {10.1051/0004-6361/201015830},
archivePrefix = {arXiv},
       eprint = {1011.3018},
 primaryClass = {astro-ph.CO},
       adsurl = {https://ui.adsabs.harvard.edu/abs/2011A&A...526A.105Z}
}


\appendix

\section{Effect of CIB contamination on the profile reconstruction}
\label{sec:cib_effect}
\begin{figure*}
\includegraphics[trim= 0mm 0mm 0mm 0mm, scale=0.36]{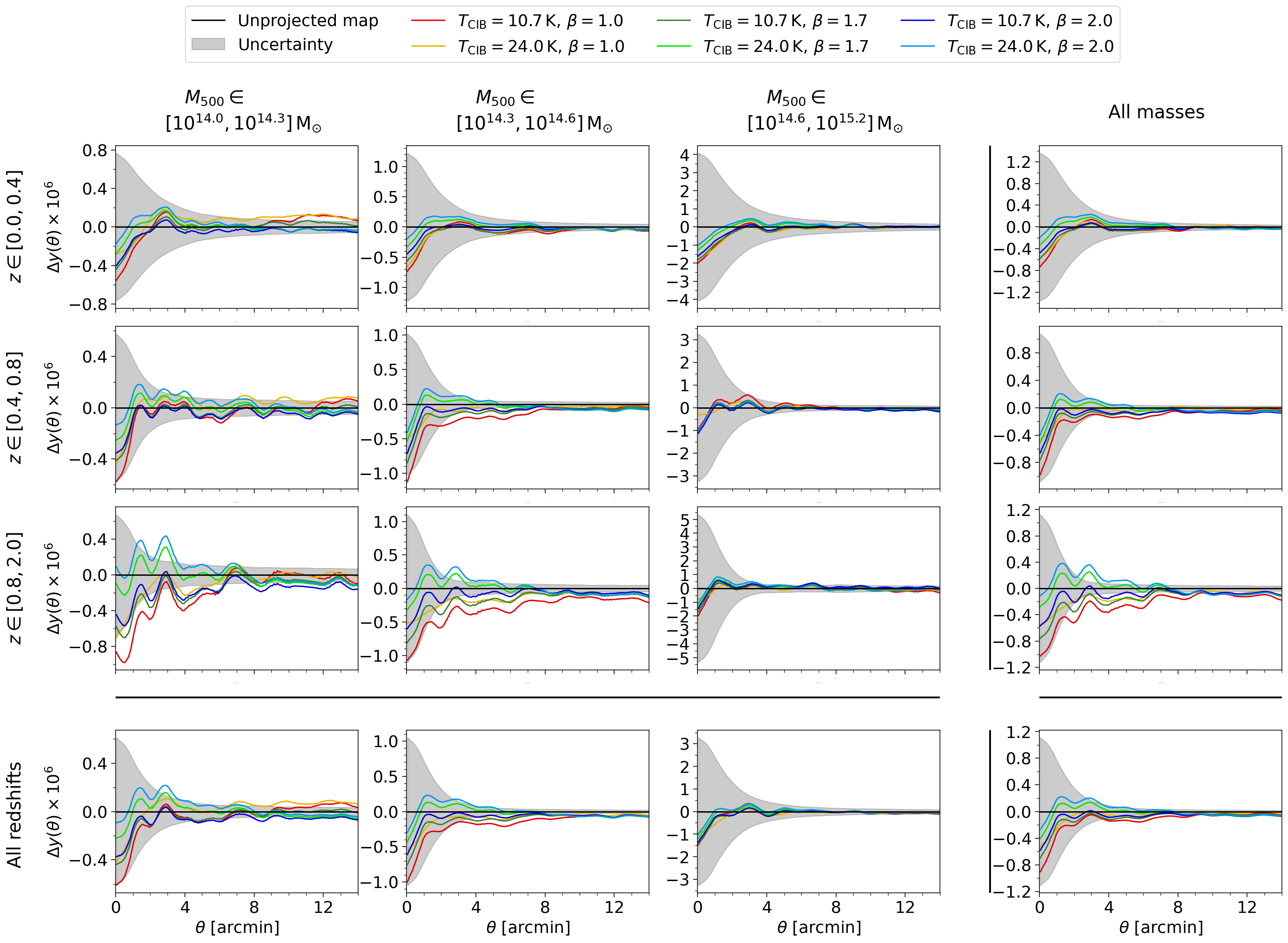}
	\caption{Effect of the CIB contamination on the reconstructed profiles, shown 
	as residuals $\Delta y(\theta)$ between the profiles obtained from the 
	CIB-deprojected maps, and the fiducial profiles obtained from the unprojected map. 
	Results are shown for all the 16 cluster samples employed in this study, and for 
	different versions of the CIB-deprojected maps (as detailed in the legend on top). 
	The uncertainties associated with the fiducial profiles are also represented as 
	shaded gray regions.}
\label{fig:cib_test}
\end{figure*}

Residual contamination from the CIB, a diffuse remnant of various astrophysical 
processes occured throughout cosmic history (particularly dust emission and redshifted light 
from distant galaxies) is a potential source of bias of any analysis based on Compton parameter 
maps. In order to assess whether the presence of CIB residuals in the ACT DR6 Compton parameter 
map constitutes a source of significant bias, the same procedure for cluster stacking and profile 
reconstruction, as described in Sec.~\ref{sec:measurement}, was repeated using different versions 
of the ACT $y$-map obtained deprojecting the CIB component. As anticipated in Sec.~\ref{ssec:ymap}, 
the CIB SED is modeled as a modified blackbody, whose functional form includes the CIB temperature 
$T_{\rm CIB}$ and an emissivity parameter $\beta$. \citet{coulton24} adopted the fiducial values  
$T_{\rm CIB}=10.7\,{\rm K}$ and $\beta=1.7$, as derived in~\citet{mccarthy24}; the previous release 
of the ACT $y$-maps~\citep{madhavacheril20}, instead, adopted the values $T_{\rm CIB}=24\,{\rm K}$ 
and $\beta=1.2$. 

The ACT-DR6 release includes deprojected maps for different values of the CIB temperature and 
emissivity; given the existing uncertainty on these parameters, the effect of CIB contaminations 
on the Compton profiles reconstructed in this study is assessed considering all six combinations 
for the values $T_{\rm CIB}=10.7\,{\rm K}$, $24.0\,{\rm K}$ and $\beta=1.0$, $1.7$, $2.0$. 
For each of these maps and each of the cluster samples considered in this study, the Compton profile 
$y_{\rm CIB}(\theta)$ is reconstructed and compared with the corresponding fiducial profile $y(\theta)$; 
the residuals $\Delta y(\theta)=y_{\rm CIB}(\theta)-y(\theta)$ are then computed and plotted in 
Fig.~\ref{fig:cib_test}. 

The figure shows that CIB deprojection does indeed have an effect on the reconstructed profiles, 
with higher values for the CIB temperature and for the emissivity generally yielding higher $y$ 
amplitudes (i.e., less contamination). We can also see, however, that these variations are generally 
within the uncertainties of the unprojected profiles, and while they always result in a decrement at 
the profile peak, they tend to be centred around the fiducial profiles at higher angular separations. 
We can also notice that the fiducial CIB parameter values adopted in~\citet{coulton24} do not correspond 
to any of the extreme cases, yielding a profile relatively close to the unprojected one. From these 
considerations, it is possible to conclude that CIB contaminations are not an important source of 
systematics in this study, and while their presence should be acknowledged, their impact on the 
final results of the present analysis is expected to be negligible.


\section{Contour plots}
\label{sec:contours}
This section presents the parameter posterior distributions as obtained from the multi-stage
MCMC runs for all pressure models. As described in Sec.~\ref{ssec:practical_implementation}, an 
initial MCMC run was launched to obtain a set of baseline best-fit values 
for all parameters. In the subsequent run, all parameters except $c_{500}$ were fixed to those baseline 
values; the resulting one-dimensional posterior distributions on $c_{500}$ are plotted in 
Fig.~\ref{fig:c500_posteriors}, for all combinations of data sets and pressure profile models. 
The following figures, instead, show the posterior distributions obtained in the last MCMC run, in 
which $c_{500}$ was fixed to its constraints from the previous round, while the other parameters 
were allowed to vary. 
In order, results are shown for the UPP (Fig.~\ref{fig:upp_contours}),
the BMP (Fig.~\ref{fig:bmp_contours}), the PTP (Fig.~\ref{fig:ptp_contours}), 
and the EUP (Fig.~\ref{fig:eup_contours}) pressure profile models.

\begin{figure*}
\includegraphics[trim= 0mm 0mm 0mm 0mm, scale=0.3]{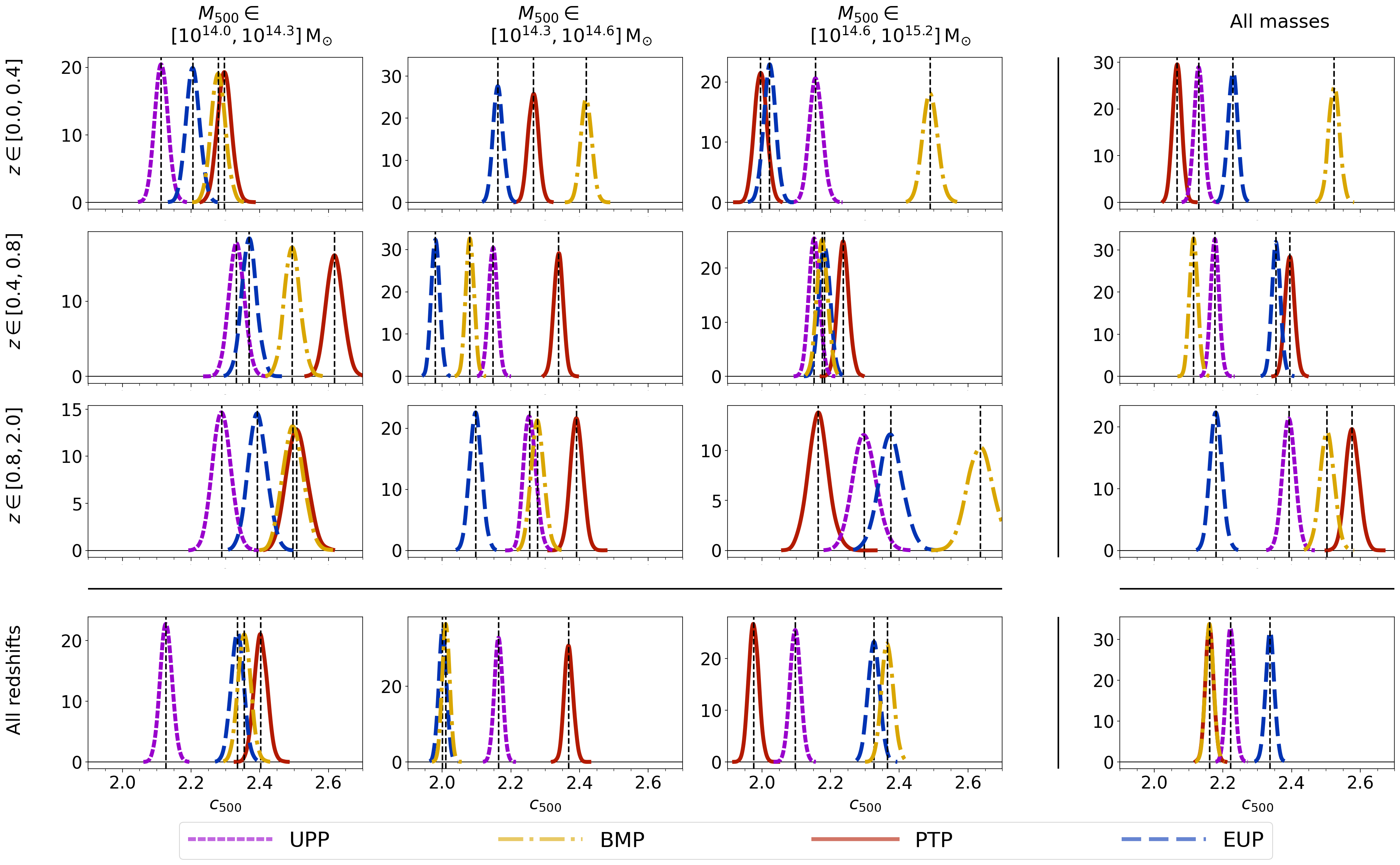}
	\caption{Posterior one-dimensional probability distributions for the fit on $c_{500}$ while keeping 
	the remaining parameters fixed. Results are shown for all mass and redshift bins, and for all four 
	models; vertical dashed lines mark the best-fit values quoted in Tables~\ref{tab:upp_bestfits} to~\ref{tab:eup_bestfits}.}
	\label{fig:c500_posteriors}
\end{figure*}

\begin{figure*}
\includegraphics[trim= 0mm 0mm 0mm 0mm, scale=0.24]{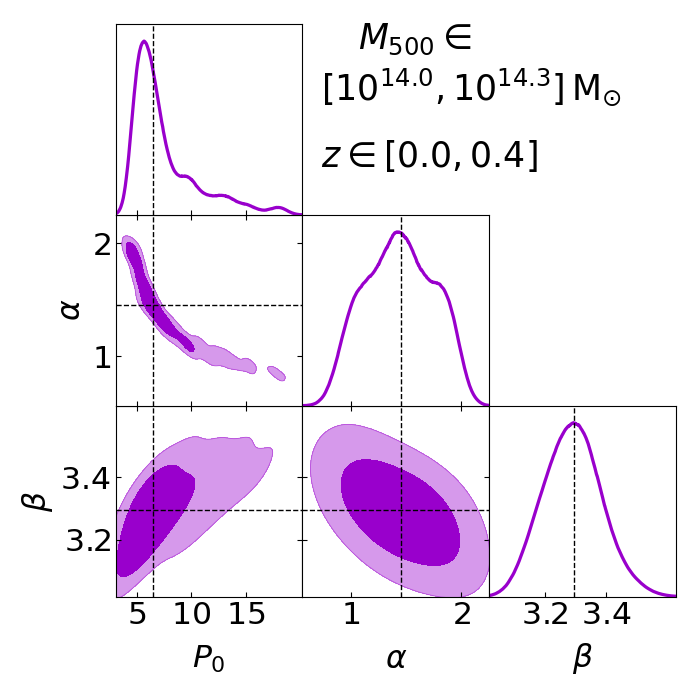}
\includegraphics[trim= 0mm 0mm 0mm 0mm, scale=0.24]{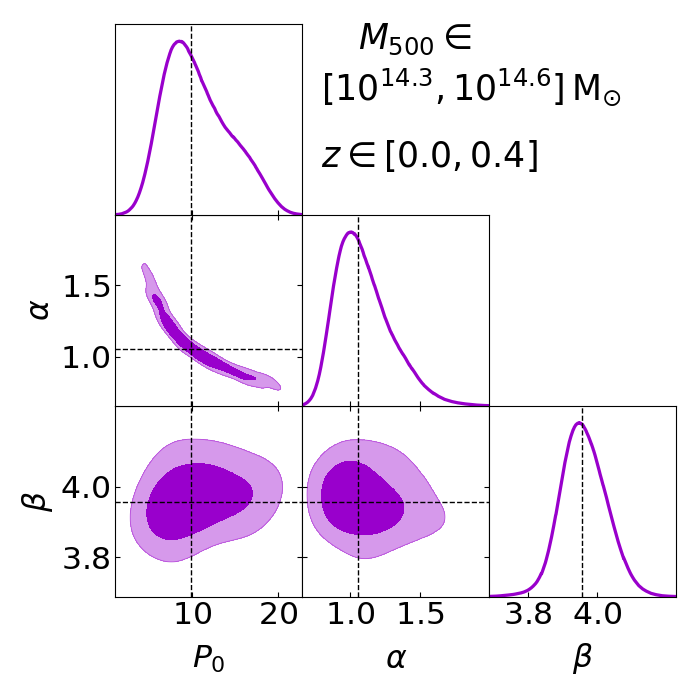}
\includegraphics[trim= 0mm 0mm 0mm 0mm, scale=0.24]{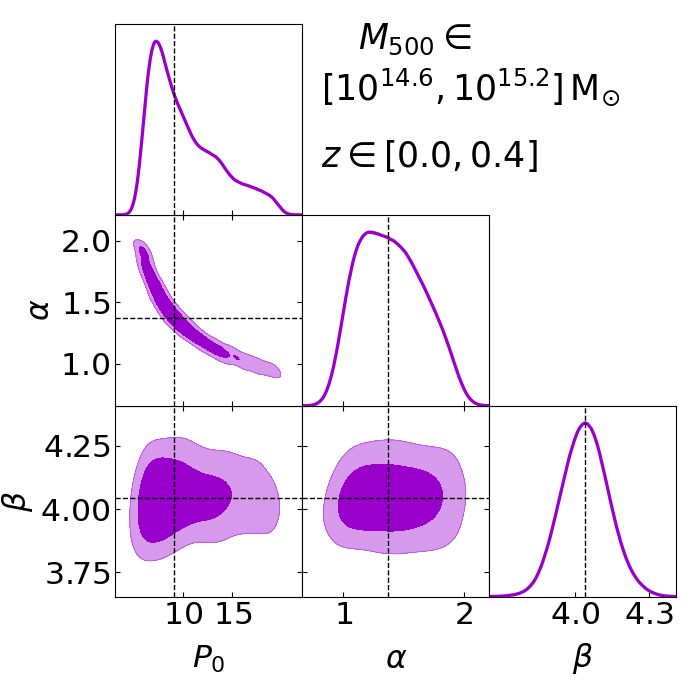}
\includegraphics[trim= 0mm 0mm 0mm 0mm, scale=0.24]{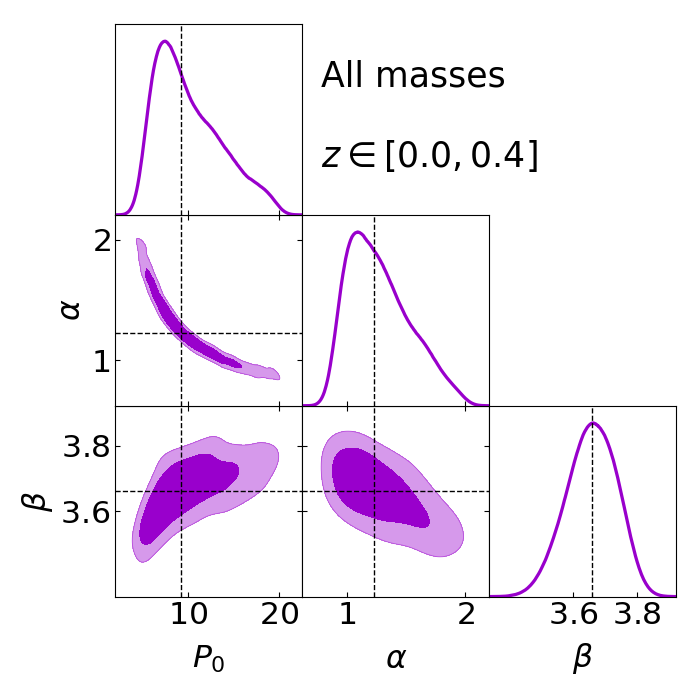} \\
\includegraphics[trim= 0mm 0mm 0mm 0mm, scale=0.24]{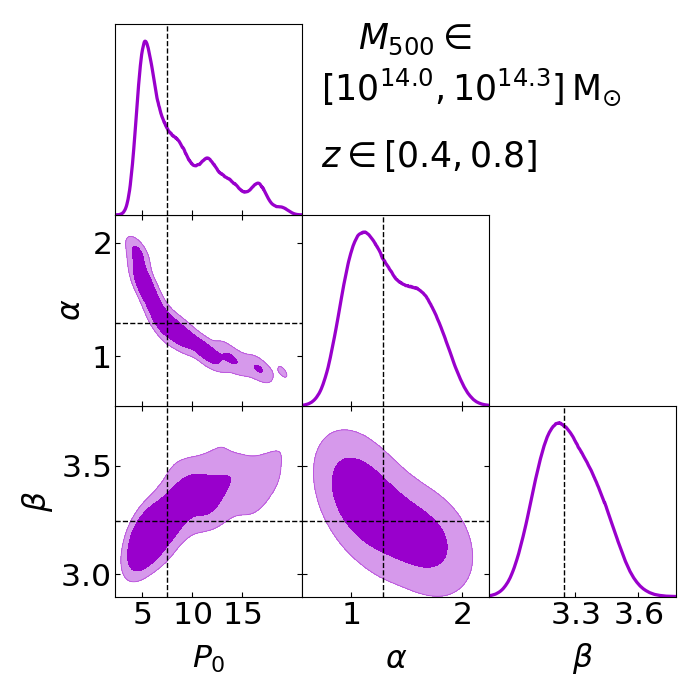}
\includegraphics[trim= 0mm 0mm 0mm 0mm, scale=0.24]{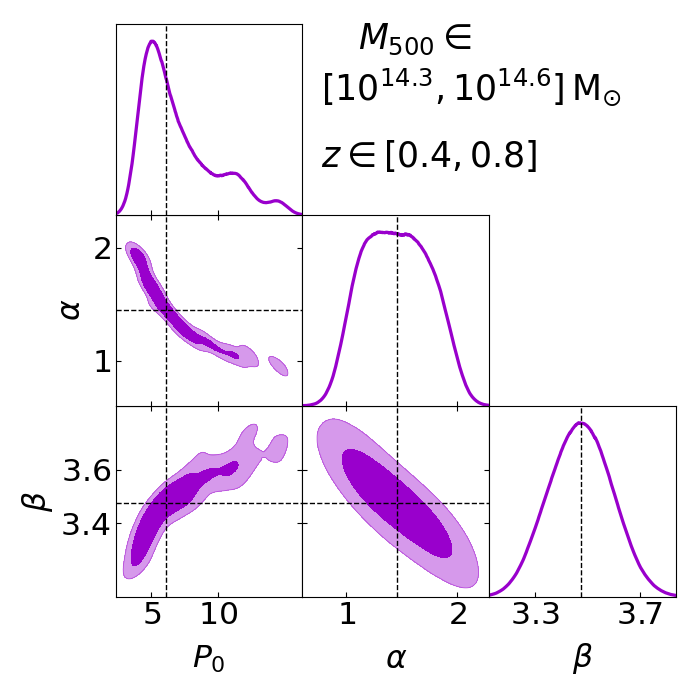}
\includegraphics[trim= 0mm 0mm 0mm 0mm, scale=0.24]{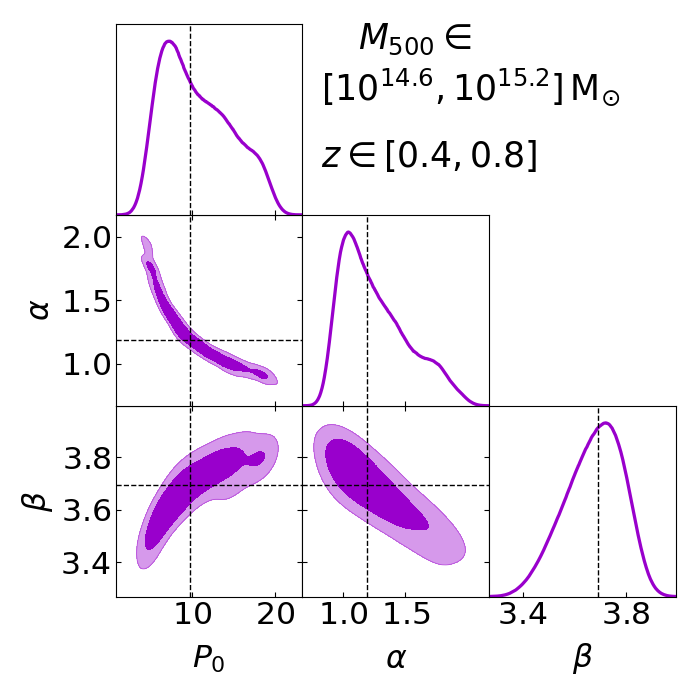}
\includegraphics[trim= 0mm 0mm 0mm 0mm, scale=0.24]{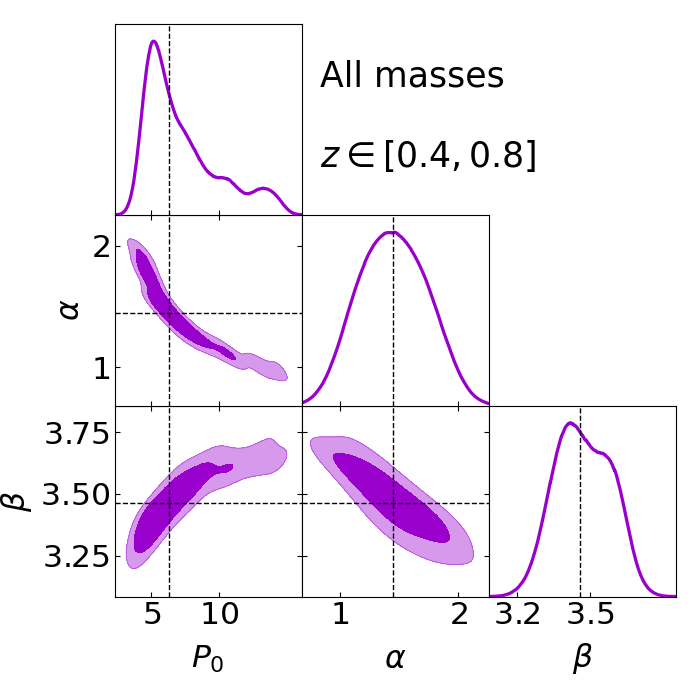} \\
\includegraphics[trim= 0mm 0mm 0mm 0mm, scale=0.24]{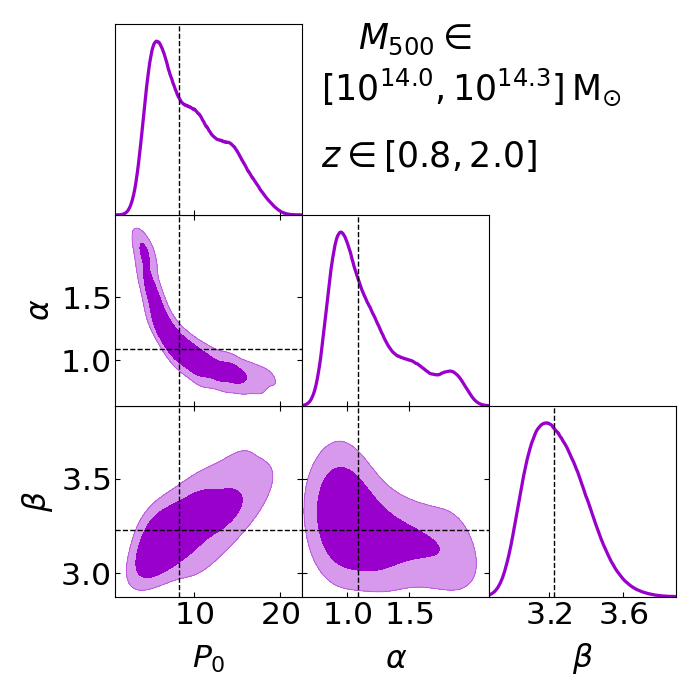}
\includegraphics[trim= 0mm 0mm 0mm 0mm, scale=0.24]{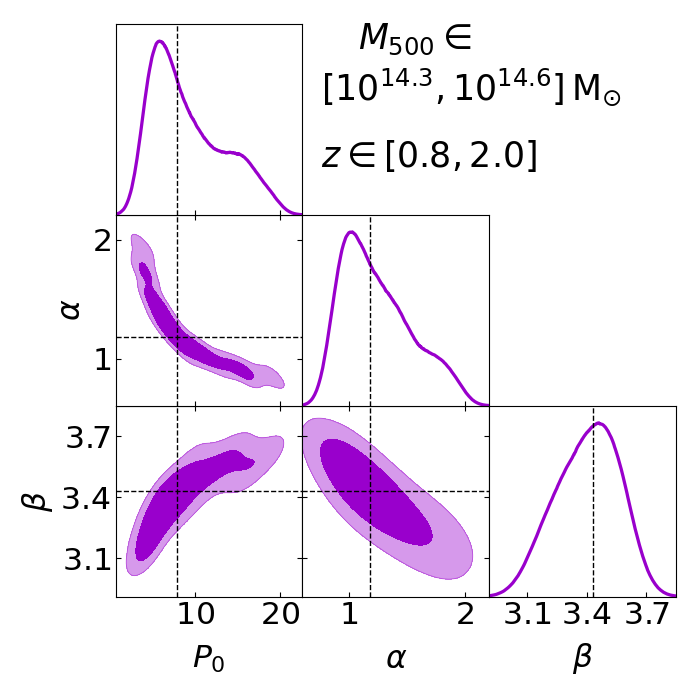}
\includegraphics[trim= 0mm 0mm 0mm 0mm, scale=0.24]{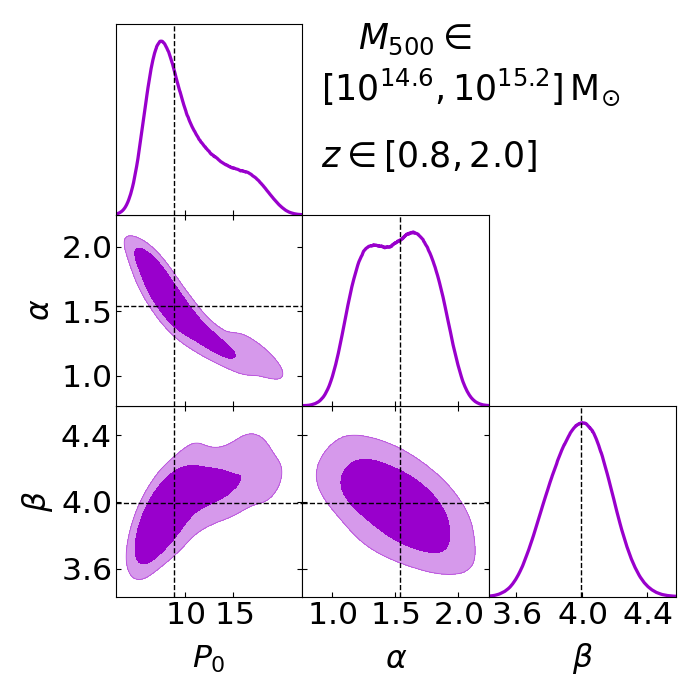}
\includegraphics[trim= 0mm 0mm 0mm 0mm, scale=0.24]{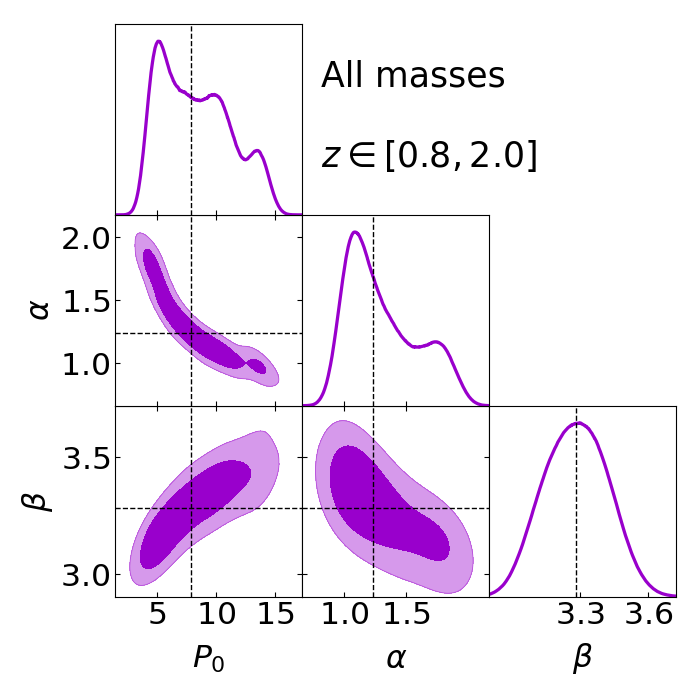} \\
\includegraphics[trim= 0mm 0mm 0mm 0mm, scale=0.24]{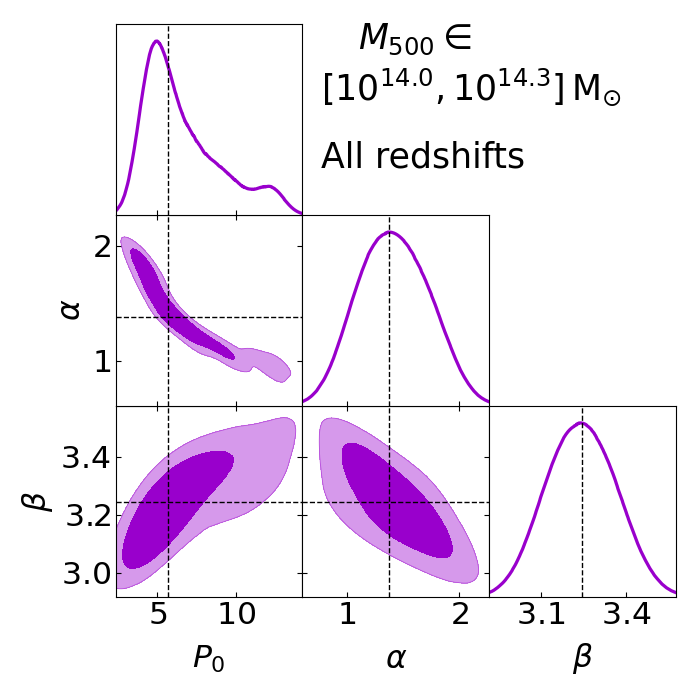}
\includegraphics[trim= 0mm 0mm 0mm 0mm, scale=0.24]{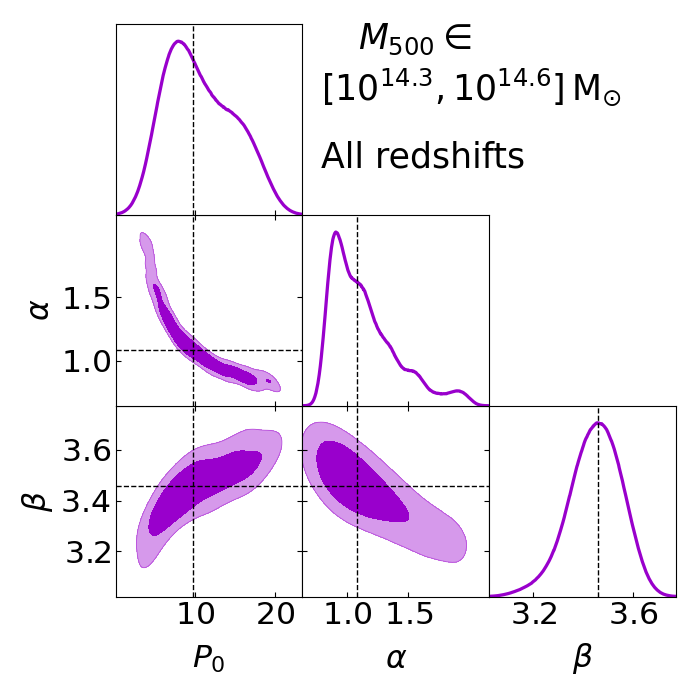}
\includegraphics[trim= 0mm 0mm 0mm 0mm, scale=0.24]{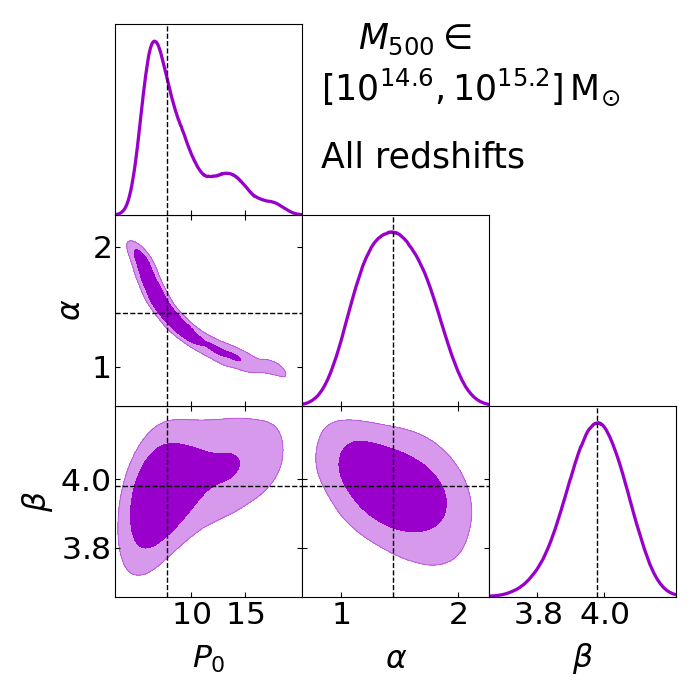}
\includegraphics[trim= 0mm 0mm 0mm 0mm, scale=0.24]{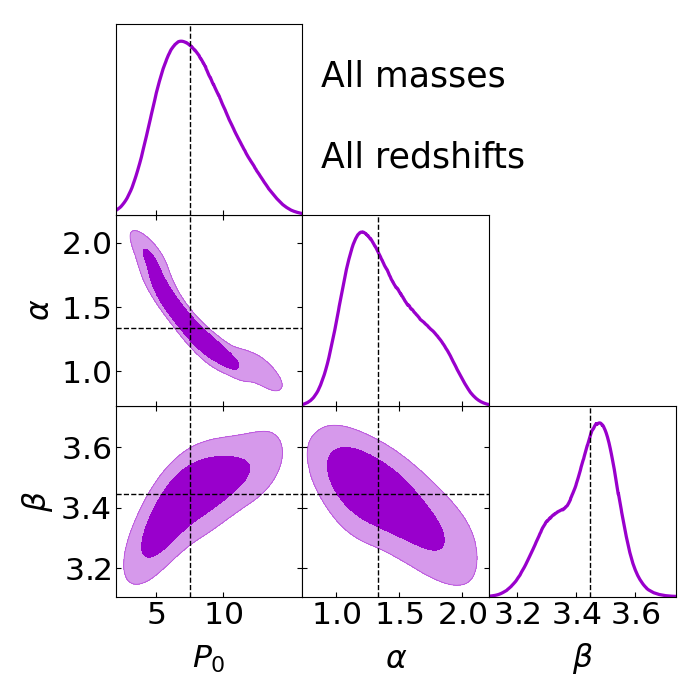} \\
\caption{Posterior contour plots for the parameters entering the UPP model; 
	the dashed lines mark the best-fit values quoted in Table~\ref{tab:upp_bestfits}.}
\label{fig:upp_contours}
\end{figure*}
\begin{figure*}
\includegraphics[trim= 0mm 0mm 0mm 0mm, scale=0.24]{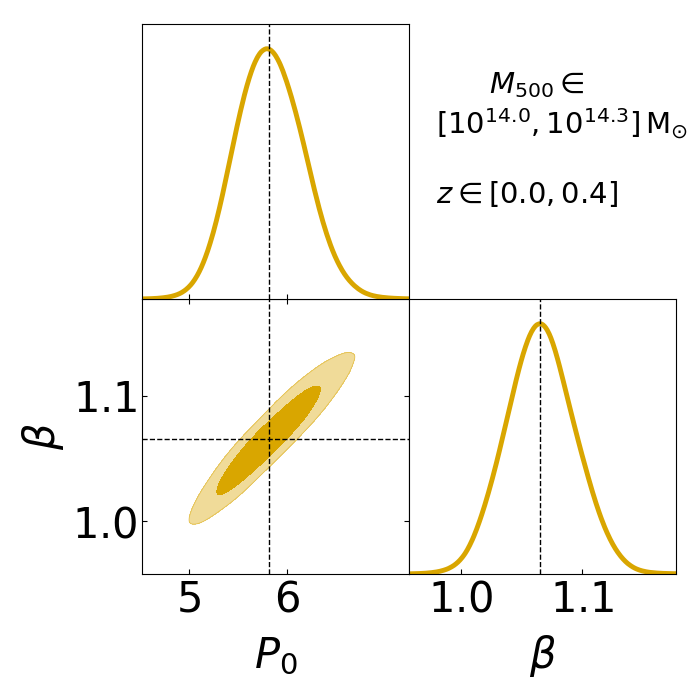}
\includegraphics[trim= 0mm 0mm 0mm 0mm, scale=0.24]{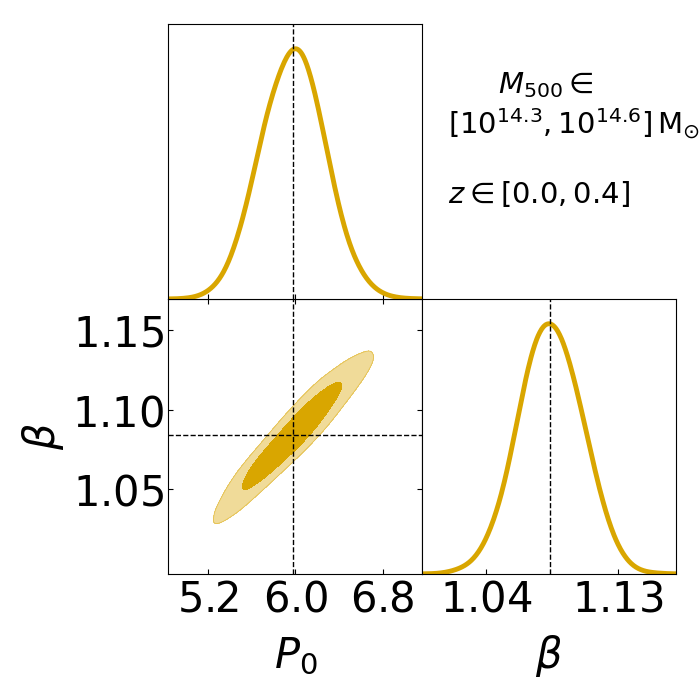}
\includegraphics[trim= 0mm 0mm 0mm 0mm, scale=0.24]{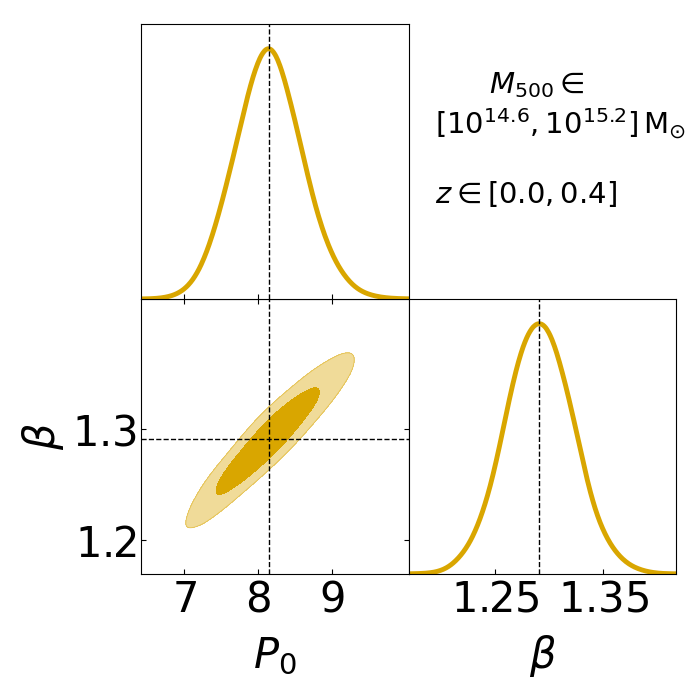}
\includegraphics[trim= 0mm 0mm 0mm 0mm, scale=0.24]{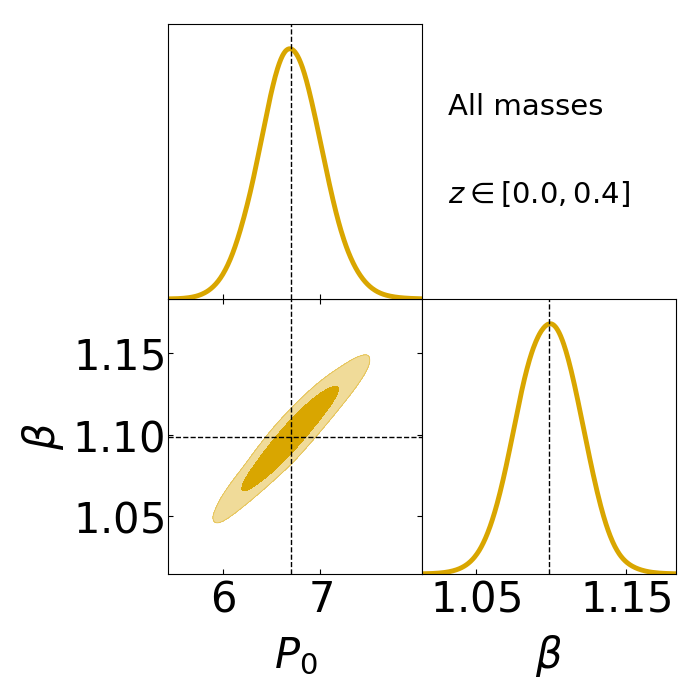} \\
\includegraphics[trim= 0mm 0mm 0mm 0mm, scale=0.24]{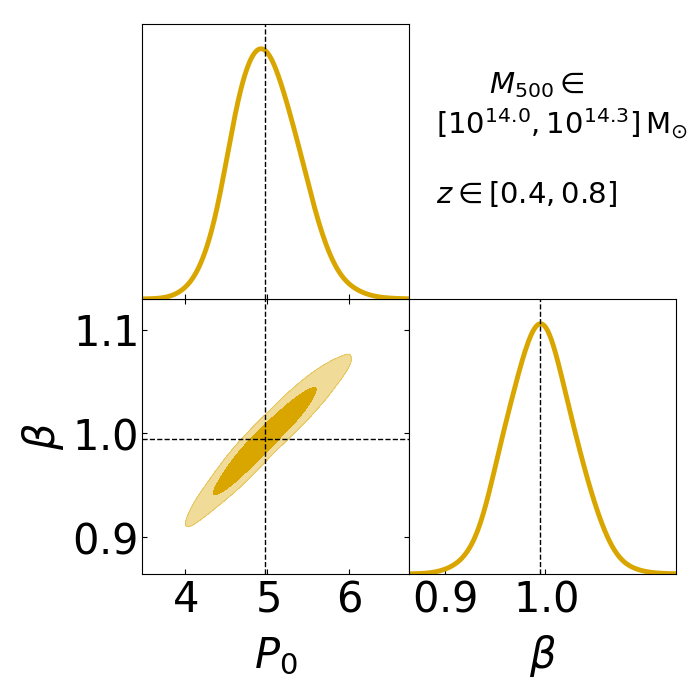}
\includegraphics[trim= 0mm 0mm 0mm 0mm, scale=0.24]{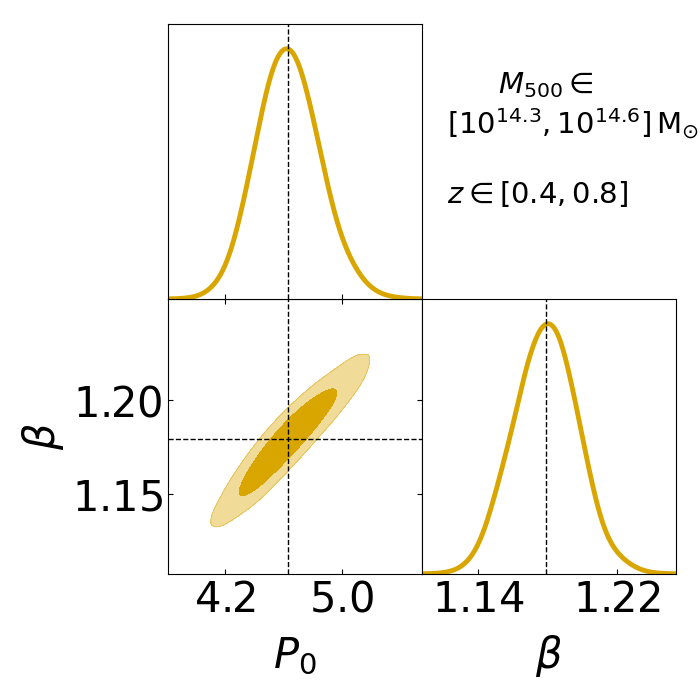}
\includegraphics[trim= 0mm 0mm 0mm 0mm, scale=0.24]{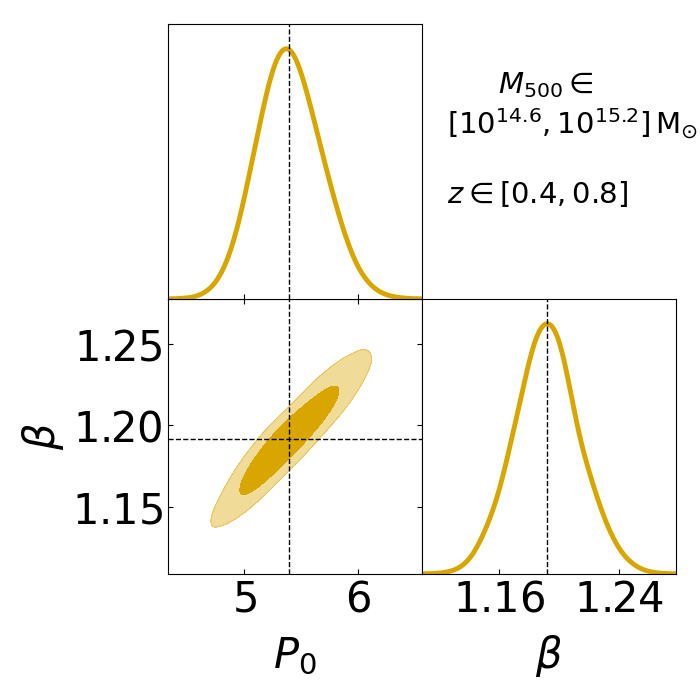}
\includegraphics[trim= 0mm 0mm 0mm 0mm, scale=0.24]{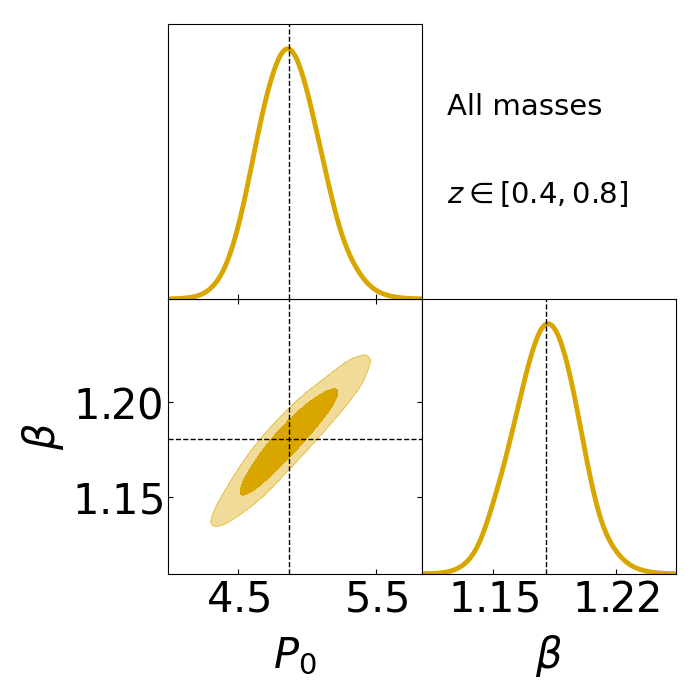} \\
\includegraphics[trim= 0mm 0mm 0mm 0mm, scale=0.24]{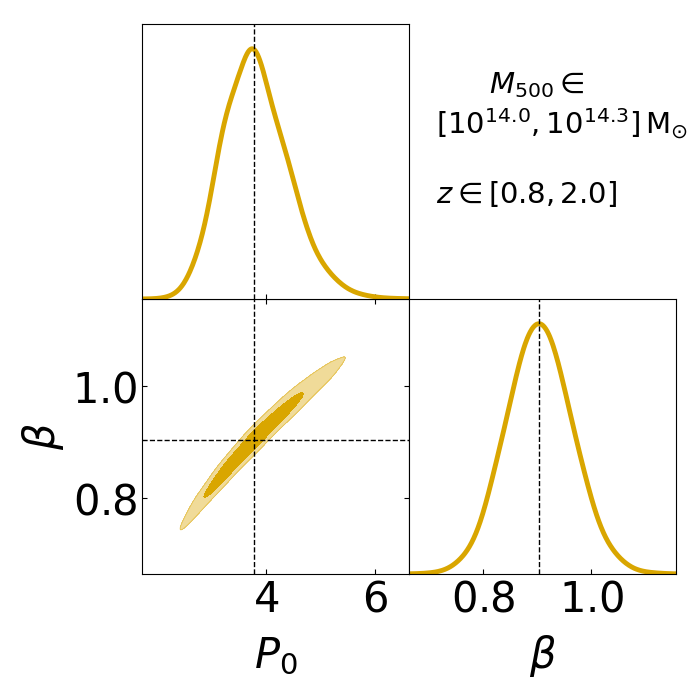}
\includegraphics[trim= 0mm 0mm 0mm 0mm, scale=0.24]{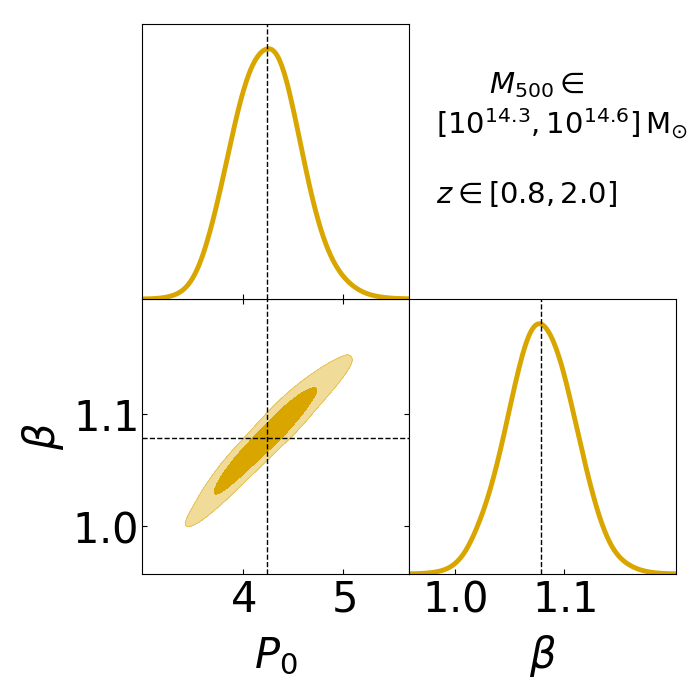}
\includegraphics[trim= 0mm 0mm 0mm 0mm, scale=0.24]{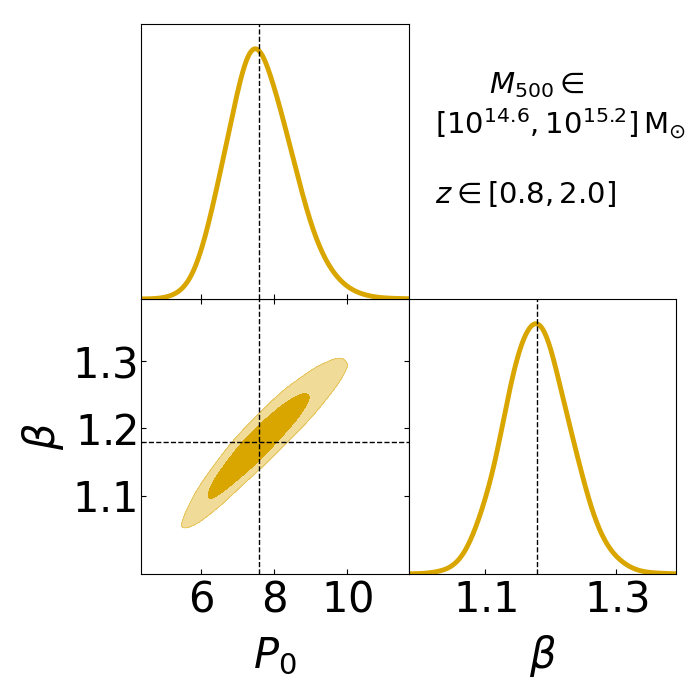}
\includegraphics[trim= 0mm 0mm 0mm 0mm, scale=0.24]{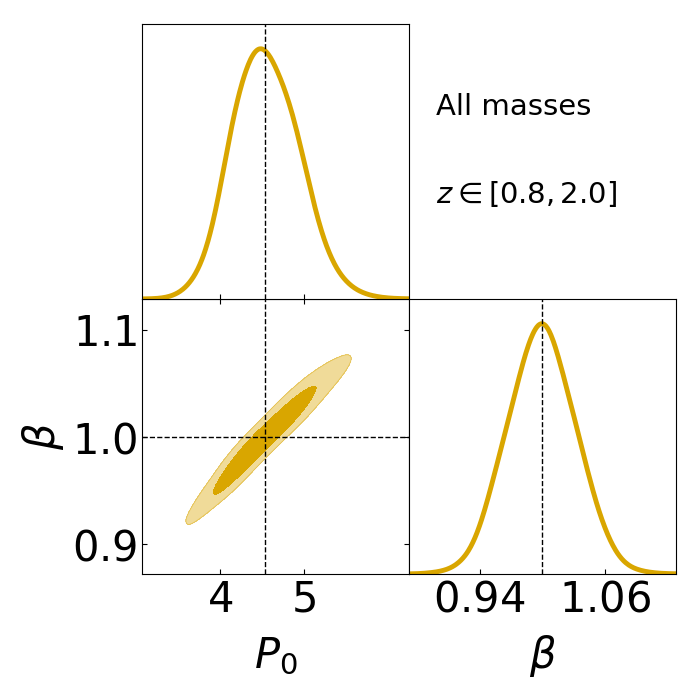} \\
\includegraphics[trim= 0mm 0mm 0mm 0mm, scale=0.24]{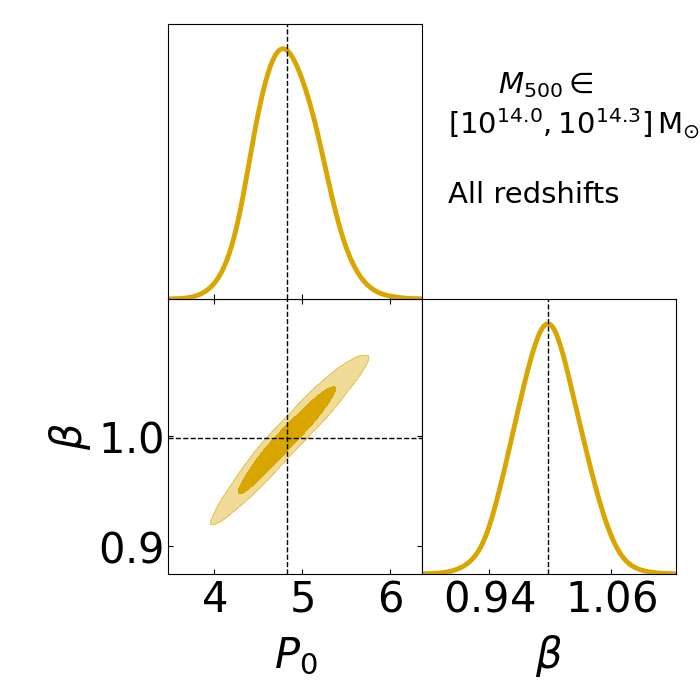}
\includegraphics[trim= 0mm 0mm 0mm 0mm, scale=0.24]{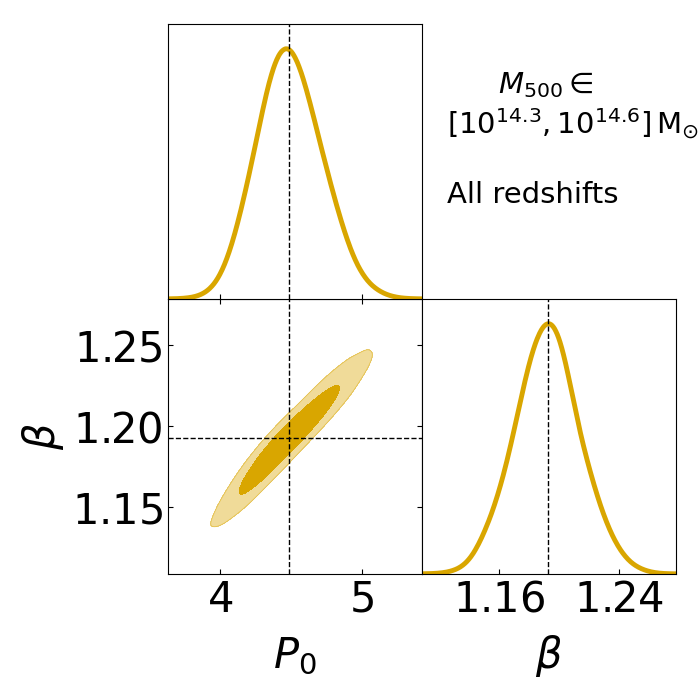}
\includegraphics[trim= 0mm 0mm 0mm 0mm, scale=0.24]{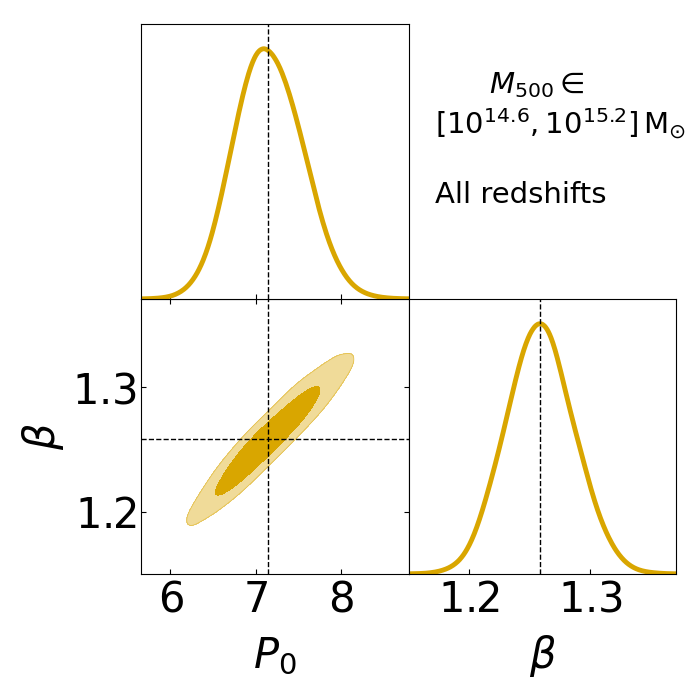}
\includegraphics[trim= 0mm 0mm 0mm 0mm, scale=0.24]{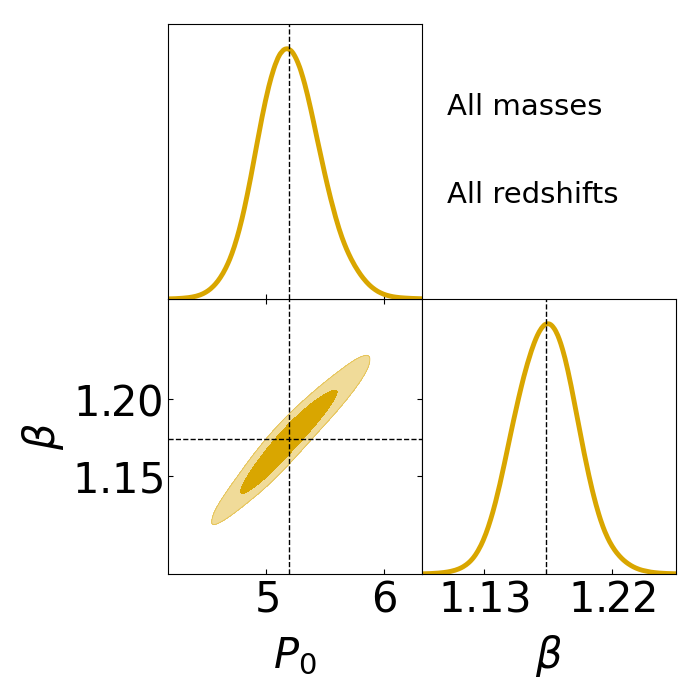} \\
\caption{Posterior contour plots for the parameters entering the BMP model; 
	the dashed lines mark the best-fit values quoted in Table~\ref{tab:bmp_bestfits}.}
\label{fig:bmp_contours}
\end{figure*}
\begin{figure*}
\includegraphics[trim= 0mm 0mm 0mm 0mm, scale=0.24]{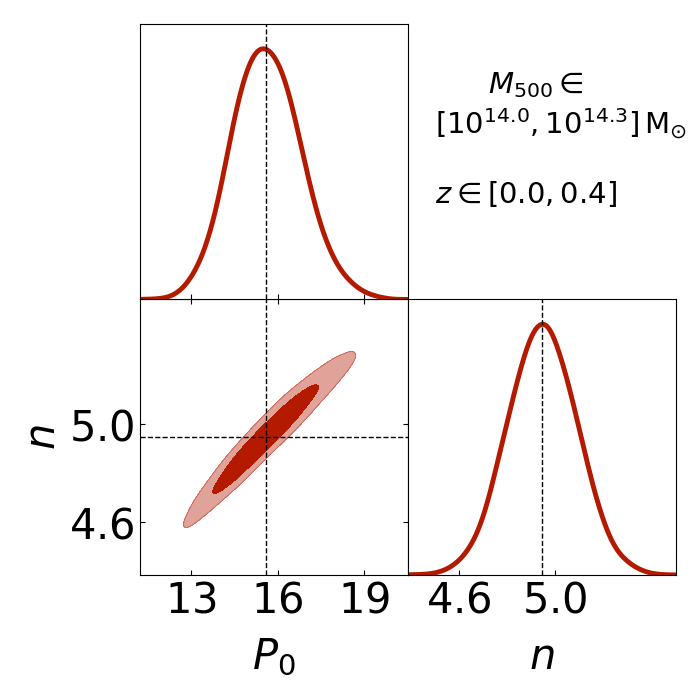}
\includegraphics[trim= 0mm 0mm 0mm 0mm, scale=0.24]{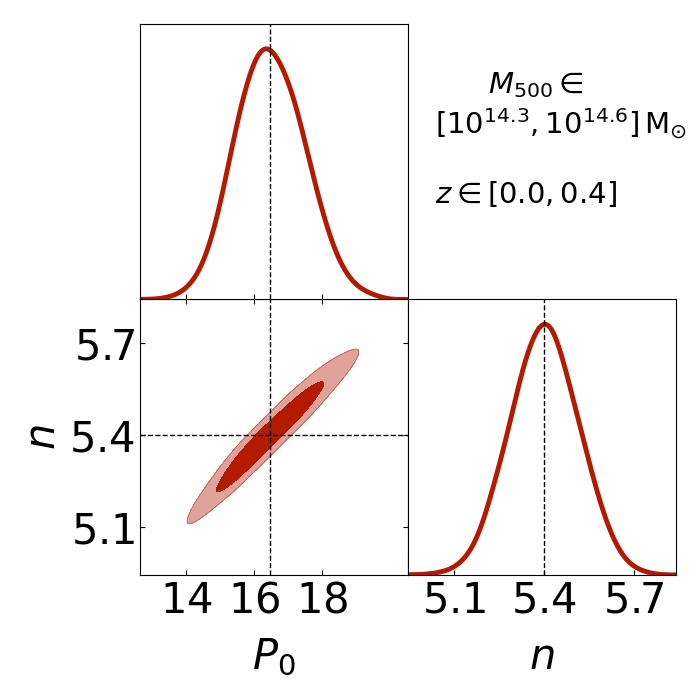}
\includegraphics[trim= 0mm 0mm 0mm 0mm, scale=0.24]{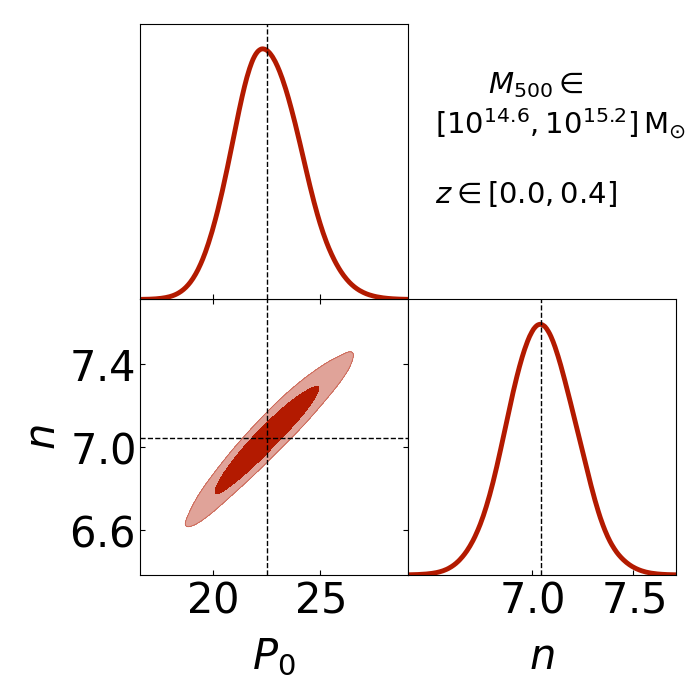}
\includegraphics[trim= 0mm 0mm 0mm 0mm, scale=0.24]{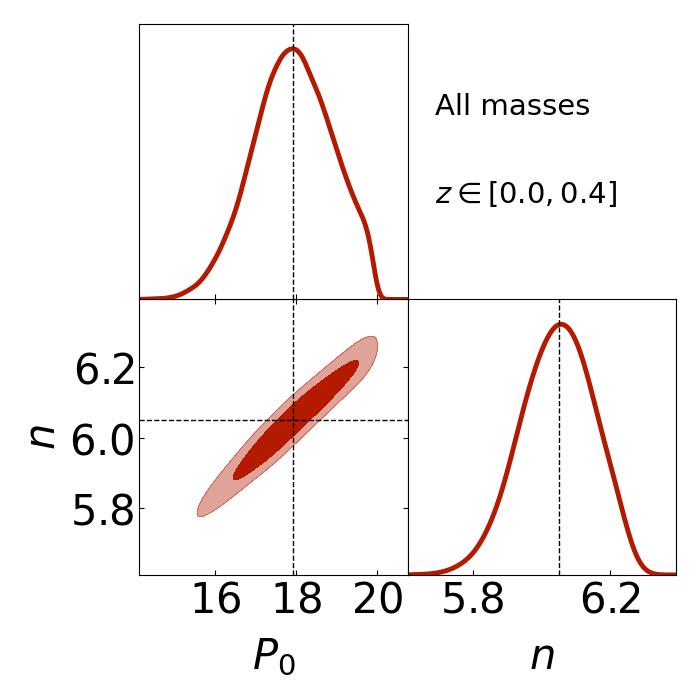} \\
\includegraphics[trim= 0mm 0mm 0mm 0mm, scale=0.24]{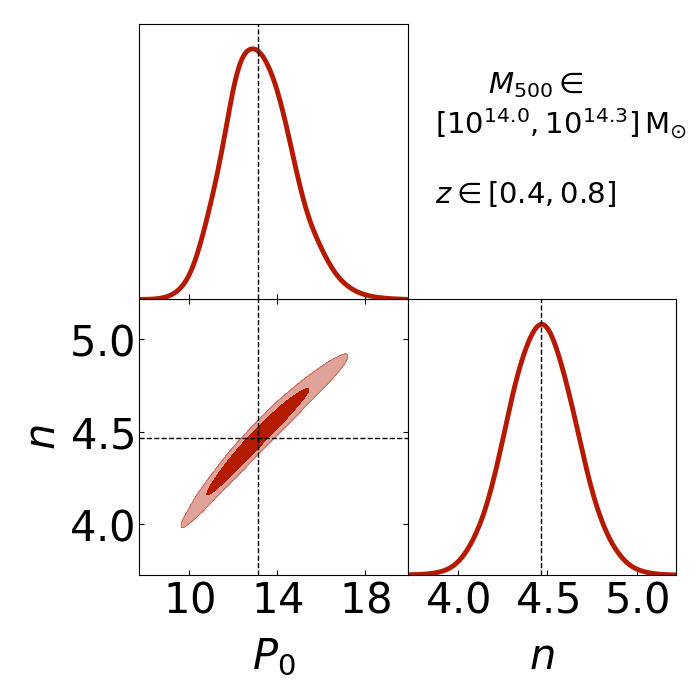}
\includegraphics[trim= 0mm 0mm 0mm 0mm, scale=0.24]{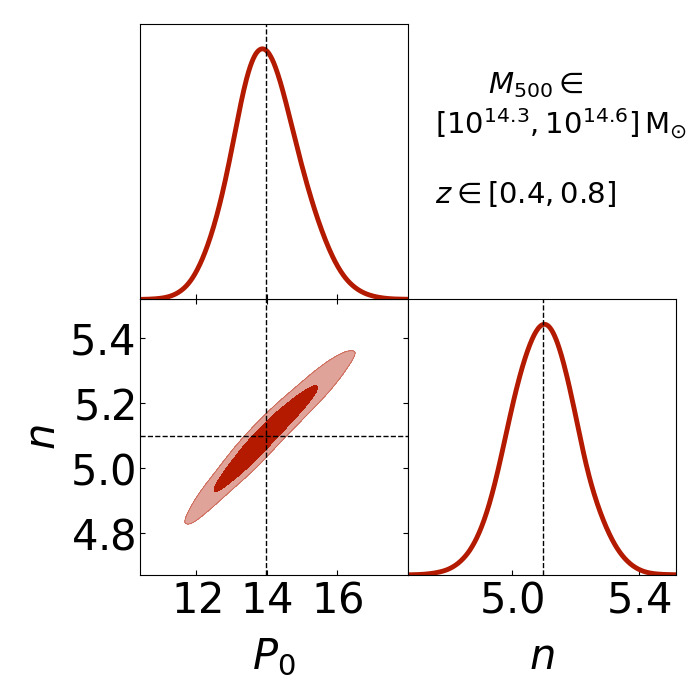}
\includegraphics[trim= 0mm 0mm 0mm 0mm, scale=0.24]{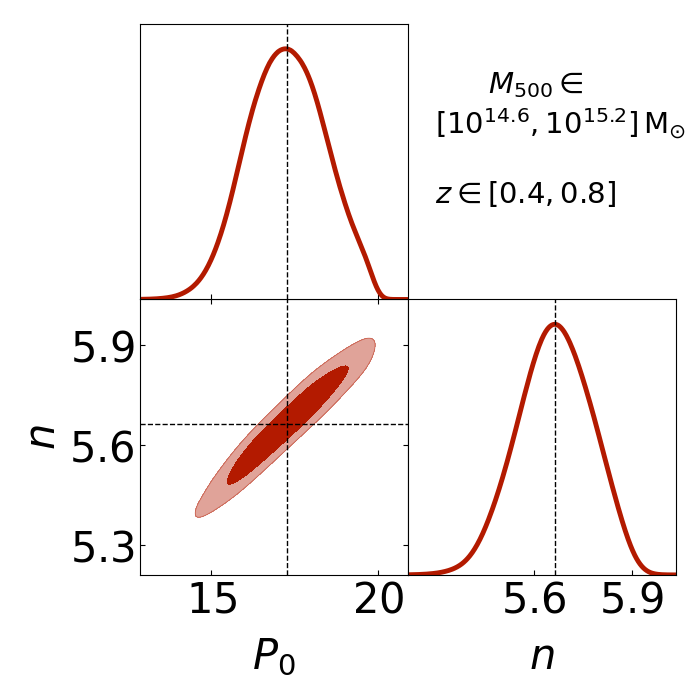}
\includegraphics[trim= 0mm 0mm 0mm 0mm, scale=0.24]{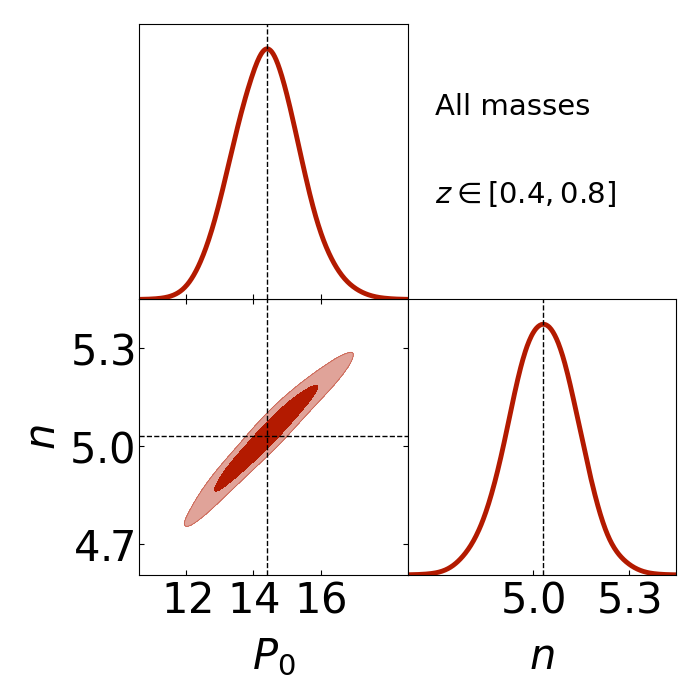} \\
\includegraphics[trim= 0mm 0mm 0mm 0mm, scale=0.24]{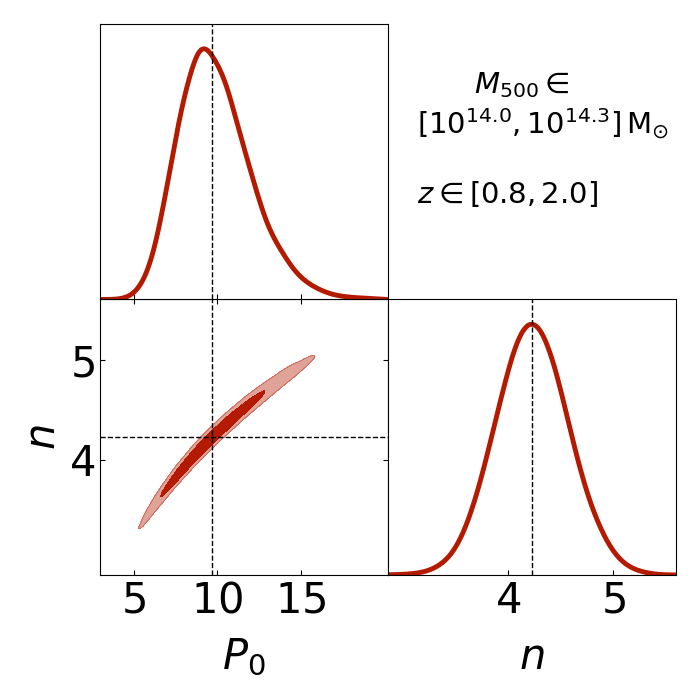}
\includegraphics[trim= 0mm 0mm 0mm 0mm, scale=0.24]{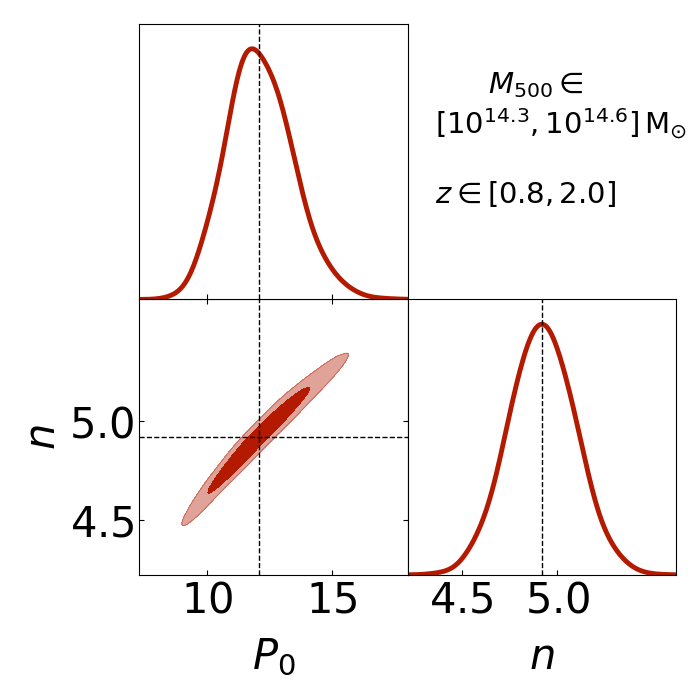}
\includegraphics[trim= 0mm 0mm 0mm 0mm, scale=0.24]{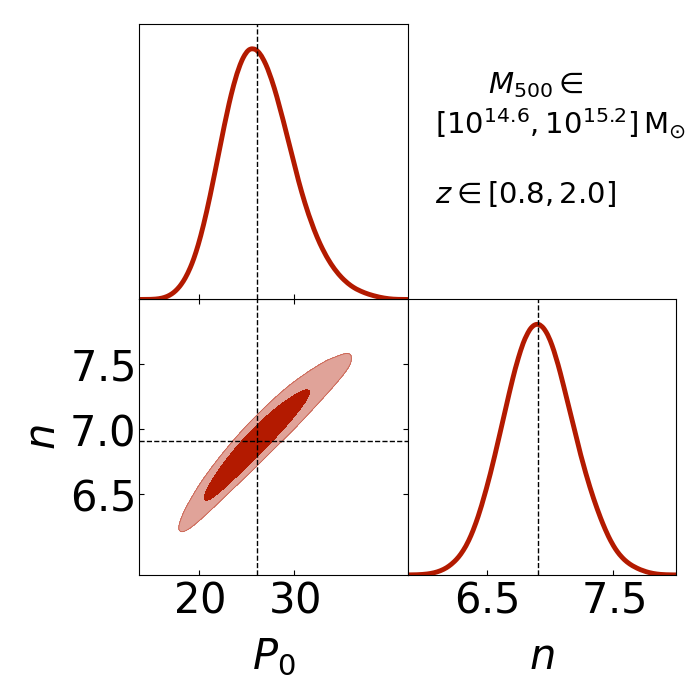}
\includegraphics[trim= 0mm 0mm 0mm 0mm, scale=0.24]{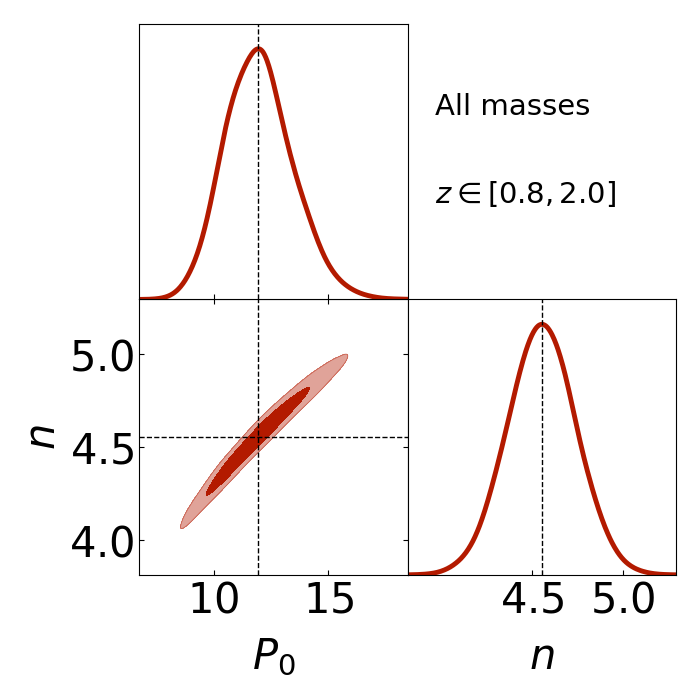} \\
\includegraphics[trim= 0mm 0mm 0mm 0mm, scale=0.24]{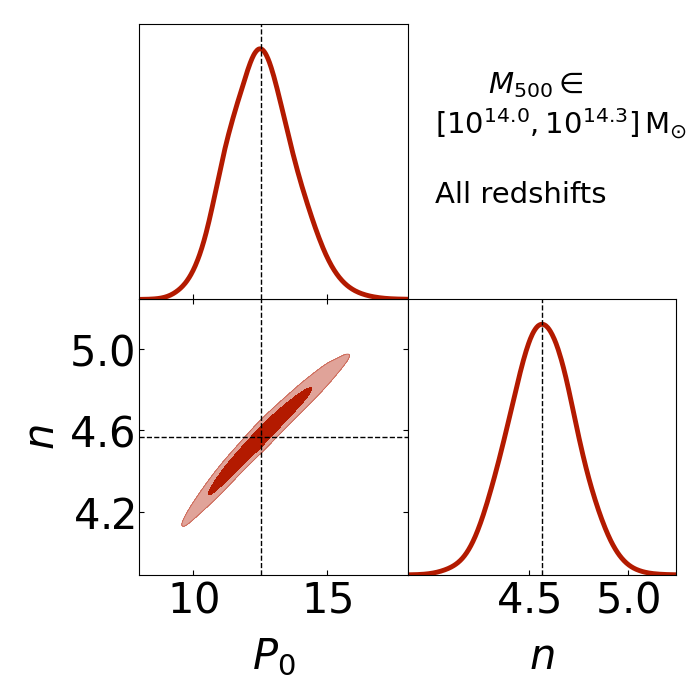}
\includegraphics[trim= 0mm 0mm 0mm 0mm, scale=0.24]{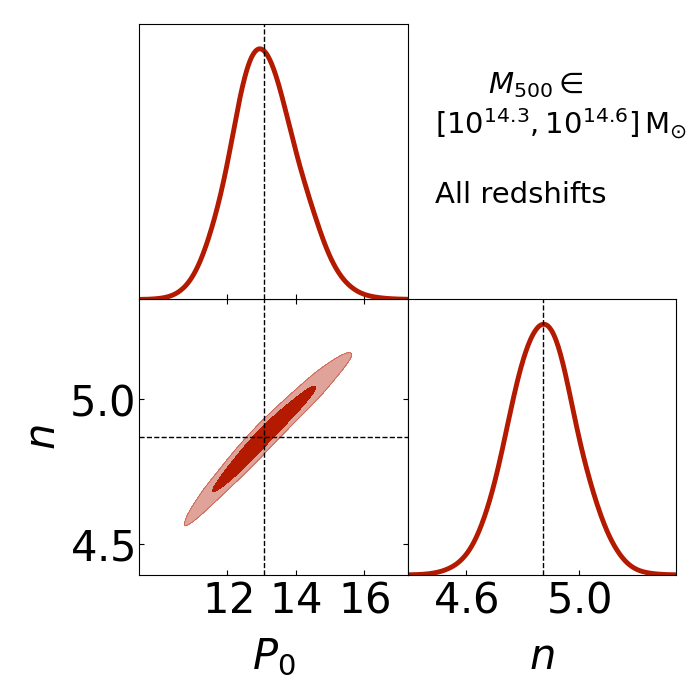}
\includegraphics[trim= 0mm 0mm 0mm 0mm, scale=0.24]{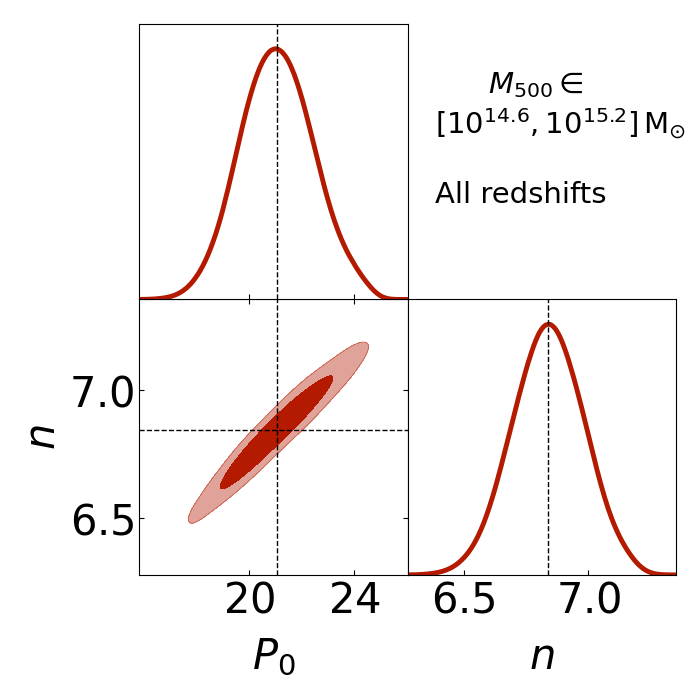}
\includegraphics[trim= 0mm 0mm 0mm 0mm, scale=0.24]{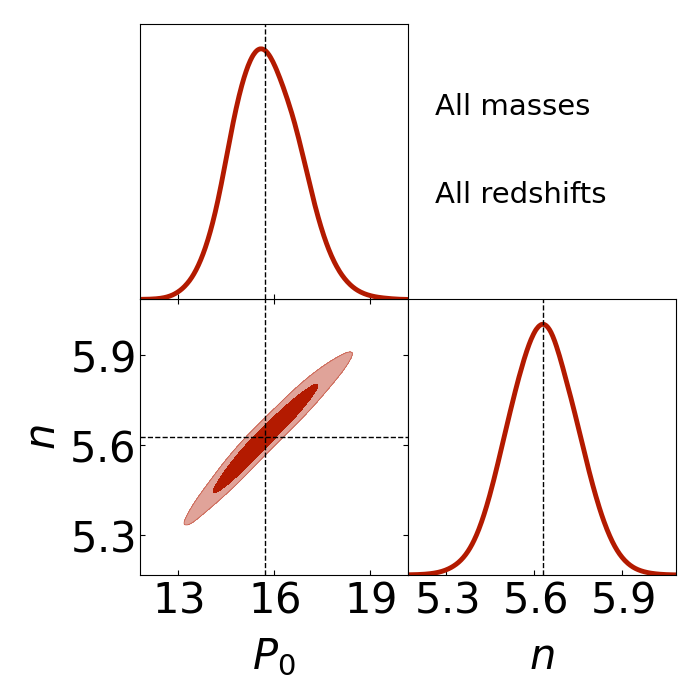} \\
\caption{Posterior contour plots for the parameters entering the PTP model; 
	the dashed lines mark the best-fit values quoted in Table~\ref{tab:ptp_bestfits}.}
\label{fig:ptp_contours}
\end{figure*}
\begin{figure*}
\includegraphics[trim= 0mm 0mm 0mm 0mm, scale=0.24]{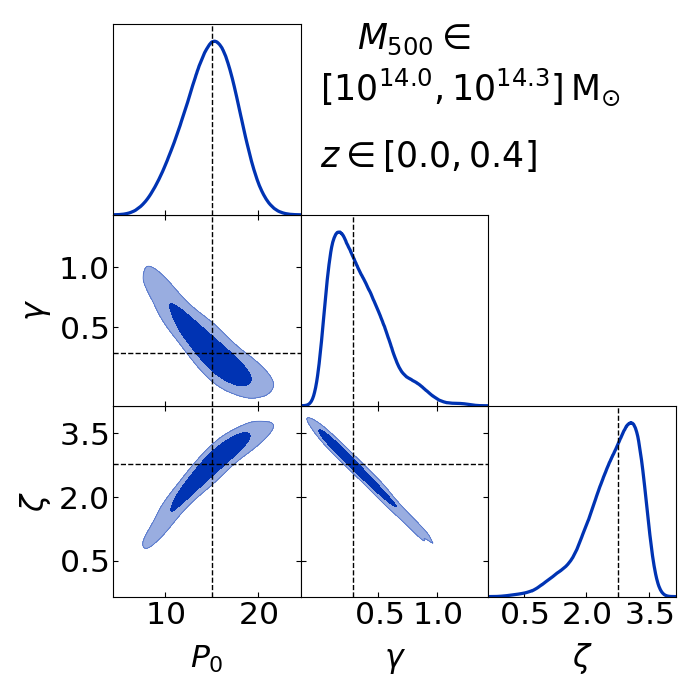}
\includegraphics[trim= 0mm 0mm 0mm 0mm, scale=0.24]{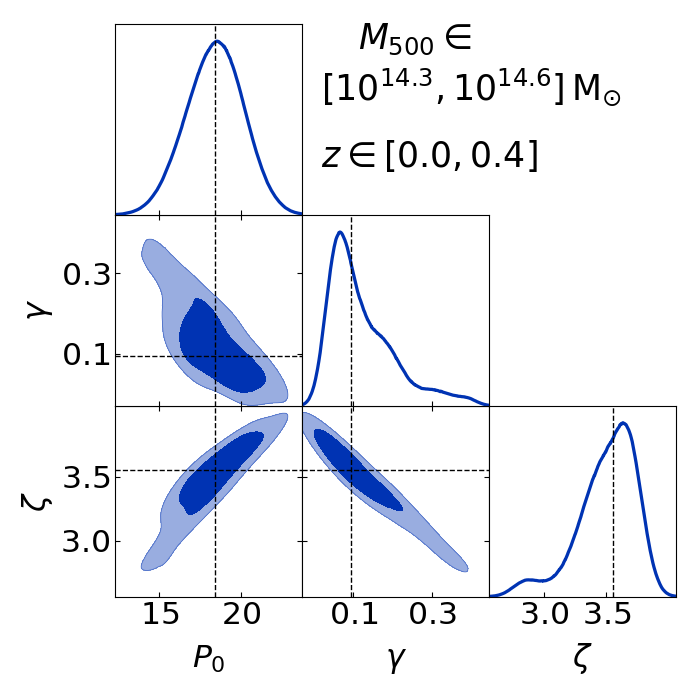}
\includegraphics[trim= 0mm 0mm 0mm 0mm, scale=0.24]{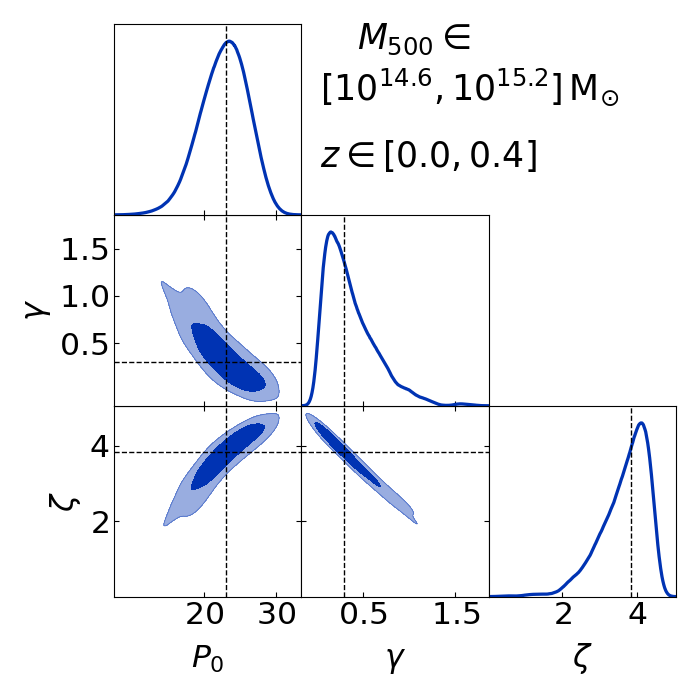}
\includegraphics[trim= 0mm 0mm 0mm 0mm, scale=0.24]{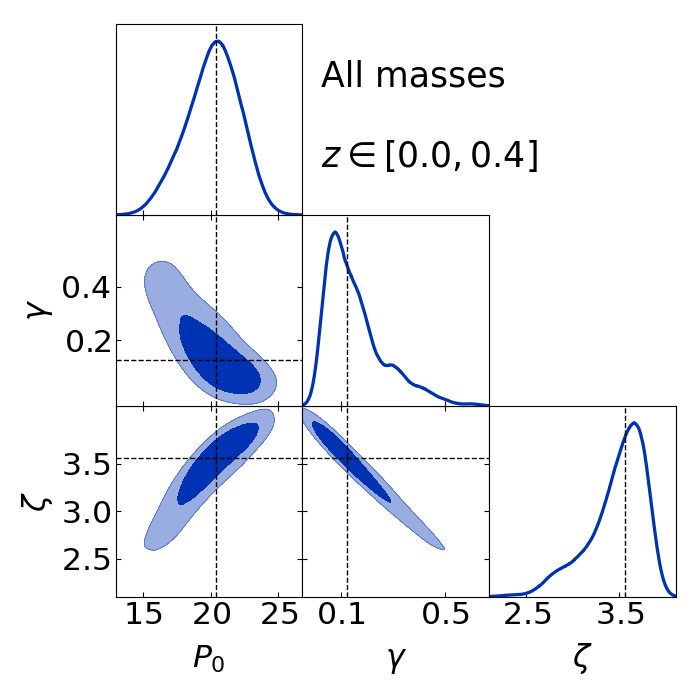} \\
\includegraphics[trim= 0mm 0mm 0mm 0mm, scale=0.24]{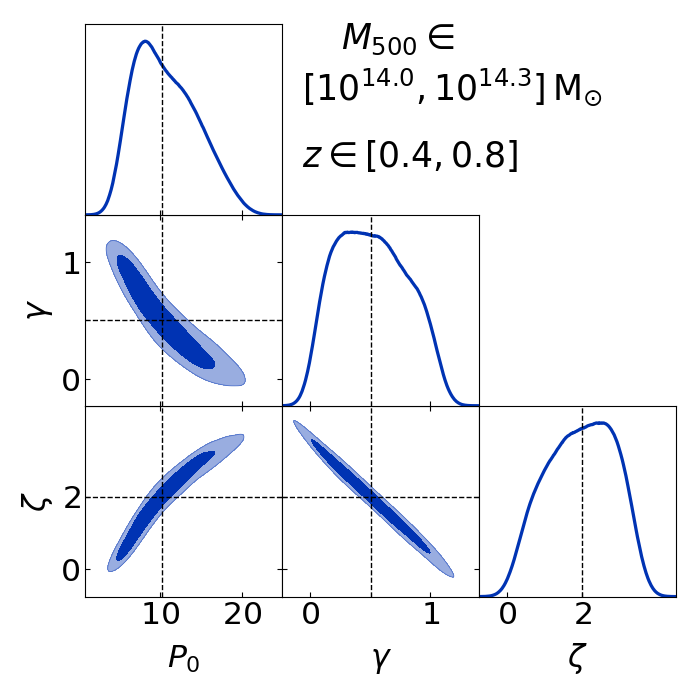}
\includegraphics[trim= 0mm 0mm 0mm 0mm, scale=0.24]{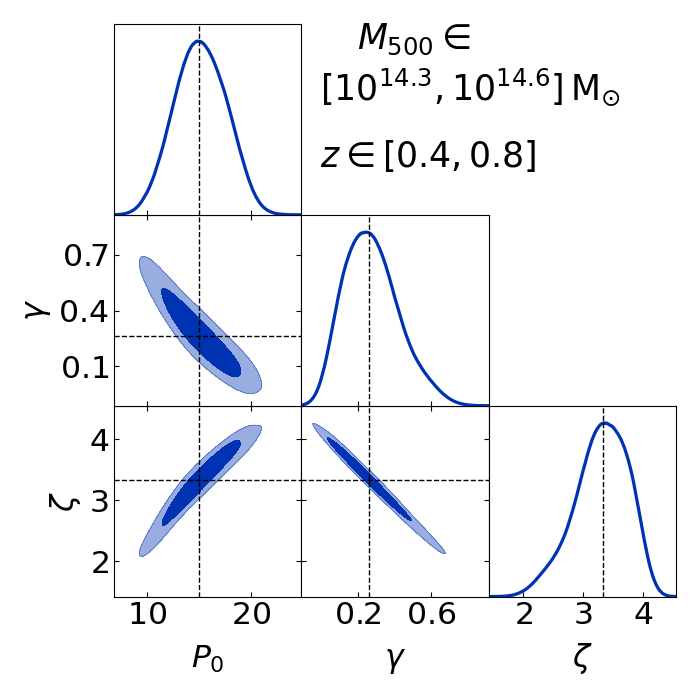}
\includegraphics[trim= 0mm 0mm 0mm 0mm, scale=0.24]{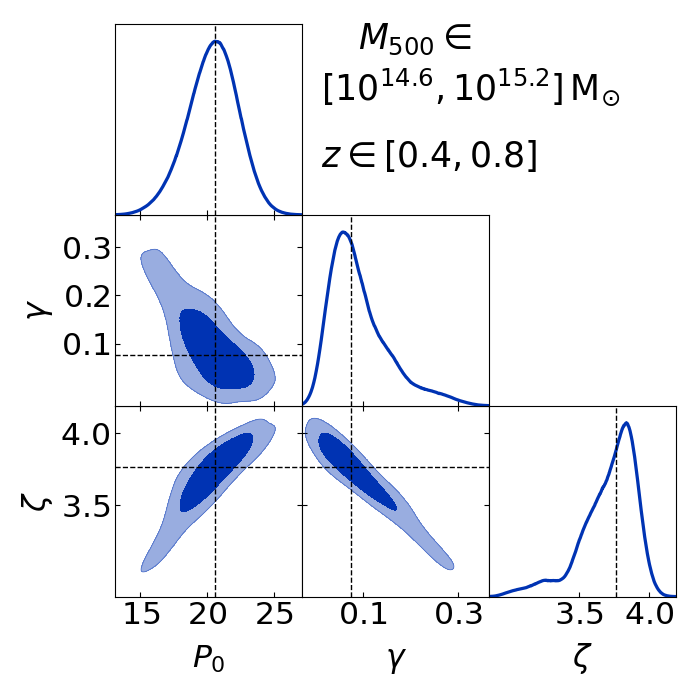}
\includegraphics[trim= 0mm 0mm 0mm 0mm, scale=0.24]{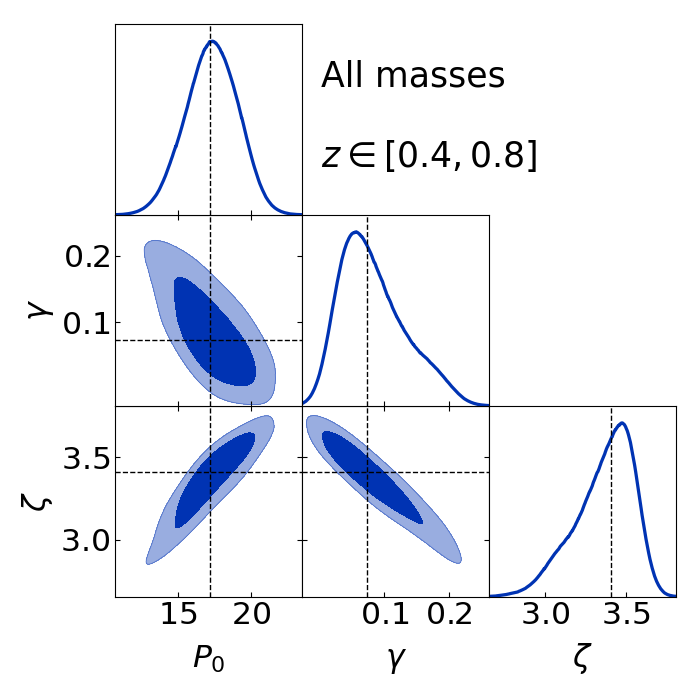} \\
\includegraphics[trim= 0mm 0mm 0mm 0mm, scale=0.24]{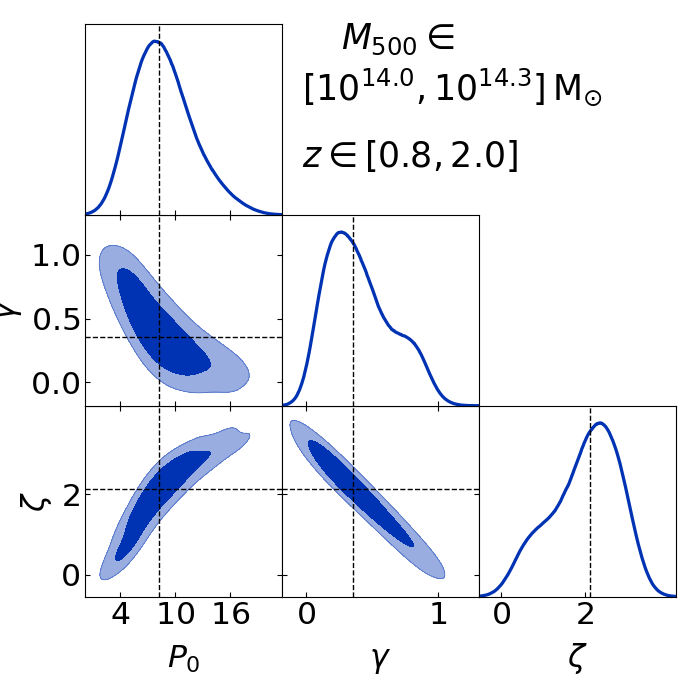}
\includegraphics[trim= 0mm 0mm 0mm 0mm, scale=0.24]{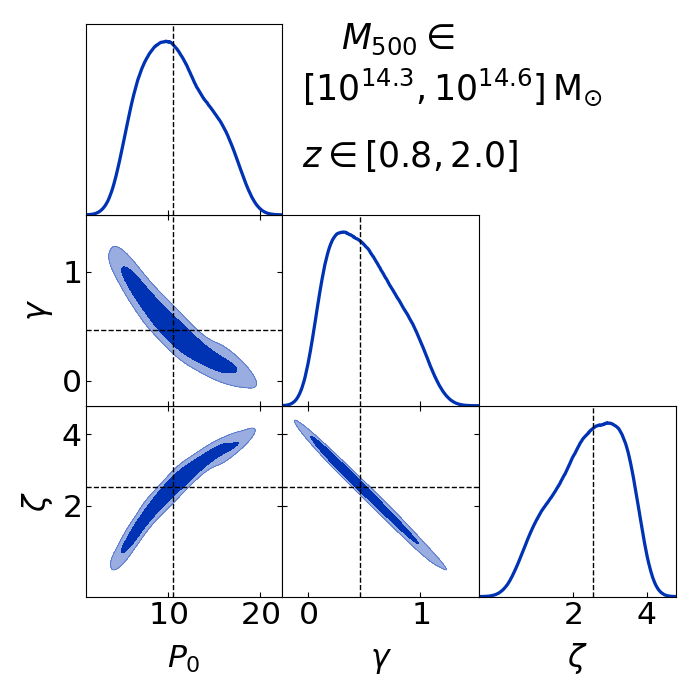}
\includegraphics[trim= 0mm 0mm 0mm 0mm, scale=0.24]{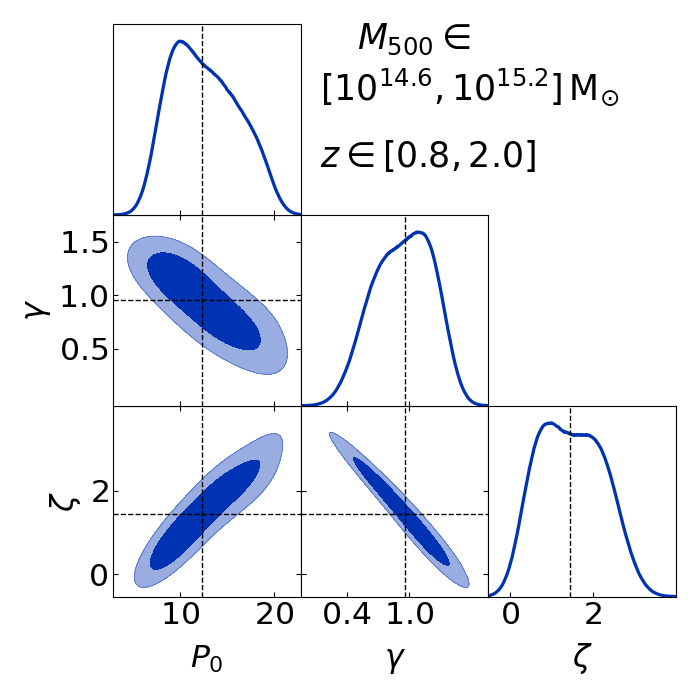}
\includegraphics[trim= 0mm 0mm 0mm 0mm, scale=0.24]{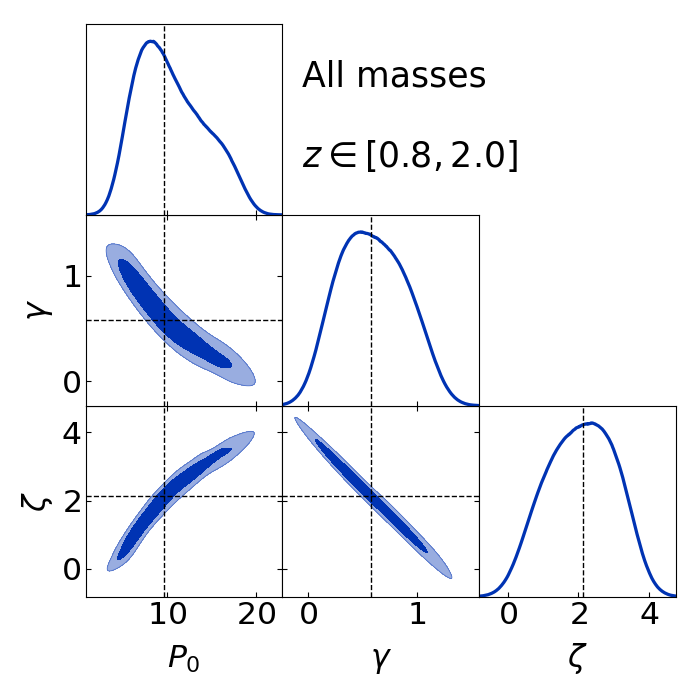} \\
\includegraphics[trim= 0mm 0mm 0mm 0mm, scale=0.24]{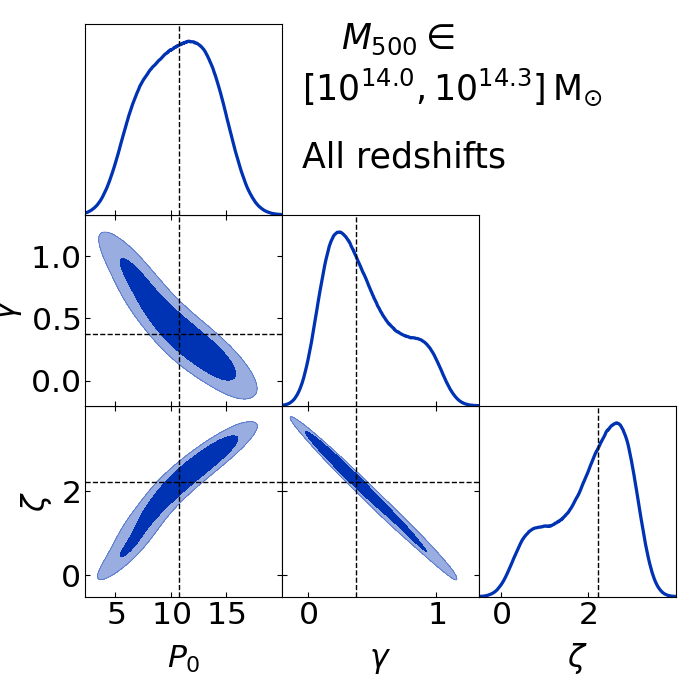}
\includegraphics[trim= 0mm 0mm 0mm 0mm, scale=0.24]{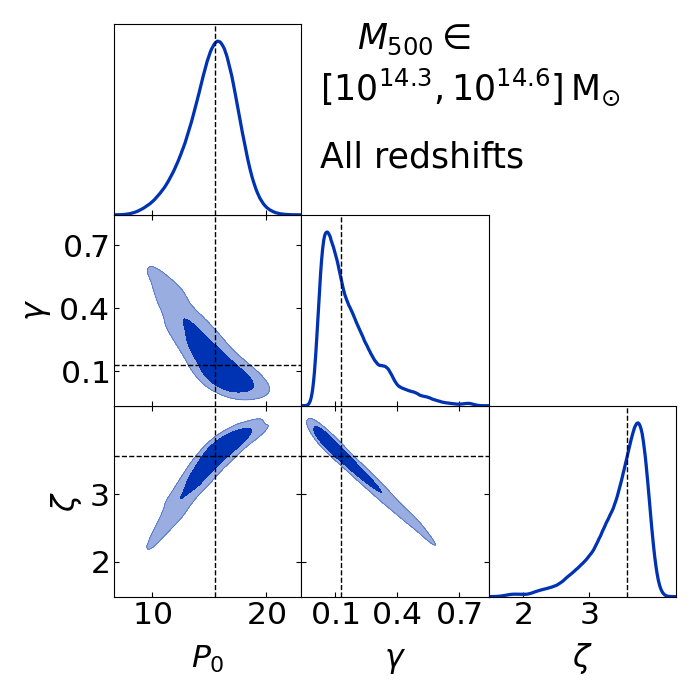}
\includegraphics[trim= 0mm 0mm 0mm 0mm, scale=0.24]{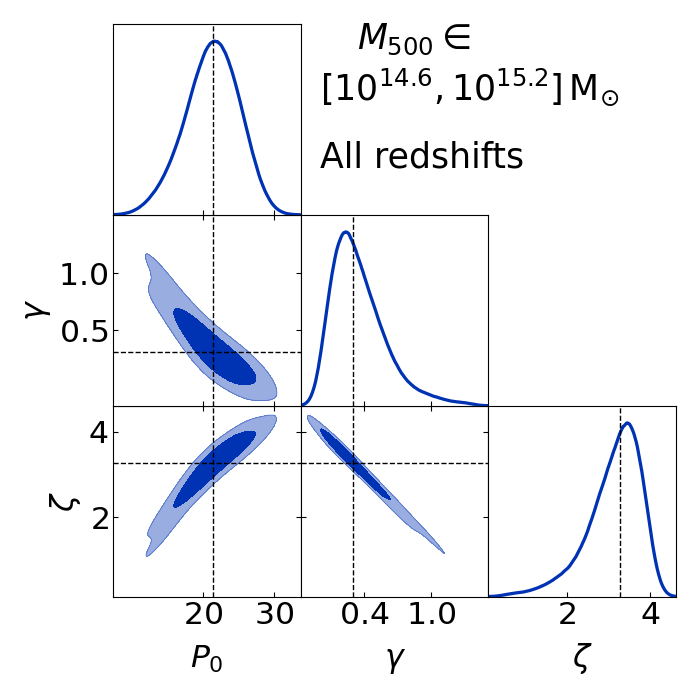}
\includegraphics[trim= 0mm 0mm 0mm 0mm, scale=0.24]{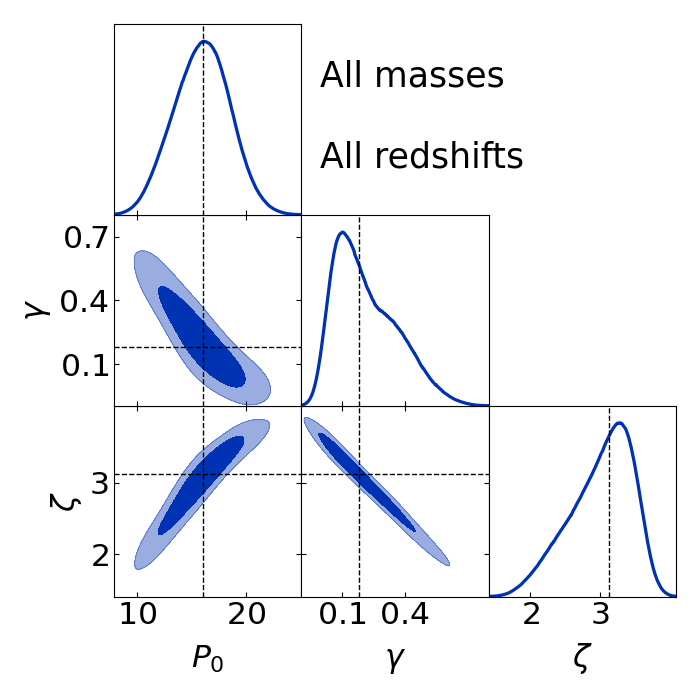} \\
\caption{Posterior contour plots for the parameters entering the EUP model; 
	the dashed lines mark the best-fit values quoted in Table~\ref{tab:eup_bestfits}.}
\label{fig:eup_contours}
\end{figure*}
\end{document}